\documentclass[letterpaper,12pt]{article}

\usepackage[left=1in,right=1in,top=1in,bottom=1in]{geometry}

\usepackage{setspace}
\usepackage[usenames]{color}
\usepackage[fleqn]{amsmath}
\usepackage{amssymb}
\usepackage{graphicx}
\usepackage{url}
\usepackage{verbatim}
\usepackage{appendix}
\usepackage{indentfirst}
\usepackage{booktabs}
\usepackage{multirow}
\usepackage[table, x11names]{xcolor}
\usepackage{ragged2e}
\usepackage{upgreek}
\usepackage{lscape}
\usepackage{longtable}
\usepackage[flushleft, referable]{threeparttablex}
\usepackage{rotating}
\usepackage[T1]{fontenc}
\usepackage[titles]{tocloft}
\usepackage{xspace}
\usepackage{ifthen}
\usepackage{cancel}
\usepackage{rotating}
\usepackage{array}
\usepackage{tabulary}
\usepackage{authblk}

\usepackage{hyperref}
\hypersetup{pdfborder={0 0 0}, colorlinks=true, urlcolor=black, linkcolor=black, citecolor=black}
\usepackage[capitalize]{cleveref}

\usepackage[right, mathlines]{lineno}
\usepackage{caption}
\DeclareCaptionLabelFormat{noSpace}{{#1}{#2}}
\DeclareCaptionListFormat{figList}{Figure {#2}.}
\DeclareCaptionListFormat{sFigList}{Figure S{#2}.}
\usepackage{subfig}

\usepackage[numbers,sort&compress]{natbib}
\input{macros.tex}

\title{\dppTitle}

\author[1]{Jamie R.\ Oaks\thanks{Corresponding author: \href{mailto:joaks1@gmail.com}{\tt joaks1@gmail.com}}\thanks{Current address: Department of Biology, University of Washington, Seattle, Washington 98195}}
\affil[1]{Department of Ecology and Evolutionary Biology, University of Kansas, Lawrence, Kansas 66045}

\date{\parbox{\linewidth}{\centering%
    \today\endgraf\bigskip
    \textbf{Running head}: Improved Estimation of Shared Evolutionary History}}

\makeatletter
\let\msTitle\@title
\let\msAuthor\@author
\let\msDate\@date
\makeatother

\begin{document}

\maketitle

\begin{abstract}
    \ifbmc{\parttitle{Background}}{}
To understand biological diversification, it is important to account for
large-scale processes that affect the evolutionary history of groups of
co-distributed populations of organisms.
Such events predict temporally clustered divergences times, a pattern
that can be estimated using genetic data from co-distributed species.
I introduce a new approximate-Bayesian method for comparative
phylogeographical model-choice that estimates the temporal distribution of
divergences across taxa from multi-locus DNA sequence data.
The model is an extension of that implemented in \msb.
\ifbmc{\parttitle{Results}}{}
By reparameterizing the model, introducing more flexible priors on
demographic and divergence-time parameters, and implementing a
non-parametric Dirichlet-process prior over divergence models, I improved
the robustness, accuracy, and power of the method for estimating shared
evolutionary history across taxa.
\ifbmc{\parttitle{Conclusions}}{}
The results demonstrate the improved performance of the new method is due
to (1) more appropriate priors on divergence-time and demographic
parameters that avoid prohibitively small marginal likelihoods for models
with more divergence events,
and (2) the Dirichlet-process providing a flexible prior on divergence
histories that does not strongly disfavor models with intermediate numbers
of divergence events.
The new method yields more robust estimates of posterior uncertainty, and
thus greatly reduces the tendency to incorrectly estimate
\change{models of shared evolutionary history} with strong support.

    \vspace{12pt}
\end{abstract}

\newpage

\section{Background}

Understanding the processes that generate biodiversity and regulate
community assembly is a major goal of evolutionary biology.
Large-scale changes to the environment, including geological and climatic
events, can affect the evolutionary history of entire communities of
co-distributed species and their associated microbiota.
For example, by partitioning communities, such an event can isolate groups of
populations and cause a temporal cluster of speciation events across
co-distributed taxa.
Given the dynamic nature of our planet, such biogeographical processes likely
play a significant role in determining diversification rates and patterns.
At recent timescales, temporal clusters of diversification caused by
biogeographical events can leave a signature in the genetic variation within
and among the affected lineages.
Thus, methods for accurately estimating models of shared evolutionary events
across co-distributed taxa from genetic data are important for better
understanding how regional and global biogeographical processes affect
biodiversity.

This inference problem is challenging due to the stochastic nature by which
mutations occur in populations and how they are inherited over generations
\cite{Hudson1990,WakeleyCoalescent}.
Thus, a method for estimating historical patterns of divergences across taxa
should explicitly model the stochastic mutational and ancestral processes that
generate and filter the genetic variation we observe in present-day genetic
data.
An appealing approach would be a comparative, Bayesian model-choice method for
inferring the probability of competing divergence histories while integrating
over uncertainty in mutational and ancestral processes via models of nucleotide
substitution and lineage coalescence.
\change{
The sample space of such a model-choice procedure would include all models
ranging from a single divergence-time parameter (i.e., simultaneous divergence
of all co-distributed taxa) to the fully generalized model in which each taxon
diverged at a unique time.}

The software package \msb implements such an approach in an
approximate-Bayesian model-choice framework \cite{Hickerson2006,Huang2011}.
The method models temporally clustered divergences across taxa caused by a
biogeographical event (or a ``divergence event'') as a single, instantaneous
occurrence.
In other words, a divergence event causes a set of taxa to share the same
moment of divergence along a continuous time scale (i.e., simultaneous
divergence).
Given aligned sequence data for \npairs{} pairs of populations, \msb 
estimates the number of divergence events shared among the pairs, the timing of
the events, and the assignment of pairs to the events, while integrating out
uncertainty in demographic parameters and the genealogical histories of the
sequences.
Thus, the method samples over all possible divergence models of differing
dimensionality (i.e., all the possible partitions of \npairs{} pairs to
$1,2,\ldots,\npairs{}$ divergence-time parameters), and, in so doing, estimates
the posterior probability of each model.

\change{
\msb has been used to address biogeographical questions in a variety of
empirical systems. Some examples include
(1) whether the rise of the Isthmus of Panama caused co-divergence among
species of echinoids co-distributed across the Pacific and Atlantic sides of
the isthmus \cite{Hickerson2006}, 
(2) if an historical seaway across the Baja Peninsula caused co-divergence
across species of squamates and mammals co-distributed both north and south of
the putative seaway \cite{Leache2007},
(3) if species of gall-wasps and their associated parasitoids share divergences
across putative glacial refugia \cite{Stone2012}, and
(4) whether repeated fragmentation of the oceanic Islands of the Philippines
during Pleistocene sea-level fluctuations caused diversification of vertebrate
taxa distributed across the islands \cite{Oaks2012}.
Such applications of the method often result in strong posterior support for
co-divergence among all or subsets of the taxa investigated (e.g.,
\cite{Barber2010,Carnaval2009,Chan2011,Hickerson2006,Leache2007,Plouviez2009,Stone2012,Voje2009,Oaks2012}).
}

For priors on divergence-time and demographic parameters, \msb uses continuous
uniform probability distributions.
\change{
This causes divergence models with more divergence-time parameters to integrate
over a \emph{much} greater parameter space with low likelihood yet high prior
density, which can result in small marginal likelihoods relative to models with
fewer divergence-time parameters \cite{Jeffreys1939,Lindley1957}.
Given that the marginal likelihood of a model weighted by its prior is what
determines its posterior probability, this can cause support for models with
fewer divergence events \cite{Oaks2012,Oaks2014reply}.
This is not a critique of Bayesian model choice in general; comparing models by
their marginal likelihoods provides a ``natural'' penalty for
over-parameterization and can be a great strength of the Bayesian approach.
However, given the sensitivity of marginal likelihoods to the prior, care is
needed when selecting prior distributions \cite{Lindley1957}.
Selecting distributions that will often place high prior density in large
regions of parameter space with low likelihood can lead to small marginal
likelihoods of parameter-rich models even if they are correct.
}

Furthermore, \msb uses a discrete uniform prior over the number of divergence
events $1,2,\ldots,\npairs{}$.
Because there are many more possible assignments of population pairs to
intermediate numbers of divergence events, this imposes a prior on divergence
models that puts most of the prior mass on models with either very few or very
many divergence-time parameters (see Figure 5 of \cite{Oaks2012}; for brevity
I will refer to this prior as ``U-shaped'').
Given that models with many divergence events can have small marginal
likelihoods due to the uniform priors on divergence-time parameters, the
U-shaped prior will effectively create a strong prior preference for models
with very few divergence events.

\change{
Recently, Oaks et al.\ \cite{Oaks2012,Oaks2014reply} found via simulation that
\msb will often strongly support models with a small number of divergence
events shared among taxa, even when divergences were random over broad
timescales.}
They suggested this behavior was due to the combination of uniform priors on
parameters causing small marginal likelihoods of richer models and the
U-shaped prior on divergence models.
Hickerson et al.\ \cite{Hickerson2013} suggested the problem was caused by
sampling error, and proposed as a solution an approximate-Bayesian model
averaging approach that samples over empirically informed uniform priors.
However, Oaks et al.\ \cite{Oaks2014reply} evaluated the approach proposed by
Hickerson et al.\ \cite{Hickerson2013} using simulations and found that it did
not mitigate the method's propensity to incorrectly infer clustered
divergences, and often preferred priors that excluded the true values of the
model's parameters.
\change{
Here, I describe a new approach that successfully mitigates spurious inference
of co-divergence while avoiding negative side effects of empirically informed
uniform priors.}

In this study, I introduce a new method, implemented in the software
\dppmsbayes, that extends the model of \msb.
I use this method to test whether alternative parameterizations and priors
improve the behavior of the approximate-Bayesian model-choice approach to
estimating shared divergence events.
The new approach uses a Dirichlet-process prior (DPP) over all possible models
of divergence, and gamma and beta probability distributions in place of uniform
priors on many of the model's parameters.
Using simulations, I show that the new implementation has improved robustness,
accuracy, and power compared to the original model.
The results confirm that the improved performance of the new model
is due to a combination of
(1) more flexible priors on divergence-time and demographic parameters that
avoid placing high prior density in improbable regions of parameter space, and
(2) a diffuse Dirichlet-process prior that does not strongly disfavor
divergence models with intermediate numbers of divergence events.
After reanalyzing sequence data from 22 pairs of taxa from the Philippines
\cite{Oaks2012} under the new model, I find a large amount of posterior
uncertainty in the number of divergence events shared among the taxa; a result
in contrast with the original \msb model and congruent with
intuition given the richness of the model and the relatively small amount of
information in the data.

\section{Methods}

\subsection{The model}
In this section, I describe the model, which is a modification of the model
implemented in \msb \cite{Huang2011,Oaks2012}.
The code implementing the new model is freely available in the open-source
software package \dppmsbayes
(\href{https://github.com/joaks1/dpp-msbayes}{\url{https://github.com/joaks1/dpp-msbayes}}).
To perform the analyses described below, I used the freely avaliable,
open-source software package \pymsbayes
(\href{https://github.com/joaks1/PyMsBayes}{\url{https://github.com/joaks1/PyMsBayes}}),
which provides a multi-processing interface to \msb and
\dppmsbayes.
I performed the work described below following the principles of
Open Notebook Science.
Using version-control software, I make progress in all aspects of the work
freely and publicly available in real-time at
\href{https://github.com/joaks1/msbayes-experiments}{\url{https://github.com/joaks1/msbayes-experiments}}.
All information necessary to reproduce my results is provided there.
I follow much of the notation of Oaks et al.\ \cite{Oaks2012}, but modify it to
aid in the description of the new model.
A summary of my notation can be found in Table \ref{tabNotation}.

I assume an investigator is interested in inferring the distribution
of divergence times among \npairs{} pairs of populations.
For each pair $i$, \popSampleSize{i}{} genome copies have been sampled, with
\popSampleSize{1}{i} copies sampled from population 1, and \popSampleSize{2}{i}
sampled from population 2.
From these genomes, let \nloci{i} be the number of DNA sequence loci collected
for population pair $i$, and \nlociTotal be the total number of unique loci
sampled across the \npairs{} pairs of populations.
I use \alignment{i}{j} to represent the multiple sequence alignment of
locus $j$ for population pair $i$.
$\alignmentVector = (\alignment{1}{1}, \ldots,
    \alignment{\npairs{}}{\nloci{\npairs{}}})$
is the full dataset,
i.e., a vector of sequence alignments for all pairs and loci.
Let \geneTree{i}{j} represent the gene tree upon which \alignment{i}{j}
evolved according to fixed \hky substitution model parameters \hkyModel{i}{j}.
The investigator must specify the parameters of all
$\hkyModelVector = (\hkyModel{1}{1}, \ldots,
\hkyModel{\npairs{}}{\nloci{\npairs{}}})$
substitution models by which the alignments evolved along the
$\geneTreeVector = (\geneTree{1}{1}, \ldots,
\geneTree{\npairs{}}{\nloci{\npairs{}}})$
gene trees.
Furthermore, the investigator must specify a vector of fixed constants
$\ploidyScalarVector = (\ploidyScalar{1}{1}, \ldots,
\ploidyScalar{\npairs{}}{\nloci{\npairs{}}})$
that scale the population-size parameters for known differences in ploidy among
loci and/or differences in generation times among population pairs.  Lastly,
the investigator must also specify a vector of fixed constants
$\mutationRateScalarConstantVector = (\mutationRateScalarConstant{1}{1},
\ldots, \mutationRateScalarConstant{\npairs{}}{\nloci{\npairs{}}})$
that scale the population-size parameters for known differences in
mutation rates among loci and/or among taxa.

\begin{linenomath}
With \alignmentVector, \hkyModelVector, \ploidyScalarVector, and
\mutationRateScalarConstantVector in hand, the joint posterior distribution
of the model is given by Bayes' rule as
\begin{equation}
    p(\geneTreeVector, \divTimeMapVector, \demographicParamVector, 
    \locusMutationRateScalarVector, \locusRateHetShapeParameter \given
    \alignmentVector, \hkyModelVector, \ploidyScalarVector,
    \mutationRateScalarConstantVector) =
    \frac{p(\alignmentVector \given \geneTreeVector, \divTimeMapVector,
        \demographicParamVector, \locusMutationRateScalarVector,
        \locusRateHetShapeParameter, \hkyModelVector, \ploidyScalarVector,
        \mutationRateScalarConstantVector)
        p(\geneTreeVector, \divTimeMapVector, \demographicParamVector,
        \locusMutationRateScalarVector, \locusRateHetShapeParameter \given
        \hkyModelVector, \ploidyScalarVector,
        \mutationRateScalarConstantVector)}{
        p(\alignmentVector \given \hkyModelVector, \ploidyScalarVector,
        \mutationRateScalarConstantVector)},
    \label{eq:fullModelCompactJoint}
\end{equation}
which can be expanded using the chain rule of probability into components that
are assumed to be independent to get
\begin{equation}
    p(\geneTreeVector, \divTimeMapVector, \demographicParamVector, 
    \locusMutationRateScalarVector, \locusRateHetShapeParameter \given
    \alignmentVector, \hkyModelVector, \ploidyScalarVector,
    \mutationRateScalarConstantVector) =
    \frac{p(\alignmentVector \given \geneTreeVector, \hkyModelVector)
        p(\geneTreeVector \given \divTimeMapVector, \demographicParamVector,
        \locusMutationRateScalarVector, \ploidyScalarVector,
        \mutationRateScalarConstantVector)
        p(\locusMutationRateScalarVector \given \locusRateHetShapeParameter)
        p(\locusRateHetShapeParameter)
        p(\divTimeMapVector)
        p(\demographicParamVector)}{
        p(\alignmentVector \given \hkyModelVector, \ploidyScalarVector,
        \mutationRateScalarConstantVector)},
    \label{eq:fullModelCompact}
\end{equation}
where
$\divTimeMapVector = (\divTimeMap{1}, \ldots, \divTimeMap{\npairs{}})$
is a vector of population divergence times for each of the \npairs{} pairs of
populations,
$\demographicParamVector = (\demographicParams{1}, \ldots,
\demographicParams{\npairs{}})$
is a vector of the demographic parameters for each of the \npairs{} population
pairs,
$\locusMutationRateScalarVector = (\locusMutationRateScalar{1}, \ldots
\locusMutationRateScalar{\nlociTotal})$
is a vector of locus-specific mutation-rate multipliers for each of the
\nlociTotal loci,
\locusRateHetShapeParameter is the shape parameter of a gamma-distributed
prior on \locusMutationRateScalar{}, and
$p(\alignmentVector \given \hkyModelVector, \ploidyScalarVector,
\mutationRateScalarConstantVector)$
is the probability of the data (or the marginal likelihood of the model) given
the fixed constants provided by the investigator.
\end{linenomath}

\begin{linenomath}
To avoid calculating the likelihood terms of Equation \ref{eq:fullModelCompact},
I distill each sequence alignment \alignment{}{} into a vector of insufficient
summary statistics \alignmentSS{}{}, thus replacing the full dataset
$\alignmentVector = (\alignment{1}{1}, \ldots,
    \alignment{\npairs{}}{\nloci{\npairs{}}})$
with vectors of summary statistics for each alignment
$\ssVectorObs = (\alignmentSSObs{1}{1}, \ldots,
    \alignmentSSObs{\npairs{}}{\nloci{\npairs{}}})$.
\change{
Optionally, for each population pair, the means of the summary statistics can
be calculated across the \nloci{} loci, and the vector can be further reduced
to $\ssVectorObs = (\alignmentSSObs{1}{}, \ldots,
\alignmentSSObs{\npairs{}}{})$.
With \ssVectorObs in hand, we can estimate the approximate joint
posterior distribution}
{\small
\begin{equation}
    p(\geneTreeVector, \divTimeMapVector, \demographicParamVector, 
    \locusMutationRateScalarVector, \locusRateHetShapeParameter \given
    \ssSpace, \hkyModelVector, \ploidyScalarVector,
    \mutationRateScalarConstantVector) =
    \frac{p(\ssSpace \given \geneTreeVector, \hkyModelVector)
        p(\geneTreeVector \given \divTimeMapVector, \demographicParamVector,
        \locusMutationRateScalarVector, \ploidyScalarVector,
        \mutationRateScalarConstantVector)
        p(\locusMutationRateScalarVector \given \locusRateHetShapeParameter)
        p(\locusRateHetShapeParameter)
        p(\divTimeMapVector)
        p(\demographicParamVector)}{
        p(\ssSpace \given \hkyModelVector, \ploidyScalarVector,
        \mutationRateScalarConstantVector)},
    \label{eq:approxModelCompact}
\end{equation}
}
where \ssSpace is the multidimensional Euclidean space around the vector of
summary statistics, the radius of which is the tolerance \tol.
The sources of approximation are the insufficiency of the statistics and the
\tol being greater than zero.
I describe the full model in detail before delving into the numerical
method of estimating the approximate model.
\end{linenomath}

\change{
\subsubsection{Likelihood and gene-tree prior terms of Equation \ref{eq:fullModelCompact}}
}
\begin{linenomath}
\change{
The likelihood and gene-tree prior terms of Equation \ref{eq:fullModelCompact}
}
can be expanded out as a product over population pairs and loci
\begin{equation}
    p(\alignmentVector \given \geneTreeVector, \hkyModelVector)
    p(\geneTreeVector \given \divTimeMapVector, \demographicParamVector,
    \locusMutationRateScalarVector, \ploidyScalarVector,
    \mutationRateScalarConstantVector) = 
    \prod_{i=1}^{\npairs{}}
    \prod_{j=1}^{\nloci{i}}
    p(\alignment{i}{j} \given \geneTree{i}{j}, \hkyModel{i}{j})
    p(\geneTree{i}{j} \given \divTimeMap{i}, \demographicParams{i},
    \locusMutationRateScalar{j}, \ploidyScalar{i}{j},
    \mutationRateScalarConstant{i}{j}).
    \label{eq:modelLikelihoodExpanded}
\end{equation}
The first term,
$p(\alignment{i}{j} \given \geneTree{i}{j}, \hkyModel{i}{j})$, is the
probability of the sequence alignment of locus $j$ for population pair $i$
given the gene tree and \hky \cite{HKY} substitution model parameters
\cite[i.e., the ``Felsenstein likelihood'']{Felsenstein1981}.
The model allows for an intra-locus recombination rate \recombinationRate,
which, for simplicity, is assumed to be zero in Equation
\ref{eq:fullModelCompact}.
If \recombinationRate is non-zero, this term requires an additional product
over the columns (sites) of each sequence alignment to allow sites to have
different genealogies.
The second term,
p(\geneTree{i}{j} \given \divTimeMap{i}, \demographicParams{i},
\locusMutationRateScalar{j}, \ploidyScalar{i}{j},
\mutationRateScalarConstant{i}{j}),
is the probability of the gene tree under a multi-population coalescent model
(i.e., species tree) where the ancestral population of pair $i$ diverges
and gives rise to the two sampled descendant populations.
Each \demographicParams{} contains the following demographic parameters: The
mutation-rate-scaled effective sizes ($\myTheta{} = 4N\mutationRate$) of the
ancestral, \ancestralTheta{}, and descendant populations, \descendantTheta{1}{}
and \descendantTheta{2}{};
the proportion of the first, \bottleScalar{1}{}, and second population,
\bottleScalar{2}{}, that persist during bottlenecks that begin immediately
after divergence in forward-time;
the proportion of time between present and divergence when the bottlenecks
end for both populations, \bottleTime{};
and the symmetric migration rate between the descendant populations,
\migrationRate{}.
Thus, the probability of the $\popSampleSize{i}{}-1$ coalescence times (node
heights) of gene tree \geneTree{i}{j} is given by a multi-population
Kingman-coalescent model \cite{Kingman1982} where the ancestral population of
size
$\ancestralTheta{i}\ploidyScalar{i}{j}\mutationRateScalarConstant{i}{j}\locusMutationRateScalar{j}$
diverges at time \divTimeMap{i} into two descendant populations of constant
size
$\descendantTheta{1}{i}\ploidyScalar{i}{j}\mutationRateScalarConstant{i}{j}\locusMutationRateScalar{j}\bottleScalar{1}{i}$
and
$\descendantTheta{2}{i}\ploidyScalar{i}{j}\mutationRateScalarConstant{i}{j}\locusMutationRateScalar{j}\bottleScalar{2}{i}$,
which, after time $\divTimeMap{i}\bottleTime{i}$, grow exponentially to their
present size 
$\descendantTheta{1}{i}\ploidyScalar{i}{j}\mutationRateScalarConstant{i}{j}\locusMutationRateScalar{j}$
and
$\descendantTheta{2}{i}\ploidyScalar{i}{j}\mutationRateScalarConstant{i}{j}\locusMutationRateScalar{j}$,
respectively.
Following divergence, the descendant populations of pair $i$ exchange migrants at a symmetric rate of
\migrationRate{i}.
\end{linenomath}

\change{
\subsubsection{Additional prior terms of Equation \ref{eq:fullModelCompact}}
}
\begin{linenomath}
The term $p(\locusRateHetShapeParameter)$ is the prior density
function for the shape parameter of the gamma-distributed prior on
rate heterogeneity among loci.
This prior is $\locusRateHetShapeParameter \sim U(1, 20)$.
The prior probability of the vector of locus-specific mutation-rate multipliers
given \locusRateHetShapeParameter then expands out as a product over
the loci
\begin{equation}
    p(\locusMutationRateScalarVector \given \locusRateHetShapeParameter) =
    \prod_{j=1}^{\nlociTotal}
    p(\locusMutationRateScalar{j} \given \locusRateHetShapeParameter),
    \label{eq:locusRateHetPrior}
\end{equation}
where each \locusMutationRateScalar{} is independently and identically
distributed (\iid) as
$\locusMutationRateScalar{} \sim Gamma(\locusRateHetShapeParameter,
1/\locusRateHetShapeParameter)$.
If the recombination rate \recombinationRate is allowed to be non-zero, the
prior term $p(\recombinationRate)$ would be added to Equation
\ref{eq:fullModelCompact}, and the prior would be $\recombinationRate \sim
Gamma(\gammaShape{\recombinationRate}, \gammaScale{\recombinationRate})$,
\change{where \gammaShape{\recombinationRate} and
\gammaScale{\recombinationRate} are specified by the investigator.}
\end{linenomath}

\begin{linenomath}
The prior term for the demographic parameters, $p(\demographicParamVector)$,
expands out into its components and as a product over the \npairs{}
pairs of populations
\begin{equation}
    p(\demographicParamVector) =
    \prod_{i=1}^{\npairs{}}
    p(\ancestralTheta{i})
    p(\descendantTheta{1}{i})
    p(\descendantTheta{2}{i})
    p(\bottleScalar{1}{i})
    p(\bottleScalar{2}{i})
    p(\bottleTime{i})
    p(\migrationRate{i}).
    \label{eq:demographicPrior}
\end{equation}
The priors for the demographic parameters are
$\ancestralTheta{} \sim Gamma(\gammaShape{\ancestralTheta{}},
\gammaScale{\ancestralTheta{}})$,
$\descendantTheta{1}{} \sim Gamma(\gammaShape{\descendantTheta{}{}},
\gammaScale{\descendantTheta{}{}})$,
$\descendantTheta{2}{} \sim Gamma(\gammaShape{\descendantTheta{}{}},
\gammaScale{\descendantTheta{}{}})$,
$\bottleScalar{1}{} \sim Beta(\betaA{\bottleScalar{}{}},
\betaB{\bottleScalar{}{}})$,
$\bottleScalar{2}{} \sim Beta(\betaA{\bottleScalar{}{}},
\betaB{\bottleScalar{}{}})$,
$\bottleTime{} \sim U(0, 1)$,
and
$\migrationRate{} \sim Gamma(\gammaShape{\migrationRate{}},
\gammaScale{\migrationRate{}})$,
\change{
where the hyper-parameters of each prior distribution can be specified by
the investigator.
By default, \ancestralTheta{}, \descendantTheta{1}{}, and \descendantTheta{2}{}
share the same prior (i.e., $\gammaShape{\ancestralTheta{}} =
\gammaShape{\descendantTheta{}{}}$ and $\gammaScale{\ancestralTheta{}} =
\gammaScale{\descendantTheta{}{}}$), but a separate gamma-distributed prior can
be assigned to \ancestralTheta{}.
Also, the \bottleScalar{1}{}, \bottleScalar{2}{}, and \migrationRate{}
parameters are optional (i.e., the investigator can assume that there has been
no migration between populations of each pair and/or the population size of
each descendant population has been constant through time).}
\end{linenomath}

\subsubsection{Priors on divergence models}
\begin{linenomath}
The prior term for the vector of divergence times for each of the \npairs{}
pairs of populations, \divTimeMapVector, can be expanded as
\begin{equation}
    p(\divTimeMapVector) = p(\divTimeIndexVector)p(\divTimeVector \given \divTimeIndexVector),
    \label{eq:divModelPrior}
\end{equation}
where \divTimeVector is an ordered set of divergence-time parameters
$\{\divTime{1}, \ldots, \divTime{\divTimeNum}\}$ whose length
\divTimeNum can range from 1 to \npairs{},
and \divTimeIndexVector is a vector of indices 
$(\divTimeIndex{1}, \ldots, \divTimeIndex{\npairs{}})$,
where
$\divTimeIndex{i} \in \{1, \ldots, \divTimeNum\}$.
These indices map each of the \npairs{} pairs of populations to a
divergence-time parameter in \divTimeVector.
Thus, \divTimeMapVector is the result of applying the mapping function
\begin{equation}
    f(\divTimeVector, \divTimeIndexVector, i) = \divTime{\divTimeIndex{i}}
    \label{eq:divTimeMapFunction}
\end{equation}
to each population pair $i$, such that
$\divTimeMapVector = (\divTimeMap{1} = f(\divTimeVector, \divTimeIndexVector,
1), \ldots, \divTimeMap{\npairs{}} = f(\divTimeVector, \divTimeIndexVector,
\npairs{}))$.
\end{linenomath}

Biologically speaking, \divTimeVector contains the times of divergence events,
the length of which \divTimeNum is the number of divergence events shared across
the \npairs{} pairs of populations.
For example, if \divTimeVector contains a single divergence-time parameter
\divTime{1}, all \npairs{} pairs of populations are constrained to diverge at
this time (i.e., \divTimeIndexVector would contain the index 1 repeated
\npairs{} times, and \divTimeMapVector would contain the value \divTime{1}
repeated \npairs{} times), whereas if it contains \npairs{} divergence-time
parameters, the model is fully generalized to allow all of the pairs to diverge
at unique times.

Unlike the model implemented in \msb, here I place priors on
\divTimeIndexVector and \divTime{}, rather than \divTimeNum and
\divTime{}.
As a result, \divTimeIndexVector determines the number of divergence-time
parameters (\divTimeNum) in the model.
Below, I first describe the prior used for \divTime{} and the timescale it
imposes on the model before discussing the priors implemented for
\divTimeIndexVector.

\begin{linenomath}
Each \divTime{} within \divTimeVector is \iid as $\divTime{} \sim
Gamma(\gammaShape{\divTime{}}, \gammaScale{\divTime{}})$,
\change{
where \gammaShape{\divTime{}} and \gammaScale{\divTime{}} are specified by the
investigator}.
Thus, given the number of unique divergence-time classes in
\divTimeIndexVector, this determines the probability of prior term
$p(\divTimeVector \given \divTimeIndexVector)$.
The divergence times are in coalescent units relative to the size of
a constant reference population, which I denote \myTheta{C}, that is equal to
the expectation of the prior on the size of the descendant populations
\begin{equation}
    \myTheta{C} = \mathbb{E}(\descendantTheta{}{}),
    \label{eq:thetaC}
\end{equation}
Given the size of the descendant populations are \iid as
$\descendantTheta{1}{} \sim Gamma(\gammaShape{\descendantTheta{}{}},
\gammaScale{\descendantTheta{}{}})$ and
$\descendantTheta{2}{} \sim Gamma(\gammaShape{\descendantTheta{}{}},
\gammaScale{\descendantTheta{}{}})$,
this becomes
\begin{equation}
    \myTheta{C} = \gammaShape{\descendantTheta{}{}}\gammaScale{\descendantTheta{}{}}.
    \label{eq:thetaCGamma}
\end{equation}
More specifically, the \divTime{} parameters are in units of
$\myTheta{C}/\mutationRate$ generations, which I denote as \globalcoalunit
generations.
Thus, each \divTime{} within \divTimeVector is proportional to time and can be
converted to the number of generations of the reference population, which I
denote \divTime{G_C}, by assuming a mutation rate and multiplying by the
effective size of the reference population
\begin{equation}
    \divTime{G_C} = \divTime \times \frac{\myTheta{C}}{\mutationRate} = \divTime
    \times \frac{
    \gammaShape{\descendantTheta{}{}}\gammaScale{\descendantTheta{}{}}}
    {\mutationRate}.
    \label{eq:divTimeGenerations}
\end{equation}
Thus, for each of the divergence times in \divTimeVector to be on the same
scale, the relative mutation rates among the pairs of populations are assumed
to be known and fixed according to the user-provided values in
\mutationRateScalarConstantVector.
\end{linenomath}

\begin{linenomath}
As described by Oaks et al.\ \cite{Oaks2012}, to get the divergence times in
units proportional to the expected number of mutations, they must be scaled by
the realized population size for locus $j$ of population-pair $i$ 
\begin{equation}
    \divTimeScaled{i}{j} = \divTimeMap{i} \times \frac{\myTheta{C}}{
        \descendantThetaMean{i} \ploidyScalar{i}{j}},
    \label{eq:divTimeScaled}
\end{equation}
where \descendantThetaMean{i} is the mean of \descendantTheta{1}{} and
\descendantTheta{2}{} for pair $i$.
This gives us the vector of scaled divergence times
$\divTimeScaledVector = (\divTimeScaled{1}{1}, \ldots,
\divTimeScaled{\npairs{}}{\nloci{\npairs{}}})$.
\end{linenomath}

\begin{linenomath}
\change{
As for the prior term $p(\divTimeIndexVector)$, the total sample space of
\divTimeIndexVector is all the possible partitionings of the \npairs{} pairs of
populations into 1 to \npairs{} divergence-time classes, where each
partitioning consists of non-overlapping and non-empty subsets whose union is
the \npairs{} pairs.
Hereinafter, I refer to these partitionings as ``ordered'' divergence models or
partitions.
}
The total number of possible partitions is a sum of the Stirling numbers of
the second kind over all possible numbers of categories \divTimeNum
\begin{equation}
    B_{\npairs{}}=\sum_{\divTimeNum=1}^{\npairs{}} \left[
    \frac{1}{\divTimeNum!} \sum_{j=0}^{\divTimeNum-1} (-1)^{j}
    \binom{\divTimeNum}{j} (\divTimeNum-j)^{\npairs{}} \right],
    \label{eq:bell}
\end{equation}
which is the Bell number \cite{Bell1934}.
The original \msb model samples over the unordered realizations of
\divTimeIndexVector, such that the sample space is reduced to all the possible
integer partitions of \npairs{} \cite{Oaks2012,Huang2011,OeisPartitionNumber,
OeisPartitionTriangle,Malenfant2011} (Table~S\ref{tabSampleSpace}).
\change{
I denote the set of all possible integer partitions of the \npairs{} pairs of
populations as \integerPartitionSet{\npairs{}} and the length of that set as
\integerPartitionNum{\npairs{}}, and I hereinafter refer to these integer
partitions as ``unordered'' divergence models or partitions.}
The advantages, disadvantages, and justification of ignoring the order
of \divTimeIndexVector is discussed in detail below.
\end{linenomath}

\begin{linenomath}
I implement two prior probability distributions over the space
of all possible divergence models (\divTimeIndexVector).
The first simply gives all possible unordered partitions of \npairs{} elements
equal probability
\begin{equation}
    p(\divTimeIndexVector) = \frac{1}{\integerPartitionNum{\npairs{}}},
    \label{eq:divModelPriorUniform}
\end{equation}
i.e., a discrete uniform prior over all the integer partitions of \npairs{}
(unordered divergence models).
I denote this prior as
$\divTimeIndexVector \sim \priorUniform$.
\end{linenomath}

The second prior is based on the Dirichlet process, which is a stochastic
process that groups random variables into an unknown number of discrete
parameter classes \cite{Ferguson1973,Antoniak1974}.
The Dirichlet process has been used as a non-parametric Bayesian approach to
many inference problems in evolutionary biology
\cite{Lartillot2004,Huelsenbeck2007,HuelsenbeckSuchard2007,Ane2007,Heath2011,Heath2012}.
Here, I use the Dirichlet process to place a prior over all possible ordered
partitions of \npairs{} population pairs into divergence-time parameter classes
(i.e., ``divergence events'').
As discussed above, the time of each divergence-time parameter is drawn from
the base distribution $\divTime{} \sim Gamma(\gammaShape{\divTime{}},
\gammaScale{\divTime{}})$.
The partitioning of the population pairs to divergence-time classes is
controlled by the concentration parameter \concentrationParam, which determines
how clustered the process will be.
I take a hierarchical approach and use a prior probability
distribution (i.e., hyperprior) for \concentrationParam \cite{Escobar1995}.
More specifically, I use a gamma-distributed prior $\concentrationParam \sim
Gamma(\gammaShape{\concentrationParam}, \gammaScale{\concentrationParam})$,
where \gammaShape{\concentrationParam} and \gammaScale{\concentrationParam} are
specified by the investigator.
I use $\divTimeIndexVector \sim \priorDPP{}$ to denote this Dirichlet-process
prior.

\begin{linenomath}
This provides a great deal of flexibility for specifying the prior uncertainty
regarding divergence models.
The concentration parameter \concentrationParam determines the prior
probability that any two pairs of populations $i$ and $j$ will be assigned to
the same divergence-time parameter
\begin{equation}
    p(\divTimeIndex{i} = \divTimeIndex{j}) = \frac{1}{1 + \concentrationParam},
    \label{eq:dppPriorSameClass}
\end{equation}
and also the prior probability of the number of divergence-time parameters
\begin{equation}
    p(\divTimeNum \boldsymbol{\mid} \concentrationParam, \npairs{}) = 
    \frac{\stirlingFirst{\npairs{}}{\divTimeNum} \concentrationParam^{\divTimeNum}}
    {\prod_{i=1}^{\npairs{}}(\concentrationParam + i - 1)},
    \label{eq:dppPriorNumClasses}
\end{equation}
where \stirlingFirst{\cdot}{\cdot} are the unsigned Stirling numbers of the
first kind.
Equations \ref{eq:dppPriorSameClass} and \ref{eq:dppPriorNumClasses} show that
smaller values of \concentrationParam will favor fewer divergence-time
parameters, and thus more clustered models of divergence, whereas larger values
will favor more divergence-time parameters, and thus less clustered models of
divergence.
\end{linenomath}

\subsection{Differences between this model and the original \texttt{msBayes} model}
\subsubsection{The prior on divergence models}
One of the key differences between my model and that of \msb \cite{Huang2011}
is the prior distribution on divergence models.
As discussed in Oaks et al.\ \cite{Oaks2012}, in \msb the prior used for
\divTimeIndexVector is a combination of a discrete uniform prior over the
possible number of divergence events \divTimeNum from 1 to \npairs{} with a
multinomial distribution on the number of times each index of \divTimeVector
appears in \divTimeIndexVector, with the constraint that all \divTime{}
parameters are represented at least once (see Equation 2 of \cite{Oaks2012}).
I denote this prior used by \msb as $\divTimeIndexVector \sim \priorOld$.
Oaks et al.\ \cite{Oaks2012} discuss how placing a uniform prior over the
number of divergence parameters (denoted \divTimeNum here, and as \numt{} in
\cite{Huang2011}) imposes an ``U-shaped'' prior over divergence
models (\divTimeIndexVector; see
Figure 5B of \cite{Oaks2012}).
To avoid this, I place priors directly on the sample space of divergence
models, thus eliminating the parameter \numt{} from the model.
I introduce two priors on divergence models:
(1) a prior that is uniform over all unordered divergence models, and
(2) a Dirichlet-process prior on all ordered divergence models.
The latter provides an investigator with a great deal of flexibility in
expressing their prior beliefs about models of divergence.

\subsubsection{Estimating ordered divergence models}
As mentioned above, \msb samples over unordered divergence models
(i.e., unordered partitions of the \npairs{} pairs of populations).
That is, the identity of each population pair, and all the information
associated with it, is discarded.
In my implementation, inference can be done on either unordered or ordered
models of divergence.
This is discussed in more detail in the description of the ABC implementation
below.

\subsubsection{The priors on nuisance parameters}
I have replaced the use of continuous uniform distributions for priors on many
of the model's parameters (\divTime{}, \ancestralTheta{},
\descendantTheta{1}{}, \descendantTheta{2}{}, \bottleScalar{1}{},
\bottleScalar{2}{}, \recombinationRate, \migrationRate{}) with more flexible
parametric distributions from the exponential family.
I introduce gamma-distributed priors for rate parameters that have a sample
space of all positive real numbers (\divTime{}, \ancestralTheta{},
\descendantTheta{1}{}, \descendantTheta{2}{}, \recombinationRate,
\migrationRate{}), and beta-distributed priors for parameters that are
proportions bounded by zero and one (\bottleScalar{1}{} and
\bottleScalar{2}{}).
These priors provide an investigator with much greater flexibility in
expressing prior uncertainty regarding the parameters of the model.

In addition, I have modified the prior on the sizes of the descendant
populations of each pair.
As described by Oaks et al.\ \cite{Oaks2012}, \msb uses the joint prior
\begin{equation}
    \descendantTheta{1}{}, \descendantTheta{2}{} \sim
    Beta(1,1) \times 2 \times U(\uniformMin{\myTheta{}},
    \uniformMax{\descendantTheta{}{}}),
    \label{eq:jointThetaPrior}
\end{equation}
such that the user-specified uniform prior on descendant population
size is a prior on the \emph{mean} size of the two descendant
populations of each pair.
Under my model, the sizes of the descendant populations of each
pair are \iid as
$\descendantTheta{1}{} \sim Gamma(\gammaShape{\descendantTheta{}{}},
\gammaScale{\descendantTheta{}{}})$
and
$\descendantTheta{2}{} \sim Gamma(\gammaShape{\descendantTheta{}{}},
\gammaScale{\descendantTheta{}{}})$.
This relaxes the assumption that the sizes of the two descendant populations
are interdependent and negatively correlated.

\subsubsection{Flexibility in parameterizing the model}
In the new implementation, I provide the ability to control the richness of the
model.
\change{
For the \myTheta{} parameters, by default, the model is fully generalized to
allow each population pair to have three parameters:
\ancestralTheta{}, \descendantTheta{1}{}, and \descendantTheta{2}{}.
Furthermore, if an investigator prefers to reduce the number of parameters, any
model of \myTheta{} parameters nested within this
general model can also be specified, including the most restricted model
where the ancestral and descendant populations of each pair share
a single \myTheta{} parameter.}

I also provide the option of eliminating the parameters associated with the
post-divergence bottlenecks in the descendant populations of each pair
(\bottleTime{}, \bottleScalar{1}{}, and \bottleScalar{2}{}),
which constrains the descendant populations to be of
constant size from present back to the divergence event.
Also, rather than eliminate the bottleneck parameters,
I allow \bottleScalar{1}{} and \bottleScalar{2}{} to be constrained to be
equal, which removes one free parameter from the model for each of the
population pairs.

Overall, my implementation allows an investigator to specify a model that has
as many as seven parameters per population pair
(\ancestralTheta{}, \descendantTheta{1}{}, \descendantTheta{2}{},
\bottleTime{}, \bottleScalar{1}{}, \bottleScalar{2}{}, and
\migrationRate{})
or as few as one parameter per pair
(\myTheta{}),
in addition to the $\popSampleSize{i}{} - 1$ coalescence-time parameters
(i.e., the node heights of the gene tree).

\subsubsection{Time scale}
As described above, divergence times are in units of
$\myTheta{C}/\mutationRate$ generations, where \myTheta{C} is the expectation
of the prior on descendant-population size.
As described by Oaks et al.\ \cite{Oaks2012}, in \msb, \myTheta{C} is half of
the upper limit of the continuous uniform prior on the mean of the descendant
population sizes.
This is only equal to the expectation of the prior if the lower limit of the
prior is zero.

\subsection{ABC estimation of the posterior of the model}
\subsubsection{Sampling from the prior}
To estimate the approximate posterior of Equation \ref{eq:approxModelCompact},
I use an ABC rejection algorithm.
The first step of this algorithm entails collecting a random sample of
parameter values from the joint prior and their associated summary
statistics.
Each sample is generated by
\change{
(1) drawing values of all the model's parameters, which I denote \hpvector{},
from their respective prior distributions;
(2) simulating gene trees $\geneTreeVector = (\geneTree{1}{1}, \ldots, 
\geneTree{\npairs{}}{\nloci{\npairs{}}})$
for each locus of each population pair by drawing coalescent times from
a multi-population Kingman-coalescent model given the demographic parameters;
(3) simulating sequence alignments 
$\alignmentVector = (\alignment{1}{1}, \ldots, \alignment{\npairs{}}{\nloci{\npairs{}}})$
along the gene trees under the
\hky substitution parameters
$\hkyModelVector = (\hkyModel{1}{1}, \ldots, \hkyModel{\npairs{}}{\nloci{\npairs{}}})$
that have the same number of sequences and sequence lengths as the observed
dataset;
and
(4) calculating population genetic summary statistics
$\ssVector{} = (\alignmentSS{1}{1}, \ldots, \alignmentSS{\npairs{}}{\nloci{\npairs{}}})$
from the simulated sequence alignments.
Optionally, an additional step can be performed to reduce the summary
statistics to the means across loci for each population pair to get
$\ssVector{} = (\alignmentSS{1}{}, \ldots, \alignmentSS{\npairs{}}{})$.
Either way, \ssVector{} contains the same summary statistics as those estimated
from the observed data \ssVectorObs.
}
After repeating this procedure \numPriorSamples times, we have a random
sample of parameter vectors
$\paramSampleMatrix = (\paramSampleVector{1}, \ldots, \paramSampleVector{\numPriorSamples})$
from the model prior and their associated vectors of summary statistics
$\ssMatrix = (\ssVector{1}, \ldots, \ssVector{\numPriorSamples})$.

For all of the analyses below, I use four summary statistics for each pair of
populations:
$\pi$ \cite{Tajima1983}, $\theta_W$ \cite{Watterson1975}, $\pi_{net}$
\cite{Takahata1985}, and $SD(\pi-\theta_W)$ \cite{Tajima1989}.
Furthermore, in addition to model parameters, each sample \hpvector{}
also contains four statistics that summarize \divTimeMapVector:
the mean (\divTimeMean), variance (\divTimeVar), dispersion index
($\divTimeDispersion = \divTimeVar/\divTimeMean$), and the number
of divergence time parameters (\divTimeNum).
Previously, these have been denoted as \meant{}{}, \vart{}{}, \vmratio{}, and
\numt{}, respectively \cite{Hickerson2006,Huang2011,Oaks2012}.
I use \divTimeMean and \divTimeVar in place \meant{}{} and \vart{}{} to make
clear that these values do not represent the prior or posterior
expectation/variance of divergence times.
I use \divTimeDispersion in place of \vmratio{} to clarify that this is a
statistic rather than a parameter of the model.
Lastly, I use \divTimeNum in place of \numt{}, because the number of
divergence-time parameters is no longer a parameter in the new implementation.

\subsubsection{Obtaining an approximate posterior from the prior samples}
I use a rejection algorithm to retain an approximate posterior sample of
\paramSampleVector{} from the prior sample
$\paramSampleMatrix = (\paramSampleVector{1}, \ldots, \paramSampleVector{\numPriorSamples})$.
First, the observed summary statistics \ssVectorObs, and the summary statistics
of the prior samples 
$\ssMatrix = (\ssVector{1}, \ldots, \ssVector{\numPriorSamples})$,
are standardized using the means and standard deviations of the statistics from
the prior sample (i.e., the prior mean is subtracted from each statistic, and the
difference is divided by the prior standard deviation).
After all statistics are standardized, the Euclidean distance between
\ssVectorObs and each \ssVector{} within \ssMatrix is calculated.
The samples that fall within a range of tolerance \tol around \ssVectorObs
are retained.
The range of tolerance is determined by specifying the desired number of
posterior samples to be retained.
Post-hoc adjustment of the posterior sample can also be performed with a number
of regression techniques \cite{Beaumont2002,Blum2009,Leuenberger2010}.
For analyses below, I use the general linear model (GLM) regression adjustment
\cite{Leuenberger2010} as implemented in \abctoolbox v1.1
\cite{ABCtoolbox}, which Oaks et al.\ \cite{Oaks2012} showed performs
very similarly to weighted local-linear regression and multinomial logistic
regression adjustments \cite{Beaumont2002} for \msb posteriors.

\subsubsection{Ordering of taxon-specific summary statistics}
As alluded to in the model description, \msb does not maintain the order of the
taxon-specific summary statistics \alignmentSS{}{} within each \ssVector{}.
Rather, the summary statistics are re-ordered by descending values of average
pairwise differences between the descendant populations
($\pi_b$) \cite{NeiLi1979,Huang2011}.
This has the advantage of reducing the sample space of possible divergence
models \divTimeIndexVector, but there are at least two disadvantages.
First, additional information in the data is lost.
By discarding the identity of the \npairs{} pairs of populations, all
pair-specific information about the amount of data (e.g., the number of gene
copies collected from each of the populations [\popSampleSize{1}{} and
\popSampleSize{2}{}], the number of loci, and the length of the loci), and the
taxon- and locus-specific parameters (\hkyModel{}{},
\mutationRateScalarConstant{}{}, \ploidyScalar{}{}, and
\locusMutationRateScalar{}) is lost.
\change{
Second, the results are more difficult to interpret, because divergence models
and parameter estimates cannot be directly associated to the taxa under study.
}

\change{
The re-ordering of the summary statistic vectors also has an important
implication for the ABC algorithm.
When calculating the Euclidean distance between the observed data and each
simulated dataset, the summary statistics being compared often represent
sequence alignments of \emph{different} taxon pairs and/or loci.
More specifically, the summary statistics calculated from the observed sequence
alignments are being compared to summary statistics calculated from datasets
simulated with potentially \emph{different} 
(1) numbers of sequences (\popSampleSize{1}{} and \popSampleSize{2}{}),
(2) length of alignments,
(3) numbers of loci (\nloci{}),
(4) \hky model parameters (\hkyModel{}{}),
(5) mutation-rate multipliers (\mutationRateScalarConstant{}{}),
and
(6) ploidy multipliers (\ploidyScalar{}{}).}

In the original descriptions of the \msb method \cite{Hickerson2006,Huang2011},
this re-ordering is justified by the fact that the expected value of $\pi_b$ is
unrelated to sample size \popSampleSize{1}{} and \popSampleSize{2}{} and thus
exchangeable among pairs.
\change{
This is incorrect for two reasons.}
First, the entire vector of summary statistics \alignmentSS{}{} for each pair
of populations is re-ordered across pairs, which implies that the justification
for re-ordering $\pi_b$ applies to all the statistics within each
\alignmentSS{}{}.
\change{
However, the expectations for statistics that estimate gross diversity (e.g.,
$\pi$ and $\theta_W$) are not independent of sample size for structured
populations (e.g., the divergent pairs of populations modeled by \msb), and
other statistics are not independent of sample size in general (e.g.,
$SD(\pi-\theta_W)$).
Second, and more importantly, having the same expectation does not ensure
random variables are exchangeable.
Rather, for variables to be exchangeable their marginal distributions must be
the same (i.e., they must be identically distributed).
\emph{None} of the summary statistics used by \msb, including $\pi_b$, have
this property when there is any variation among taxa or loci in the
(1) numbers of sequences (\popSampleSize{1}{} and \popSampleSize{2}{}),
(2) length of alignments,
(3) numbers of loci (\nloci{}),
(4) \hky model parameters (\hkyModel{}{}),
(5) mutation-rate multipliers (\mutationRateScalarConstant{}{}),
or
(6) ploidy multipliers (\ploidyScalar{}{}).
Whenever such variation is present (i.e., nearly all empirical applications),
the taxon-specific summary statistics \alignmentSS{}{} are not exchangeable,
and the reshuffling of the summary statistic vectors is not mathematically
valid.

The magnitude of the affect of this violation of exchangeability is not known.
Huang et al.\ \cite{Huang2011} demonstrated that the reordering of the summary
statistic vectors can greatly increase the method's tendency to infer a single
divergence event.
By definition, if the summary statistic vectors were exchangeable, the
reordering would not change the likelihood or posterior (barring sampling
error).
Thus, the results of Huang et al.\ \cite{Huang2011} suggest the reordering of
the statistics is potentially introducing sizeable error to the analysis.
}

\change{
For comparability with \msb, I maintain the option for re-ordering
taxon-specific summary statistics by $\pi_b$.
However, by default, the order is preserved, and ordered divergence models are
estimated.
In all of the simulation-based analyses described below, the summary statistic
vectors \emph{are} exchangeable, because the simulated datasets have the same 
(1) numbers of sequences,
(2) length of sequences,
(3) numbers of loci,
(4) \hky model parameters,
(5) mutation-rate multipliers,
and
(6) ploidy multipliers.}

\subsection{Assessing model-choice behavior and robustness}
Following the simulation-based approach of Oaks et al.\ \cite{Oaks2012}, I
characterize the behavior of several models under the ideal conditions where
the data are generated from parameters drawn from the same prior distributions
used for analysis (i.e., the prior is correct).
I selected the following four model priors for these analyses
(Table~\ref{tabPriors}).
\begin{enumerate}
    \item The \modelOld model represents the original \msb implementation with
        the U-shaped prior on unordered divergence models and uniform priors on
        divergence-time and demographic parameters; $\divTimeIndexVector \sim
        \priorOld$, $\divTime{} \sim U(0,10)$, $\ancestralTheta{} \sim U(0,
        0.05)$, and $\descendantThetaMean{} \sim U(0, 0.05)$.
    \item The \modelUshaped model with the U-shaped prior of \msb on unordered
        divergence models, but with exponential priors on divergence-time and
        demographic parameters; $\divTimeIndexVector \sim \priorOld$,
        $\divTime{} \sim Exp(mean=2.887)$, $\ancestralTheta{} \sim
        Exp(mean=0.025)$, $\descendantTheta{1}{} \sim Exp(mean=0.025)$, and
        $\descendantTheta{2}{} \sim Exp(mean=0.025)$.
    \item The \modelUniform model with a uniform prior over unordered
        divergence models and exponential priors on divergence-time and
        demographic parameters; $\divTimeIndexVector \sim \priorUniform$,
        $\divTime{} \sim Exp(mean=2.887)$, $\ancestralTheta{} \sim
        Exp(mean=0.025)$, $\descendantTheta{1}{} \sim Exp(mean=0.025)$, and
        $\descendantTheta{2}{} \sim Exp(mean=0.025)$.
    \item The \modelDPP model with a Dirichlet-process prior on ordered
        divergence models and exponential priors on divergence-time and
        demographic parameters; $\divTimeIndexVector \sim \priorDPP{\sim
        Gamma(2,2)}$, $\divTime{} \sim Exp(mean=2.887)$, $\ancestralTheta{}
        \sim Exp(mean=0.025)$, $\descendantTheta{1}{} \sim Exp(mean=0.025)$,
        and $\descendantTheta{2}{} \sim Exp(mean=0.025)$.
\end{enumerate}
I selected the exponential prior on divergence time used in models \modelDPP,
\modelUniform, and \modelUshaped to have the same variance as the uniform prior
in model \modelOld.
I selected the exponential prior on population size used in models \modelDPP,
\modelUniform, and \modelUniform to have the same mean as the uniform prior in
model \modelOld, so that all four models have the same \myTheta{C} and thus the
same units of time.
All of the models were the same in other respects, with three free \myTheta{}
parameters for each population pair, two uniformly distributed ($beta(1,1)$)
\bottleScalar{}{} parameters per pair, no migration, no recombination, and
re-sorting of taxon-specific summary statistics by $\pi_b$ (i.e., sampling
unordered divergence models).
For all simulations, I used a data structure of eight population pairs, with a
single 1000 base-pair locus sampled from 10 individuals from each population.

For each of the four models, I simulated $1\e6$ samples from the prior and
50,000 datasets, also drawn from the prior.
I then analyzed each of the simulated datasets, retaining a
posterior of 1000 samples from the respective prior.
A GLM-regression adjusted posterior was also estimated from each of the
posterior samples \cite{Leuenberger2010}.
To assess the robustness of each of the four models, I also analyzed the
datasets simulated under the other three models.
Overall, for each model, I produced 200,000 posterior estimates,
50,000 from the datasets simulated under that model,
and 150,000 from the datasets simulated under the
other three models.

For each set of 50,000 simulated datasets, I used the posterior estimates
to assess the model-choice behavior of each model.
I did this by assigning the 50,000 estimates of the posterior probability
of one-divergence event to 20 bins of width 0.05, and plotted
the estimated posterior probability of each bin against the proportion of
replicates in that bin with a true value consistent with one divergence
event \cite{Huelsenbeck2004,Oaks2012}.
\change{
Ideally, the estimated posterior probability of the one-divergence model
should estimate the probability that the one-divergence model is correct.
For large numbers of simulation replicates, the proportion of the replicates in
each bin for which the one-divergence model is true will approximate the
probability that the one-divergence model is the correct model.
Thus, if the method has the desirable behavior such that the estimated
posterior probability of the one-divergence model is an unbiased estimate of
the probability that the one-divergence model is correct, the points should
fall near the identity line.
For example, let us say the method estimates a posterior probability of 0.90
for 1000 datasets simulated from the prior.
If the method is accurately estimating the probability that the one-divergence
model is correct given the data, then the one-divergence model should be the
true model in approximately 900 of the 1000 replicates.
Any trend away from the identity line indicates the method is biased in the
sense that it is not accurately estimating the probability that the
one-divergence model is the correct model.

I constructed these plots using two criteria for the one-divergence model:
}
(1) the number of divergence-time parameters ($\divTimeNum = 1$) and
(2) the dispersion index of divergence times ($\divTimeDispersion < 0.01$).
For the latter, $\divTimeDispersion < 0.01$ has been commonly used as an
arbitrary criterion for a single ``simultaneous'' divergence event
(e.g., \cite{Hickerson2006,Leache2007,Stone2012}).
I focused on the one-divergence model to assess model-choice behavior, because
it is often of biogeographic interest and is easily comparable among
the three different priors used on divergence models.

In addition to the four models above, I also assessed the behavior of a model
that samples over ordered divergence models (i.e., the order of the
taxon-specific summary statistic vectors were maintained for the observed and
simulated datasets); all other settings were identical to the \modelDPP model.
I denote this model as \modelDPPOrdered.
I simulated $1\e6$ prior samples and 50,000 datasets, and
analyzed them as above.
I was not able to analyze the simulated datasets of the other models under
the ordered model, because the identity of the population pairs is not
contained in the simulations of the other models.

\subsection{Assessing power}
I evaluated the power of the same four models (Table~\ref{tabPriors}) to
detect random variation in divergence times using methods similar to
Oaks et al.\ \cite{Oaks2012}.
For all power simulations, I used a data structure identical to that of the
empirical dataset of Philippine vertebrates analyzed by Oaks et al.\
\cite{Oaks2012}, which consists of 22 pairs of populations.
Due to the larger number of pairs, I used a different hyperprior on the
concentration parameter for the \modelDPP model; I used a prior of
$\divTimeIndexVector \sim \priorDPP{\sim Gamma(1.5,18.1)}$ over divergence
models for the model \modelDPP.
All other aspects of the four models in Table~\ref{tabPriors} were identical to
those used in the validation analyses described above.
For each of the four models, I generated $2\e6$ samples from the prior.

Next, I simulated datasets from three series of models in which the divergence
times of the 22 pairs were random (i.e., no clustering; $\divTimeNum = 22$).
The models comprising each series differ in the variance of the distribution
from which the divergence times are randomly drawn.
When the variance of random divergence times is small, all of the models in
Table~\ref{tabPriors} are expected to struggle to detect this variation and
will often incorrectly estimate highly clustered models of divergence (i.e.,
few divergence events).
The goal is to assess how much temporal variation in random divergence times is
necessary before the behavior of the models of Table~\ref{tabPriors} begins to
improve.
\change{
This will determine the timescales over which the models can reliably detect
random variation in divergence times and avoid spurious inference of clustered
divergence models.}

Specifically, I simulated datasets from the following three series of six
models (Table~\ref{tabPowerModels}).
\begin{enumerate}
    \item The \powerSeriesOld models are identically distributed as \modelOld
        except the divergence times for each of the 22 pairs of
        populations are randomly drawn from a series of uniform distributions,
        $U(0, \divt{max})$, where \divt{max} was set to: 0.2, 0.4, 0.6, 0.8,
        1.0, and 2.0, in \globalcoalunit generations.
    \item The \powerSeriesUniform models are identically distributed as
        \modelUniform and \modelDPP except the 22 divergence times are
        randomly drawn from the same series of uniform priors as above.
    \item The \powerSeriesExp models are also identically distributed as
        \modelUniform and \modelDPP except the 22 divergence times are
        randomly drawn from a series of of exponential distributions:
        $Exp(mean=0.058)$, $Exp(mean=0.115)$, $Exp(mean=0.173)$,
        $Exp(mean=0.231)$, $Exp(mean=0.289)$, and $Exp(mean=0.577)$. These
        exponential distributions have the same variance as their uniform
        counterparts in the first two series of models.
\end{enumerate}
For each of the six models in each of the three series of models, I simulated
1000 datasets (18,000 datasets in total).
I then analyzed each simulated dataset under all four prior models
(Table~\ref{tabPriors}), producing 72,000 posterior estimates, each with 1000
samples.
I also estimated a GLM-regression adjusted posterior from each of the
posterior samples \cite{Leuenberger2010}.

\subsection{An empirical application}
I also assessed the behavior of the newly implemented models when applied to
the empirical dataset of Oaks et al.\ \cite{Oaks2012}, which is comprised of
sequence data from 22 pairs of taxa from the Philippine Islands
(\cite{Oaks2012dryad}; Dryad DOI: 10.5061/dryad.5s07m).
I analyzed these data under five different models, which are detailed in
Table~\ref{tabEmpiricalModels}.
All of these models except one (\empModelDPPSimple) have six free demographic
parameters per pair of taxa (\ancestralTheta{},
\descendantTheta{1}{},
\descendantTheta{2}{},
\bottleTime{},
\bottleScalar{1}{}, and
\bottleScalar{2}{}),
in addition to the $\popSampleSize{i}{} - 1$ coalescent times.
Three of these models use a Dirichlet-process prior on divergence models:
\empModelDPP, \empModelDPPInform, and \empModelDPPSimple.
The \empModelDPP model represents the priors that Oaks et al.\ \cite{Oaks2012}
would have selected to reflect their prior uncertainty about the parameters of
the model if provided the more flexible distributions that are now implemented.
To assess prior sensitivity, the \empModelDPPInform model uses a more
informative exponentially distributed prior on divergence times, but otherwise
is identical to \empModelDPP.
To assess sensitivity to parameterization, I also applied the simplest
possible model under the new implementation (\empModelDPPSimple) with only a
single demographic parameter (\myTheta{}) per taxon pair, in addition to the
$\popSampleSize{i}{} - 1$ coalescent times.
I also applied the original \msb model (\empModelOld) with priors selected to
make the results directly comparable to those of the \empModelDPP model;
the uniform prior on divergence times was selected to have the same variance as
the exponential prior of the \empModelDPP model, and the prior on population
size was selected to have the same mean so that the models are on the same
timescale.
I also applied a model with a uniform distribution over divergence models
(\empModelUniform).
For each of these models, I simulated 2\e{7} samples from the prior, and
retained an approximate posterior of the 10,000 samples with the smallest
Euclidean distance from the summary statistics calculated from the empirical
sequence alignments.

To compare models that sample over ordered versus unordered models of
divergence, I also analyzed the data from the subset of nine-taxon pairs that
are sampled from the Islands of Negros and Panay in the Philippines.
The model I used for these analyses had a Dirichlet-process prior over
divergence models and two demographic parameters (\ancestralTheta{} and
\descendantTheta{}{}) for each pair of taxa, in addition to the
$\popSampleSize{i}{} - 1$ coalescent times (see Table~\ref{tabEmpiricalModels}
for details).
One of the models, which I denote \npModelDPPOrdered, maintained the identity
of the taxon pairs and sampled over ordered models of divergence, while the
other (\npModelDPP) re-sorted the summary statistics of the pairs by $\pi_b$,
losing the identity of the taxa and thus sampled over unordered models of
divergence.
For both analyses, I simulated 5\e{7} samples from the prior and retained an
approximate posterior of 10,000 samples.

\section{Results}
\subsection{Validation analyses: Estimation accuracy}
In terms of estimating the variance of divergence times (\divTimeDispersion),
the models with exponentially distributed priors (\modelUshaped, \modelUniform,
and \modelDPP) perform similarly when applied to datasets generated under all
four of the models in Table \ref{tabPriors}
(Figure~S\ref{figValidationAccuracyOmega}).
The \modelOld model performs similarly to these models when applied to its
own datasets, but is sensitive to model violations and is more biased when
applied to data generated under the other three models
(Figure~S\ref{figValidationAccuracyOmega}).
Results are similar for the GLM-adjusted estimates of \divTimeDispersion,
albeit the regression adjustment tends to improve estimates of this continuous
statistic for all the models
(Figure~S\ref{figValidationAccuracyOmegaGlm}).

The same general pattern is seen for estimates of \divTimeMean, with
(1) all four models performing similarly when applied to the data generated
under the \modelOld model,
(2) the models with exponentially distributed priors performing similarly when
applied to data generated under the other three models, and
(3) the \modelOld model is sensitive to model violations and is more biased
whenever applied to data generated under other models
(Figure~S\ref{figValidationAccuracyTime}).
Also, the regression adjustment tends to slightly improve estimates of
this continuous statistic for all of the models
(Figure~S\ref{figValidationAccuracyTimeGlm}).

In terms of estimating the number of divergence events (\divTimeNum), the
\modelDPP model has the lowest root mean square error (RMSE) when applied
to data generated under most of the models of Table \ref{tabPriors}
(Figure~S\ref{figValidationAccuracyPsi}).
The \modelOld model performs slightly better when applied to its own data,
but is the worst performer when applied to data generated under other models
(Figure~S\ref{figValidationAccuracyPsi}).
There is a trend of $\modelDPP > \modelUniform > \modelUshaped > \modelOld$
in terms of estimation accuracy as measured by RMSE when the models are applied
to data generated under most of the models
(Figure~S\ref{figValidationAccuracyPsi}).
Unlike for the continuous statistics, regression adjustment of this discrete
statistic tends to increase estimation bias; all of the models tend to
underestimate \divTimeNum after the GLM-adjustment
(Figure~S\ref{figValidationAccuracyPsiGlm}).

\subsection{Validation analyses: Model-choice accuracy}
The \msb model, and my modification of it, is a model-choice method
with the primary purpose of estimating the probabilities of models
of divergence across taxa.
Thus, it is critical to assess the method's ability to accurately
estimate the posterior probabilities of divergence models.
Consistent with the findings of Oaks et al.\ \cite{Oaks2012}, my results
demonstrate that the unadjusted estimates of divergence-model posterior
probabilities are generally more accurate than regression-adjusted estimates
(compare the plots along the upper-left to lower-right diagonal for
Figure~\ref{figValidationModelChoicePsi} versus
S\ref{figValidationModelChoicePsiGlm} and
Figure~\ref{figValidationModelChoiceOmega} versus
S\ref{figValidationModelChoiceOmegaGlm}).
Regression adjustment results in biased estimates of the posterior probability
of the one-divergence model when all model assumptions are satisfied, which is
well illustrated in Figure S\ref{figValidationModelChoiceOmegaGlm}.
As a result, I will focus my discussion of the results on the unadjusted
estimates.

I find that all four models accurately estimate the posterior probability of
the one-divergence model when applied to their own datasets (i.e., when the
prior is correct; see diagonal of Figures~\ref{figValidationModelChoicePsi} \&
\ref{figValidationModelChoiceOmega}).
The \modelUniform and \modelDPP models show robustness to prior violations and
perform well when applied to data generated under other models
(Figures~\ref{figValidationModelChoicePsi} \&
\ref{figValidationModelChoiceOmega}).
However, both are less accurate and tend to underestimate the probability of the
one-divergence model when applied to the data generated under \modelUshaped
(Figures~\ref{figValidationModelChoicePsi} \&
\ref{figValidationModelChoiceOmega}).
In contrast, the \modelOld model is biased toward overestimating the posterior
probability of the one-divergence model when applied to data generated under
the other three models
(Figures~\ref{figValidationModelChoicePsi} \&
\ref{figValidationModelChoiceOmega}).
This bias is particularly strong whenever divergence models are not
distributed under its U-shaped prior
(Figure~\ref{figValidationModelChoicePsi}C--D).
The other model with the U-shaped prior on divergence models, but exponential
priors on parameters (\modelUshaped), performs similarly to the \modelOld model
in that it performs well when applied to its own data, but overestimates the
probability of the one-divergence model when applied to data generated by the
other models
(Figures~\ref{figValidationModelChoicePsi} \&
\ref{figValidationModelChoiceOmega}).
However, the bias is stronger in the \modelOld model than \modelUshaped.

\change{
Overall, the results suggest that the \modelDPP and \modelUniform models are
relatively robust in terms of model-choice accuracy, and when model violations
do cause them to be biased, they tend to under-estimate the probability of the
model with a single, shared divergence event.}
In contrast, the \modelOld model is very sensitive to model violations, and
strongly over-estimates the probability of the one-divergence model whenever
the model is misspecified.
Furthermore, the results suggest that using exponentially distributed priors on
nuisance parameters rather than uniform priors helps the \modelUshaped model
perform better than \modelOld, but it is still hindered by the U-shaped prior
on divergence models and tends to over-estimate the probability of the
one-divergence model whenever there are violations of the model.

\subsection{Validation analyses: Ordered divergence models}
The results show that the method performs similarly when sampling over ordered
models of divergence
(Figures S\ref{figValidationAccuracyDPPOrderedDPPOrdered} \&
S\ref{figValidationModelChoiceDPPOrderedDPPOrdered}).
This suggests that the method is not adversely affected by the increase
in the number of possible discrete models (from 22 unordered to 4140 ordered
models) when there are eight pairs of populations.
\change{
This is encouraging, because, as discussed above, estimating unordered models
of divergence by shuffling the summary statistic vectors calculated from the
sequence alignments is not valid for most empirical datasets.
Given these results, estimation of unordered divergence models should be
avoided for empirical applications of the method.}

\subsection{Power analyses: Estimation accuracy}
All of the models I evaluated (Table \ref{tabPriors}) struggle to estimate the
variance of divergence times \divTimeDispersion regardless of which of the
three series of models (Table \ref{tabPowerModels}) the data were generated
under
(Figures
S\labelcref{figPowerAccuracyOld,figPowerAccuracyUniform,figPowerAccuracyExp}).
The models with the U-shaped prior on divergence models (\modelOld and
\modelUshaped) tend to underestimate the variance in divergence times 
(Plots A--L of Figures
S\labelcref{figPowerAccuracyOld,figPowerAccuracyUniform,figPowerAccuracyExp}).
whereas the models with Uniform or Dirichlet-process priors over divergence
models tend to overestimate variance in divergence times
(Plots M--X of Figures
S\labelcref{figPowerAccuracyOld,figPowerAccuracyUniform,figPowerAccuracyExp}).

When the divergence times of the 22 population pairs are randomly drawn from a
series of exponential priors (\powerSeriesExp), the \modelDPP model is the
best estimator of \divTimeDispersion, followed by \modelUniform
(Figure~S\ref{figPowerAccuracyExp}).
The \modelOld model is strongly biased toward underestimating
\divTimeDispersion, estimating values of zero for most of the replicates across
all the data models of \powerSeriesExp (Figure~S\ref{figPowerAccuracyExp}).
The results of the \modelUshaped model are intermediate between those of
\modelOld and the new models \modelDPP and \modelUniform
(Figure~S\ref{figPowerAccuracyExp}).

Similarly, when the true divergence times are randomly drawn from a series of
uniform priors (\powerSeriesUniform), the \modelDPP and \modelUniform models
tend to over-estimate the variance in divergence times, whereas the \modelOld
model underestimates \divTimeDispersion, estimating values of zero for most
replicates across all the data models of \powerSeriesUniform
(Figure~S\ref{figPowerAccuracyUniform}).
Again, the performance of the \modelUshaped model is intermediate between the
\modelOld and \modelDPP/\modelUniform models (Figure~S\ref{figPowerAccuracyUniform}).
The results are very similar when the four models are applied to the data
simulated under the \powerSeriesOld series of models
(Figure~S\ref{figPowerAccuracyOld}).

\subsection{Power analyses: Model choice}
The modifications of the \msb model decrease the method's bias toward clustered
divergences when applied to data generated under random divergence times
(Figure~\ref{figPowerPsiOld4} \& 
S\labelcref{figPowerPsiOld,figPowerPsiUniform,figPowerPsiExp}).
The \modelOld model performs the worst of the four models across
all three series of data-generating models, inferring a single divergence event across
most of the 18,000 simulations
(Figure~\ref{figPowerPsiOld4}A--D \& plots A--F of Figures
S\labelcref{figPowerPsiOld,figPowerPsiUniform,figPowerPsiExp}).
Importantly, the \modelOld \change{model} tends to strongly support these
estimates of one divergence across most of the simulations
(Figure~\ref{figPowerPsiProbOld4}A--D \& plots A--F of Figures
S\labelcref{figPowerPsiProbOld,figPowerPsiProbUniform,figPowerPsiProbExp}).
The \modelDPP model also prefers the one-divergence model 
when divergences are random over narrow windows of time, but
performs much better when divergences are random over a timescale
of 1--2 coalescent units
(Figure~\ref{figPowerPsiOld4}M--P \& plots S--X of Figures
S\labelcref{figPowerPsiOld,figPowerPsiUniform,figPowerPsiExp}).
However, even when \modelDPP infers the one-divergence model over narrow
timescales, the posterior probability support is always low
(Figure~\ref{figPowerPsiProbOld4}M--P \& plots S--X of Figures
S\labelcref{figPowerPsiProbOld,figPowerPsiProbUniform,figPowerPsiProbExp}).
The \modelUniform model never infers the one-divergence model in any of the
simulation replicates but still tends to infer relatively few (4--6) divergence
events when divergences are random over longer periods 
(Figure~\ref{figPowerPsiOld4} I--L \& plots M--R of Figures
S\labelcref{figPowerPsiOld,figPowerPsiUniform,figPowerPsiExp}).
Using exponential priors on divergence-time and demographic parameters does
increase the power of the \modelUshaped model compared to \modelOld across all
three series of data models, but the U-shaped prior still prevents the model
from performing as well as the \modelDPP and \modelUniform models 
(Figure~\ref{figPowerPsiOld4} \& 
S\labelcref{figPowerPsiOld,figPowerPsiUniform,figPowerPsiExp}).

The improved power of the new models is even more pronounced when looking at
estimates of the variance of divergence times (\divTimeDispersion) across the
simulations
(Figure~\ref{figPowerOmegaOld4} \& 
S\labelcref{figPowerOmegaOld,figPowerOmegaUniform,figPowerOmegaExp}).
The performance among the models is so different, that the histograms of
\divTimeDispersion estimates cannot be plotted along a shared x-axis.
The \modelDPP and \modelUniform models perform similarly across all three
series of data models, inferring values of \divTimeDispersion consistent with
one divergence event ($\divTimeDispersion < 0.01)$ in almost none of the
replicates across all the simulations.
In contrast, the \modelOld model infers values consistent with a single
divergence event in most of the replicates across all the simulations.
Using exponential priors on divergence-time and demographic parameters greatly
increases the power of the \modelUshaped model to detect variation in
divergence times relative to \modelOld, but it still has less power than the
models with Dirichlet-process or uniform priors across divergence models 
(Figure~\ref{figPowerOmegaOld4} \& 
S\labelcref{figPowerOmegaOld,figPowerOmegaUniform,figPowerOmegaExp}).
Although the \divTimeDispersion threshold of 0.01 is arbitrary, Oaks et al.\
\cite{Oaks2012} did show via simulation that the true value of
\divTimeDispersion will almost always be greater than 0.01 when divergences are
random over periods of 0.1 coalescent units or more (see Figure~S4 of
\cite{Oaks2012}).

As mentioned above, the increased power of the new models is also evident when
looking at the estimated posterior probability of the one-divergence model
across the power analyses 
(Figure~\ref{figPowerPsiProbOld4} \& 
S\labelcref{figPowerPsiProbOld,figPowerPsiProbUniform,figPowerPsiProbExp}).
The \modelDPP and \modelUniform models estimate low posterior probability of
$\divTimeNum = 1$ across all of the simulations.
This is in contrast to the \modelOld model, which infers high posterior probabilities of a single
divergence for most replicates across all simulations
(Figure~\ref{figPowerPsiProbOld4} \& 
S\labelcref{figPowerPsiProbOld,figPowerPsiProbUniform,figPowerPsiProbExp}).
The exponential priors on divergence-time and demographic parameters (model
\modelUshaped) result in lower estimates of the probability of one divergence
when compared to \modelOld, but higher estimates when compared to \modelUniform
and \modelDPP
(Figure~\ref{figPowerPsiProbOld4} \& 
S\labelcref{figPowerPsiProbOld,figPowerPsiProbUniform,figPowerPsiProbExp}).
The \modelDPP and \modelUniform models do frequently support the one-divergence
model according to a Bayes factor criterion of greater than 10, but still
less frequently than the \modelOld model.
This result is not surprising given the extremely small prior probability of
the one-divergence model under the \modelDPP and \modelUniform models (i.e.,
very few posterior samples of the one-divergence model will result in a
large Bayes factor under these models).
However, the small posterior probability of the one-divergence model estimated
under \modelDPP and \modelUniform should prevent an investigator from
overinterpreting the Bayes factor as strong support for clustered divergences.

Lastly, when looking at the estimated posterior probability of
\divTimeDispersion being consistent with one shared divergence
($p(\divTimeDispersion < 0.01 | \ssSpace)$), I find the same pattern of model
behavior, with \modelDPP and \modelUniform inferring low probabilities across
all simulations, \modelOld inferring high probabilities, and \modelUshaped
inferring intermediate values
(Figure~\ref{figPowerOmegaProbOld4} \& 
S\labelcref{figPowerOmegaProbOld,figPowerOmegaProbUniform,figPowerOmegaProbExp}).

\subsection{Empirical results}
As expected based on the results of Oaks et al.\ \cite{Oaks2012}, when the
Philippines data are analyzed under the \empModelOld model, there is strong
support for very few divergence events shared among all 22 pairs of taxa, with
a maximum \emph{a posteriori} (MAP) estimate of one-shared divergence
(Figure~\ref{figPhilippines}A).
When these data are analyzed using models allowed by the new implementation,
there is much less support for highly clustered models and much greater
uncertainty regarding the number of divergence events shared among the taxa,
especially under the DPP models (Figure~\ref{figPhilippines}B--E).
Figure \ref{figPhilippines} also shows the prior distribution across the number
of divergence events (\divTimeNum) for each model, as well as the
average prior probability of an unordered and ordered model of divergence
(\divTimeIndexVector) across \divTimeNum.
Estimates under the new models tend to be similar to the prior, which is
expected under such a parameter-rich model when there is limited information
from the data (four summary statistics from a single locus for each pair of
taxa).

The disparity between the results of the \empModelOld model and the new models
is even more pronounced when looking at the 10 divergence models
(\divTimeIndexVector) estimated to have the highest probability under each of the
models
(Figures~S\labelcref{figDivModelsDPP,figDivModelsDPPInform,figDivModelsDPPSimple,figDivModelsUniform,figDivModelsOld}).
Again, the new models estimate more divergences, a large amount of posterior
uncertainty, and an order of magnitude smaller probability for their respective
MAP-divergence model when compared to the \empModelOld model
(Figures~S\labelcref{figDivModelsDPP,figDivModelsDPPInform,figDivModelsDPPSimple,figDivModelsUniform,figDivModelsOld}).

Figure \ref{figNegrosPanay} shows the estimated posterior probability
distribution over the number of divergence events when the data from the
nine-taxon pairs from the Islands of Negros and Panay are analyzed under DPP
models that sample over unordered (\npModelDPP) and ordered
(\npModelDPPOrdered) models of divergence.
The results are similar under both models and, again, yield a large amount of
uncertainty about the number of divergence events that is similar to the prior
uncertainty.
\change{
The small difference between the results of the \npModelDPP and
\npModelDPPOrdered models is consistent across multiple analyses, and thus
could be due to error introduced to the \npModelDPP model by the invalid
shuffling of the summary statistic vectors.
}
Both models estimate a similar set of 10 unordered divergence models with the
highest posterior probability
(Figures~S\labelcref{figDivModelsNP,figDivModelsNPOrdered}).

\change{
The main advantages of the \npModelDPPOrdered model over the \npModelDPP are
that
(1) the incorrect shuffling of the summary statistic vectors is avoided,
(2) the identity of the taxa is maintained, and thus a fully marginalized
estimate of divergence times across the taxa can be obtained
(Figure~S\ref{figMarginalTimes}), and
(3) the probability of co-divergence among any set of taxa can be
estimated from the posterior sample.
}

\section{Discussion}
My results demonstrate that using alternative priors on parameters and
divergence models improved the behavior of the \msb model.
In the new implementation, model-choice estimation is more accurate and shows
greater robustness to model violations
(Figure~\ref{figValidationModelChoicePsi} \&
\ref{figValidationModelChoiceOmega}).
The original model is very sensitive to violations and, when present,
strongly over-estimates the probability of one-divergence event shared
across all taxa
(Figure~\ref{figValidationModelChoicePsi} \&
\ref{figValidationModelChoiceOmega}).
\change{
When more appropriate priors are used for divergence-time and demographic
parameters, and either a Dirichlet-process or uniform prior applied across
divergence models, the model is less sensitive to violations, and, when
violations do cause bias, the method tends to underestimate
the probability of models with temporally clustered divergences
(Figure~\ref{figValidationModelChoicePsi} \&
\ref{figValidationModelChoiceOmega}).
Given that clustered models are often of particular interest to biogeographers,
this behavior of the new method can be considered conservative.
}

The modifications also improve the method's power to detect random
variation in divergence times, reducing the tendency to estimate clustered
divergences
(Figures~\labelcref{figPowerPsiOld4,figPowerPsiProbOld4,figPowerOmegaOld4,figPowerOmegaProbOld4}).
My results are similar to those of Oaks et al.\ \cite{Oaks2012} in that I find
\msb will often infer strong support for clustered divergences when divergences
are random over quite broad timescales
(Figures~\labelcref{figPowerPsiOld4,figPowerPsiProbOld4,figPowerOmegaOld4,figPowerOmegaProbOld4}).
My results expand on this by showing that this behavior is consistent
across a range of conditions underlying the data.
The new method, \dppmsbayes, has greater power to detect random temporal
variation in divergences, is less prone to spurious inference of clustered
divergence models, and much less likely to incorrectly infer such models with
strong support
(Figures~\labelcref{figPowerPsiOld4,figPowerPsiProbOld4,figPowerOmegaOld4,figPowerOmegaProbOld4}).

By evaluating a model intermediate between the old and new implementation
(\modelUshaped), I was able to determine the relative affects of my 
modifications to the model.
Across all of the analyses, the results show that using better priors on
divergence-time and demographic parameters alone does improve the performance
of the method.
The magnitude of the bias toward inferring support for the one-divergence
model when there are model violations is reduced when the exponential
priors are used in place of the uniform priors
(Figure~\ref{figValidationModelChoicePsi} \&
\ref{figValidationModelChoiceOmega}).
Furthermore, using exponential priors improves the method's power to detect
temporally random divergences
(Figures~\labelcref{figPowerPsiOld4,figPowerPsiProbOld4,figPowerOmegaOld4,figPowerOmegaProbOld4}).
Throughout the analyses, the intermediate model (\modelUshaped) performs better
than the \msb model, but not as well as the models with alternative priors on
divergence models.
This suggests, as predicted by Oaks et al.\ \cite{Oaks2012,Oaks2014reply}, that
the tendency of \msb to erroneously support models of temporally clustered
divergences is caused by a combination of
(1) small marginal likelihoods of models with more \divTime{} parameters due to
uniform priors on divergence-time and demographic parameters and
(2) the U-shaped prior on divergence models giving low prior density to models
with intermediate numbers of divergence parameters.
The former essentially rules out models with many \divTime{} parameters, which
causes the latter to act like an "L-shaped" prior with a spike of
prior density on the one-divergence model.
Given the parameter richness of the model and the relatively small amount of
information in the summary statistics, it is not surprising that the
combination of these two factors can create a strong tendency to infer
clustered models of divergence.

While the modifications improve the behavior of the model, I urge caution when
using the method and interpreting its results.
The method attempts to approximate the posterior of a very parameter-rich model
using relatively little information from the data.
For example, when applied to the dataset of 22 taxon pairs from the Philippines
\cite{Oaks2012}, the model has as many as 604--625 free parameters
(depending on \divTimeNum), and samples over 1002 unordered divergence models.
Even under the simplest possible model allowed under the new implementation,
the model still has 471--492 free parameters.
Furthermore, the stochastic coalescent and mutational processes being modeled
predict a large amount of variation in possible datasets even when the
parameter values are known.
The richness and stochastic nature of the model makes for a difficult inference
problem, especially when using a small number of summary statistics calculated
from the sequence alignments of each taxon pair.
The population-genetic summary statistics used by the method contain little
information about many of the free parameters in the model.
Thus, I expect the improved method will still be sensitive to priors, and the
power, while improved, may still be low.
While there is much less prior sensitivity under the new model compared to
those observed by Oaks et al.\ \cite{Oaks2012}, there is still an effect when
comparing the results of the empirical data analyzed under a diffuse
(\empModelDPP) and informative (\empModelDPPInform) divergence-time prior
(Figure~\ref{figPhilippines} C versus D).
The fact that the posterior shifts toward the prior under the informative prior
suggests that the shift away from the prior toward fewer divergence events
under the diffuse prior might still be caused by small marginal likelihoods
of models with more divergence-time parameters (Figure~\ref{figPhilippines}).

Nonetheless, it is reassuring to see a large amount of posterior uncertainty
when the new implementation is applied to the empirical datasets
(Figure~\labelcref{figPhilippines,figNegrosPanay}).
Applications of the \msb model often result in strong posterior support for
estimated scenarios
(e.g.,
\cite{Barber2010,Carnaval2009,Chan2011,Hickerson2006,Leache2007,Plouviez2009,Stone2012,Voje2009,Oaks2012}),
as I found here (Figure~\ref{figPhilippines}).
Given the richness of the model, the variance of the processes being modeled,
and the relatively small amount of information in the summary statistics
calculated from the sequence data, finding strong posterior support for any
scenario is unexpected.
Based on results of the empirical and power analyses
(Figures~\labelcref{figPowerPsiProbOld4,figPowerOmegaProbOld4,figPhilippines,figNegrosPanay}),
the new implementation more accurately reflects posterior uncertainty and
avoids spurious support for biogeographical scenarios.

\change{
I also urge caution when using \dppmsbayes due to the lack of theoretical
validation of Bayesian model choice when the full data are replaced by summary
statistics that are insufficient for discriminating across models under
comparison \cite{Robert2011}, which is certainly the case here.
Robert et al.\ \cite{Robert2011} demonstrated that ABC estimates of model
posterior probabilities can be inaccurate when such across-model insufficient
statistics are used.

Given all of these caveats, I encourage investigators to view this method as a
means of exploring their data for general temporal patterns of divergences
across taxa, rather than a rigorous means of evaluating hypotheses.
As recommended by Oaks et al.\ \cite{Oaks2012}, any results from the method
should be accompanied by
(1) analyses under a variety of priors to assess the assumptions underlying
model inference and the prior sensitivity of the results, and
(2) simulation-based power analyses to provide insight into the temporal
resolution of the method.
Both approaches are important to help guide the interpretation of results.
}

Given the difficulty of this estimation problem, I anticipate that
full-likelihood methods that can leverage all of the information present in the
sequence data will become increasingly important for robustly estimating shared
evolutionary history across taxa \cite{JeetDiss}.
With improving numerical methods for sampling over models of differing
dimensionality \cite{Green1995,Lemey2009}, advances in Monte Carlo techniques
\cite{Jordan2012}, and increasing efficiency of likelihood calculations
\cite{Ayres2012}, analyzing rich comparative phylogeograpical models in a
full-likelihood Bayesian framework is becoming computationally practical,
especially when considering that simulating millions of random datasets from
the prior under the simple ABC rejection approach is inefficient and
computationally nontrivial.

\section{Conclusions}
\change{
I introduced a new model for estimating shared divergence histories across taxa
from DNA sequence data within an approximate-Bayesian model-choice framework.
}
The new method, \dppmsbayes, takes a non-parametric approach to the problem by
using a Dirichlet-process prior on the temporal distribution of divergences
across taxa.
The new method shows improved robustness, accuracy, and power compared to the
existing method, \msb.
\change{
Compared to \msb, the new approach better estimates posterior uncertainty,
which greatly reduces the chances of incorrectly estimating biogeographical
scenarios of shared divergence events.
This is important, because models of shared divergence events are often of
particular interest to researchers who employ these methods.
}
This new tool will allow evolutionary biologists to better leverage comparative
genetic data to assess the affects of regional and global biogeographical
processes on biodiversity.

\section{Acknowledgments}
\change{
I am greatful to Melissa Callahan, Mark Holder, Emily McTavish, Daniel Money,
Jordan Koch, Adam Leach\'{e}, Peter Foster, two anonymous reviewers, and
Christian Robert and his blog
(\href{http://xianblog.wordpress.com/}{\url{http://xianblog.wordpress.com/}})
for insightful comments that greatly improved this work.
}
I thank the National Science Foundation for supporting this work (DEB
1011423 and DBI 1308885).
This work was also supported by the University of Kansas (KU) Office of Graduate
Studies, Society of Systematic Biologists, Sigma Xi Scientific Research
Society, KU Department of Ecology and Evolutionary Biology, and the KU
Biodiversity Institute.
I also thank Mark Holder, the KU Information and Telecommunication Technology
Center, KU Computing Center, and the iPlant Collaborative for the computational
support necessary to conduct the analyses presented herein.

\bibliographystyle{bmc-mathphys.bst}
\bibliography{references}

\newpage
\singlespacing

\renewcommand\listfigurename{Figure Captions}
\cftsetindents{fig}{0cm}{2.2cm}
\renewcommand\cftdotsep{\cftnodots}
\setlength\cftbeforefigskip{10pt}
\cftpagenumbersoff{fig}
\listoffigures

\newpage
\singlespacing

\begin{table}[htbp]
    \sffamily
    \scriptsize
    \ifbmc{}{\rowcolors{2}{}{myGray}}
    \addtolength{\tabcolsep}{-0.1cm}
\caption{Summary of the notation used throughout this work; modified from Oaks et al.\ \cite{Oaks2012}.}
    \centering
    \begin{tabular}{ l >{\raggedright\hangindent=0.5cm}m{14cm} }
        \toprule
        \textbf{Symbol} & \textbf{Description} \tn
        \midrule
        \npairs{} & Number of population pairs. \tn
        \popSampleSize{i}{} & The number of genome copies sampled from population pair $i$, with \popSampleSize{1}{i} sampled from population 1, and \popSampleSize{2}{i} from population 2. \tn
        \nloci{i} & Number of loci sampled from population pair $i$. \tn
        \nlociTotal & Total number of unique loci sampled. \tn
        \alignment{i}{j} & Sequence alignment of locus $j$ sampled from population pair $i$. \tn
        \alignmentSSObs{i}{j} & Population genetic summary statistics calculated from \alignment{i}{j}. \tn
        \alignmentVector & Vector containing the sequence alignments of each locus from each population pair: $(\alignment{1}{1},\ldots,\alignment{\npairs{}}{\nloci{\npairs{}}})$. \tn
        \ssVectorObs & Vector containing the summary statistics of each locus from each population pair: $(\alignmentSSObs{1}{1},\ldots,\alignmentSSObs{\npairs{}}{\nloci{\npairs{}}})$. \tn
        \ssSpace & Multi-dimensional Euclidean space around the observed summary statistics, \ssVectorObs. \tn
        \tol & Radius of \ssSpace, i.e., the tolerance of the ABC estimation. \tn
        \geneTree{i}{j} & Gene tree of the sequences in \alignment{i}{j}. \tn
        \geneTreeVector & Vector containing the gene trees of each locus from each population pair: $(\geneTree{1}{1},\ldots,\geneTree{\npairs{}}{\nloci{\npairs{}}})$. \tn
        \divTimeNum & Number of population divergence-time parameters shared among the \npairs{} population pairs. \tn
        \divTime{} & Time of population divergence in \globalcoalunit generations. \tn
        \divTimeVector & Set of divergence-time parameters: $\{\divTime{1},\ldots,\divTime{\divTimeNum}\}$. \tn
        \divTimeIndex{i} & The index of the divergence-time in \divTimeVector to which population pair $i$ is mapped. \tn
        \divTimeIndexVector & Vector of divergence-time indices: $(\divTimeIndex{1},\ldots,\divTimeIndex{\npairs{}})$. \tn
        \divTimeMap{i} & Time of divergence in \globalcoalunit generations between the populations of pair $i$. \tn
        \divTimeMapVector & Vector of divergence times for each of the population pairs: $(\divTimeMap{1},\ldots,\divTimeMap{\npairs{}})$. \tn
        \divTimeScaled{i}{j} & Scaled time of divergence between the populations of pair $i$ for locus $j$. \tn
        \divTimeScaledVector & Vector containing the scaled divergence times of each locus from each population pair: $(\divTimeScaled{1}{1},\ldots,\divTimeScaled{\npairs{}}{\nloci{\npairs{}}})$. \tn
        \descendantTheta{1}{i}, \descendantTheta{2}{i} & Mutation-rate-scaled effective population size of the $1^{st}$ and $2^{nd}$ descendent population, respectively, of pair $i$. \tn
        \ancestralTheta{i} & Mutation-rate-scaled effective population size of the population ancestral to pair $i$. \tn
        \descendantThetaVector{1}, \descendantThetaVector{2} & Vectors $(\descendantTheta{1}{1},\ldots,\descendantTheta{1}{\npairs{}})$ and $(\descendantTheta{2}{1},\ldots,\descendantTheta{2}{\npairs{}})$, respectively. \tn
        \ancestralThetaVector & Vector containing the \ancestralTheta{} parameters for each population pair: $(\ancestralTheta{1},\ldots,\ancestralTheta{\npairs{}})$. \tn
        \locusMutationRateScalar{j} & Mutation-rate multiplier of locus $j$. \tn
        \locusMutationRateScalarVector & Vector containing the locus-specific mutation-rate multipliers: $(\locusMutationRateScalar{1},\ldots,\locusMutationRateScalar{\nlociTotal})$. \tn
        \locusRateHetShapeParameter & The shape parameter of the gamma prior distribution on \locusMutationRateScalar{}. \tn
        \bottleScalar{1}{i}, \bottleScalar{2}{i} & \myTheta{}-scaling parameters that determine the magnitude of the population bottleneck in the $1^{st}$ and $2^{nd}$ descendant population of pair $i$, respectively. The bottleneck in each descendant population begins immediately after divergence. \tn
        \bottleScalarVector{1}, \bottleScalarVector{2} & Vectors $(\bottleScalar{1}{1},\ldots,\bottleScalar{1}{\npairs{}})$ and $(\bottleScalar{2}{1},\ldots,\bottleScalar{2}{\npairs{}})$, respectively. \tn
        \bottleTime{i} & Proportion of time between present and \divTimeMap{i} when the bottleneck ends for the descendant populations of pair $i$. \tn
        \bottleTimeVector & Vector containing the \bottleTime{} parameters for each population pair: $(\bottleTime{1},\ldots,\bottleTime{\npairs{}})$. \tn
        \migrationRate{i} & Symmetric migration rate between the descendant populations of pair $i$. \tn
        \migrationRateVector & Vector containing the migration rates for each population pair: $(\migrationRate{1},\ldots,\migrationRate{\npairs{}})$. \tn
        \ploidyScalar{i}{j} & \myTheta{}-scaling constant provided by the investigator for locus $j$ of pair $i$. This constant is required to scale \myTheta{} for differences in ploidy among loci or differences in generation times among taxa. \tn
        \mutationRateScalarConstant{i}{j} & \myTheta{}-scaling constant provided by the investigator for locus $j$ of pair $i$. This constant is required to scale \myTheta{} for differences in mutation rates among loci or among taxa. \tn
        \ploidyScalarVector & Vector of ploidy and/or generation-time scaling constants: $(\ploidyScalar{1}{1},\ldots,\ploidyScalar{\npairs{}}{\nloci{\npairs{}}})$. \tn
        \mutationRateScalarConstantVector & Vector of mutation-rate scaling constants: $(\mutationRateScalarConstant{1}{1},\ldots,\mutationRateScalarConstant{\npairs{}}{\nloci{\npairs{}}})$. \tn
        \divTimeMean  & Mean of divergence times across the \npairs{} population pairs. \tn
        \divTimeVar  & Variance of divergence times across the \npairs{} population pairs. \tn
        \divTimeDispersion  & Dispersion index of divergence times across the \npairs{} population pairs ($\divTimeVar/\divTimeMean$). \tn
        \numPriorSamples & Number of samples from the joint prior. \tn
        \paramSampleVector{} & Vector of parameter values drawn from the joint prior. \tn
        \ssVector{} & Vector containing the summary statistics calculated from data simulated under parameter values drawn from the prior (\hpvector{}). \tn
        \paramSampleMatrix{} & Random sample of $\paramSampleVector{1}, \ldots, \paramSampleVector{\numPriorSamples}$ drawn form the prior. \tn
        \ssMatrix & Summary statistic vectors $\ssVector{1}, \ldots, \ssVector{\numPriorSamples}$ for each $\hpvector{1}, \ldots, \hpvector{\numPriorSamples}$ drawn from the prior. \tn
        \bottomrule
    \end{tabular}
    \label{tabNotation}
\end{table}

\clearpage

\begin{table}[htbp]
    \sffamily
    \scriptsize
    \addtolength{\tabcolsep}{-0.08cm}
    \ifbmc{}{\rowcolors{2}{}{myGray} \mcrowcolors}
    \caption{The models evaluated in the simulation-based analyses.  For
        the \modelDPP model, the prior on the concentration parameter,
        $\concentrationParam \sim Gamma(\cdot,\cdot)$, was set to $Gamma(2,2)$
        for the validation analyses and $Gamma(1.5,18.1)$ for the power
        analyses. The distributions of divergence times are given in units of
        \globalcoalunit generations followed in brackets by units of millions
        of generations ago (MGA), with the former converted to the latter
        assuming a per-site rate of 1\e{-8} mutations per generation. For model
        \modelOld, the priors for theta parameters are $\ancestralTheta{} \sim
        U(0, 0.05)$ and $\descendantTheta{1}{}, \descendantTheta{2}{} \sim
        Beta(1, 1) \times 2 \times U(0, 0.05)$. The later is summarized as
        $\descendantThetaMean{} \sim U(0, 0.05)$. For the \modelDPP and
        \modelUniform, and \modelUshaped models, \ancestralTheta{},
        \descendantTheta{1}{}, and \descendantTheta{2}{} are independently
        and exponentially distributed with a mean of 0.025.}
    \centering
    \begin{tabular}{ l l l l l }
        \toprule
        & \multicolumn{4}{c}{Priors} \\
        \cmidrule(){2-5}
        Model & \divTimeIndexVector & \divTime{} & \myTheta{} &  \\
        \midrule
            \modelOld & $\divTimeIndexVector \sim \priorOld$
                      & $\divTime{} \sim U(0,10 \; [25 \; MGA])$
                      & $\ancestralTheta{} \sim U(0, 0.05)$
                      & $\descendantThetaMean{} \sim U(0, 0.05)$ \\
            \modelUshaped & $\divTimeIndexVector \sim \priorOld$
                          & $\divTime{} \sim Exp(mean=2.887 \; [7.22 \; MGA])$
                          & \multicolumn{2}{l}{$\ancestralTheta{} \sim \descendantTheta{1}{} \sim \descendantTheta{2}{} \sim Exp(mean=0.025)$} \\
            \modelUniform & $\divTimeIndexVector \sim \priorUniform$
                          & $\divTime{} \sim Exp(mean=2.887 \; [7.22 \; MGA])$
                          & \multicolumn{2}{l}{$\ancestralTheta{} \sim \descendantTheta{1}{} \sim \descendantTheta{2}{} \sim Exp(mean=0.025)$} \\
            \modelDPP & $\divTimeIndexVector \sim \priorDPP{\sim Gamma(\cdot,\cdot)}$
                      & $\divTime{} \sim Exp(mean=2.887 \; [7.22 \; MGA])$
                      & \multicolumn{2}{l}{$\ancestralTheta{} \sim \descendantTheta{1}{} \sim \descendantTheta{2}{} \sim Exp(mean=0.025)$} \\
        \bottomrule
    \end{tabular}
    \label{tabPriors}
\end{table}

\clearpage

\begin{table}[htbp]
    \sffamily
    \footnotesize
    \ifbmc{}{\rowcolors{2}{}{myGray} \mcrowcolors}
    \caption{The models used to simulate pseudo-replicate datasets for
        assessing the power of the models in Table \ref{tabPriors}.  The
        distributions of divergence times are given in units of \globalcoalunit
        generations followed in brackets by units of millions of generations
        ago (MGA), with the former converted to the latter assuming a per-site
        rate of 1\e{-8} mutations per generation. For all of the
        \powerSeriesOld models, the priors for theta parameters are
        $\ancestralTheta{} \sim U(0, 0.05)$ and $\descendantTheta{1}{},
        \descendantTheta{2}{} \sim Beta(1, 1) \times 2 \times U(0, 0.05)$. The
        later is summarized as $\descendantThetaMean{} \sim U(0, 0.05)$. For
        the \powerSeriesUniform and \powerSeriesExp models, \ancestralTheta{},
        \descendantTheta{1}{}, and \descendantTheta{2}{} are independently
        and exponentially distributed with a mean of 0.025.}
    \centering
    \begin{tabular}{ l l l l l }
        \toprule
        & \multicolumn{4}{c}{Priors} \\
        \cmidrule(){2-5}
        Model series & \divTimeIndexVector & \divTime{} & \myTheta{} &  \\
        \midrule
            \powerSeriesOld & $\divTimeNum = 22$
                            & $\divTime{} \sim U(0,0.2 \; [0.5 \; MGA])$
                            & $\ancestralTheta{} \sim U(0, 0.05)$
                            & $\descendantThetaMean{} \sim U(0, 0.05)$ \\
                            & $\divTimeNum = 22$
                            & $\divTime{} \sim U(0,0.4 \; [1.0 \; MGA])$
                            & $\ancestralTheta{} \sim U(0, 0.05)$
                            & $\descendantThetaMean{} \sim U(0, 0.05)$ \\
                            & $\divTimeNum = 22$
                            & $\divTime{} \sim U(0,0.6 \; [1.5 \; MGA])$
                            & $\ancestralTheta{} \sim U(0, 0.05)$
                            & $\descendantThetaMean{} \sim U(0, 0.05)$ \\
                            & $\divTimeNum = 22$
                            & $\divTime{} \sim U(0,0.8 \; [2.0 \; MGA])$
                            & $\ancestralTheta{} \sim U(0, 0.05)$
                            & $\descendantThetaMean{} \sim U(0, 0.05)$ \\
                            & $\divTimeNum = 22$
                            & $\divTime{} \sim U(0,1.0 \; [2.5 \; MGA])$
                            & $\ancestralTheta{} \sim U(0, 0.05)$
                            & $\descendantThetaMean{} \sim U(0, 0.05)$ \\
                            & $\divTimeNum = 22$
                            & $\divTime{} \sim U(0,2.0 \; [5.0 \; MGA])$
                            & $\ancestralTheta{} \sim U(0, 0.05)$
                            & $\descendantThetaMean{} \sim U(0, 0.05)$ \\
        \midrule
            \powerSeriesUniform & $\divTimeNum = 22$
                            & $\divTime{} \sim U(0,0.2 \; [0.5 \; MGA])$
                            & \multicolumn{2}{l}{$\ancestralTheta{} \sim \descendantTheta{1}{} \sim \descendantTheta{2}{} \sim Exp(mean=0.025)$} \\
                            & $\divTimeNum = 22$
                            & $\divTime{} \sim U(0,0.4 \; [1.0 \; MGA])$
                            & \multicolumn{2}{l}{$\ancestralTheta{} \sim \descendantTheta{1}{} \sim \descendantTheta{2}{} \sim Exp(mean=0.025)$} \\
                            & $\divTimeNum = 22$
                            & $\divTime{} \sim U(0,0.6 \; [1.5 \; MGA])$
                            & \multicolumn{2}{l}{$\ancestralTheta{} \sim \descendantTheta{1}{} \sim \descendantTheta{2}{} \sim Exp(mean=0.025)$} \\
                            & $\divTimeNum = 22$
                            & $\divTime{} \sim U(0,0.8 \; [2.0 \; MGA])$
                            & \multicolumn{2}{l}{$\ancestralTheta{} \sim \descendantTheta{1}{} \sim \descendantTheta{2}{} \sim Exp(mean=0.025)$} \\
                            & $\divTimeNum = 22$
                            & $\divTime{} \sim U(0,1.0 \; [2.5 \; MGA])$
                            & \multicolumn{2}{l}{$\ancestralTheta{} \sim \descendantTheta{1}{} \sim \descendantTheta{2}{} \sim Exp(mean=0.025)$} \\
                            & $\divTimeNum = 22$
                            & $\divTime{} \sim U(0,2.0 \; [5.0 \; MGA])$
                            & \multicolumn{2}{l}{$\ancestralTheta{} \sim \descendantTheta{1}{} \sim \descendantTheta{2}{} \sim Exp(mean=0.025)$} \\
        \midrule
            \powerSeriesExp & $\divTimeNum = 22$
                            & $\divTime{} \sim Exp(mean = 0.058 \; [0.14 \; MGA])$
                            & \multicolumn{2}{l}{$\ancestralTheta{} \sim \descendantTheta{1}{} \sim \descendantTheta{2}{} \sim Exp(mean=0.025)$} \\
                            & $\divTimeNum = 22$
                            & $\divTime{} \sim Exp(mean = 0.115 \; [0.29 \; MGA])$
                            & \multicolumn{2}{l}{$\ancestralTheta{} \sim \descendantTheta{1}{} \sim \descendantTheta{2}{} \sim Exp(mean=0.025)$} \\
                            & $\divTimeNum = 22$
                            & $\divTime{} \sim Exp(mean = 0.173 \; [0.43 \; MGA])$
                            & \multicolumn{2}{l}{$\ancestralTheta{} \sim \descendantTheta{1}{} \sim \descendantTheta{2}{} \sim Exp(mean=0.025)$} \\
                            & $\divTimeNum = 22$
                            & $\divTime{} \sim Exp(mean = 0.231 \; [0.58 \; MGA])$
                            & \multicolumn{2}{l}{$\ancestralTheta{} \sim \descendantTheta{1}{} \sim \descendantTheta{2}{} \sim Exp(mean=0.025)$} \\
                            & $\divTimeNum = 22$
                            & $\divTime{} \sim Exp(mean = 0.289 \; [0.72 \; MGA])$
                            & \multicolumn{2}{l}{$\ancestralTheta{} \sim \descendantTheta{1}{} \sim \descendantTheta{2}{} \sim Exp(mean=0.025)$} \\
                            & $\divTimeNum = 22$
                            & $\divTime{} \sim Exp(mean = 0.577 \; [1.44 \; MGA])$
                            & \multicolumn{2}{l}{$\ancestralTheta{} \sim \descendantTheta{1}{} \sim \descendantTheta{2}{} \sim Exp(mean=0.025)$} \\
        \bottomrule
    \end{tabular}
    \label{tabPowerModels}
\end{table}

\begin{table}[htbp]
    \sffamily
    \footnotesize
    \addtolength{\tabcolsep}{-0.08cm}
    \ifbmc{}{\rowcolors{2}{myGray}{}}
    \caption{The models used to analyze the data from the 22 pairs of taxa from
        the Philippines ($\mathbf{M}$), and a subset of nine of those pairs
        from the Islands of Negros and Panay ($\mathbb{M}$).
        In addition to the $\popSampleSize{}{} - 1$ coalescent times, the
        \empModelDPPSimple has only a single \myTheta{} parameter for
        each taxon pair.
        The remaining $\mathbf{M}$ models have three \myTheta{}, two
        \bottleScalar{}{}, and one \bottleTime{} parameter.
        The distributions of divergence times are given in units of
        \globalcoalunit generations followed in brackets by units of millions
        of generations ago (MGA), with the former converted to the latter
        assuming a per-site rate of 1\e{-8} mutations per generation.
        The \npModelDPP model (and its \npModelDPPOrdered counterpart
        that samples over ordered divergence models) has only two
        \myTheta{} parameters (the descendant populations of each
        pair share the same \myTheta{} parameter, and there are no
        bottleneck parameters).}
    \centering
    \begin{tabulary}{\textwidth}{ l L }
        \toprule
        Model & Priors \\
        \midrule
            \empModelOld & $\divTimeIndexVector \sim \priorOld$ \tb
        $\divTime{} \sim U(0,34.64 \, [17.3 \; MGA])$ \tb
                      $\ancestralTheta{} \sim U(0, 0.01)$ \tb
                      $\descendantTheta{1}{}, \descendantTheta{2}{} \sim
                            Beta(1, 1) \times 2 \times U(0, 0.01)$ \tb
                      $\bottleScalar{1}{} \sim U(0, 1)$ \tb
                      $\bottleScalar{2}{} \sim U(0, 1)$ \\[0.25em]
            \empModelUniform & $\divTimeIndexVector \sim \priorUniform$ \tb
                      $\divTime{} \sim Exp(mean=10 \; [5 \; MGA])$ \tb
                      $\ancestralTheta{} \sim Exp(mean=0.005)$ \tb
                      $\descendantTheta{1}{} \sim Exp(mean=0.005)$ \tb
                      $\descendantTheta{2}{} \sim Exp(mean=0.005)$ \tb
                      $\bottleScalar{1}{} \sim Beta(5, 1)$ \tb
                      $\bottleScalar{2}{} \sim Beta(5, 1)$ \\[0.25em]
            \empModelDPP & $\divTimeIndexVector \sim \priorDPP{\sim Gamma(1.5,18.1)}$ \tb
                      $\divTime{} \sim Exp(mean=10 \; [5 \; MGA])$ \tb
                      $\ancestralTheta{} \sim Exp(mean=0.005)$ \tb
                      $\descendantTheta{1}{} \sim Exp(mean=0.005)$ \tb
                      $\descendantTheta{2}{} \sim Exp(mean=0.005)$ \tb
                      $\bottleScalar{1}{} \sim Beta(5, 1)$ \tb
                      $\bottleScalar{2}{} \sim Beta(5, 1)$ \tb \\
            \empModelDPPInform & $\divTimeIndexVector \sim \priorDPP{\sim Gamma(1.5,18.1)}$ \tb
                      $\divTime{} \sim Exp(mean=6 \; [3 \; MGA])$ \tb
                      $\ancestralTheta{} \sim Exp(mean=0.005)$ \tb
                      $\descendantTheta{1}{} \sim Exp(mean=0.005)$ \tb
                      $\descendantTheta{2}{} \sim Exp(mean=0.005)$ \tb
                      $\bottleScalar{1}{} \sim Beta(5, 1)$ \tb
                      $\bottleScalar{2}{} \sim Beta(5, 1)$ \tb \\
            \empModelDPPSimple & $\divTimeIndexVector \sim \priorDPP{\sim Gamma(1.5,18.1)}$ \tb
                      $\divTime{} \sim Exp(mean=10 \; [5 \; MGA])$ \tb
                      $\ancestralTheta{} = \descendantTheta{1}{} = \descendantTheta{2}{} \sim Exp(mean=0.005)$ \tb
                      $\bottleScalar{1}{} = \bottleScalar{2}{} = 1.0$ \\[0.25em]
            \npModelDPP & $\divTimeIndexVector \sim \priorDPP{\sim Gamma(1.5,5.0)}$ \tb
                      $\divTime{} \sim Exp(mean=10 \; [5 \; MGA])$ \tb
                      $\ancestralTheta{} \sim Exp(mean=0.005)$ \tb
                      $\descendantTheta{1}{} = \descendantTheta{2}{} \sim Exp(mean=0.005)$ \tb
                      $\bottleScalar{1}{} = \bottleScalar{2}{} = 1.0$ \\[0.25em]
        \bottomrule
    \end{tabulary}
    \label{tabEmpiricalModels}
\end{table}

\clearpage

\newpage

\mFigure{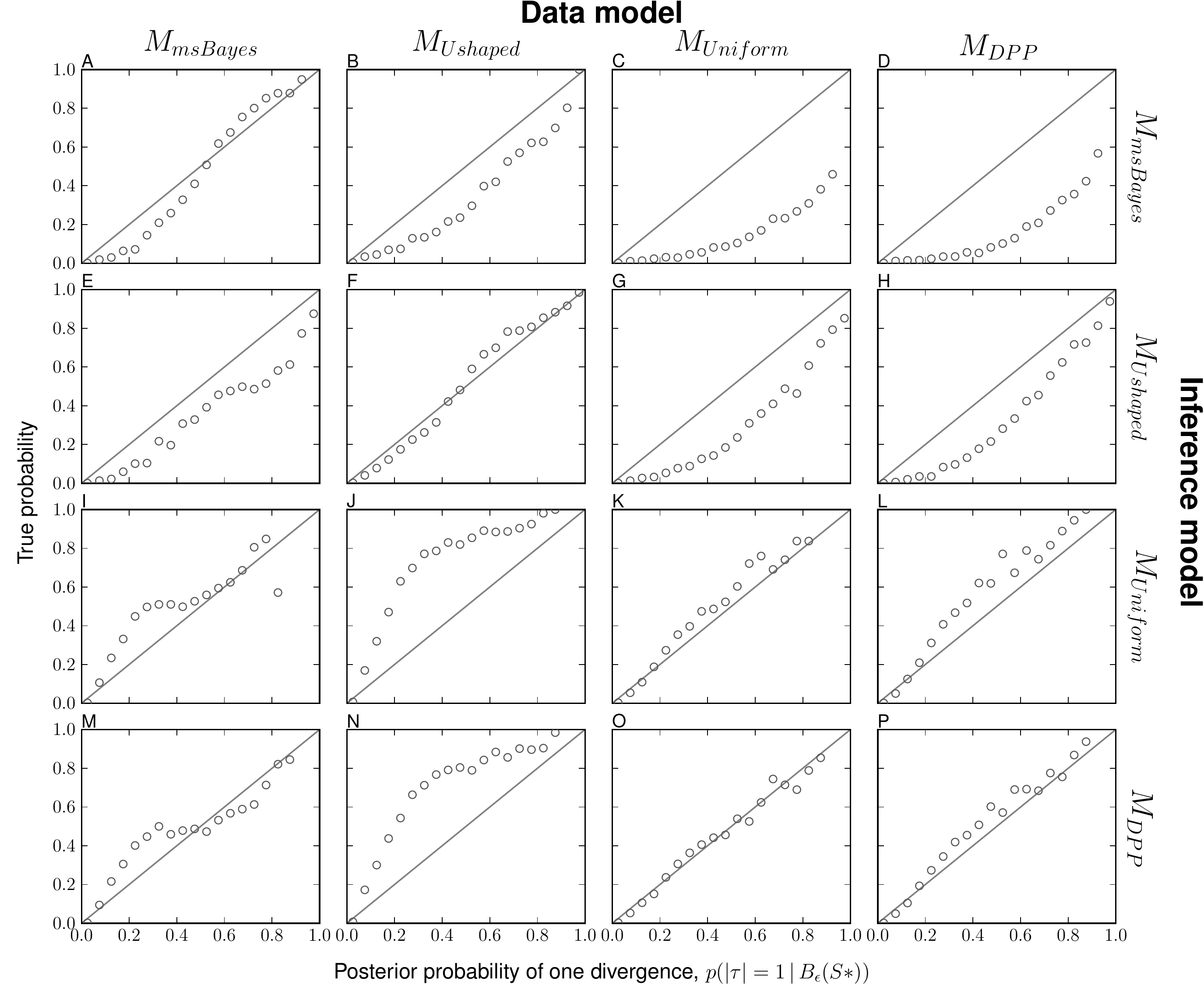}{%
    \ifbmc{\csentence{Comparison of model-choice accuracy.}}{}
    \validationModelChoiceComparisonCaption{unadjusted}{$\divTimeNum = 1$}
}{figValidationModelChoicePsi}

\mFigure{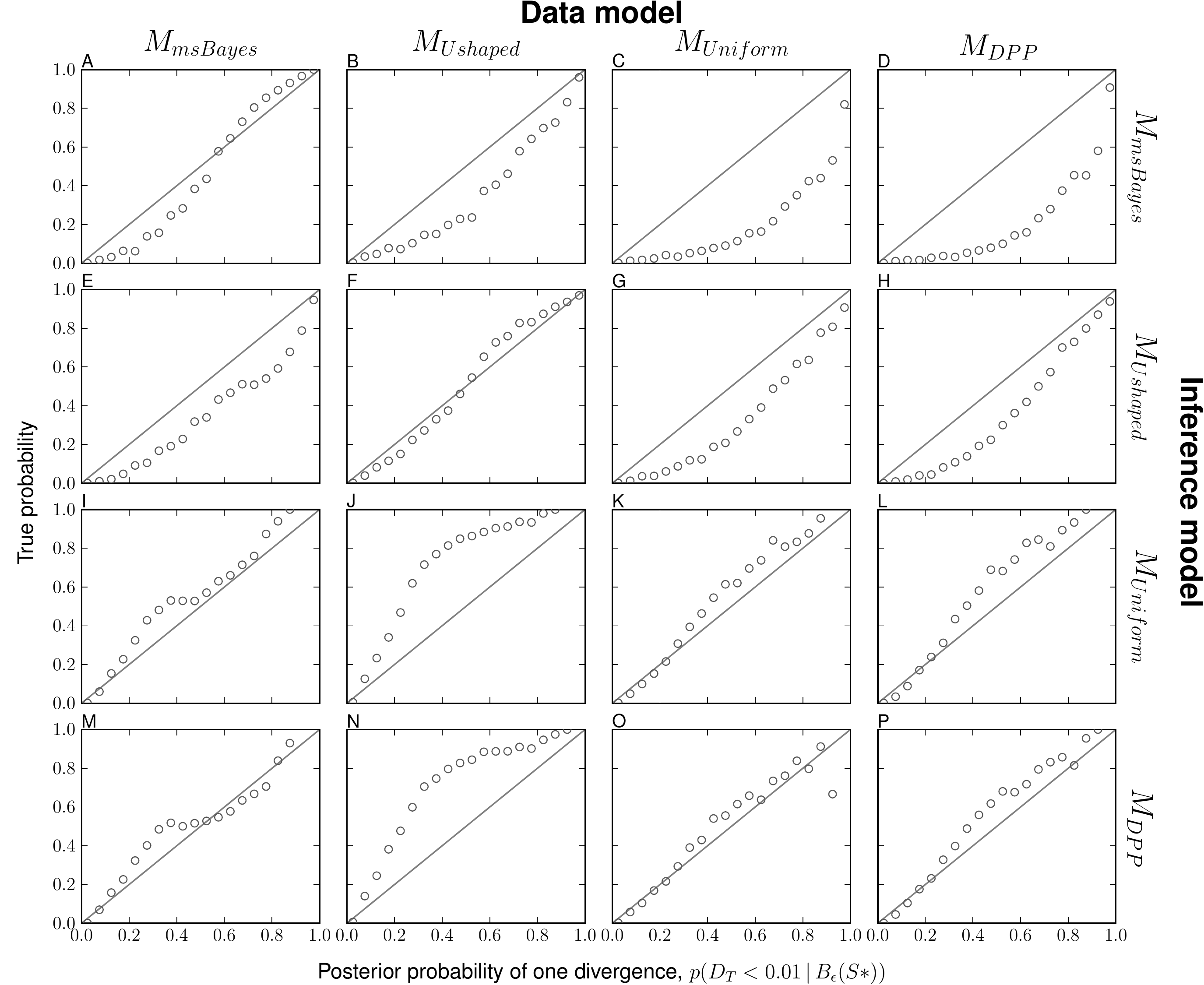}{
    \ifbmc{\csentence{Comparison of model-choice accuracy using
    \divTimeDispersion threshold.}}{}
    \validationModelChoiceComparisonCaption{unadjusted}{$\divTimeDispersion < 0.01$}
}{figValidationModelChoiceOmega}

\mFigure{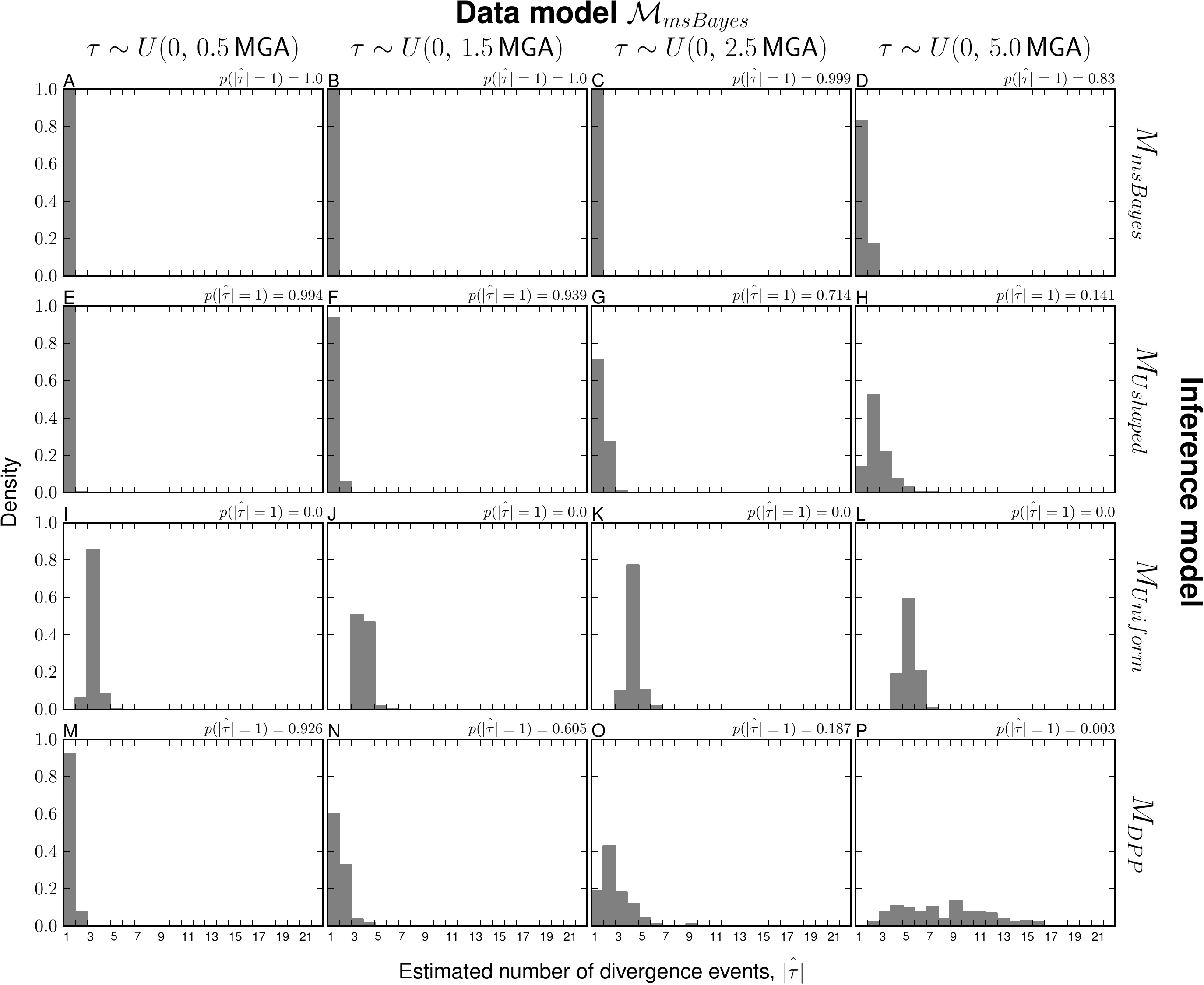}{
    \ifbmc{\csentence{Power to avoid spurious estimation of clustered
    divergences when divergence times are random.}}{}
    \powerCommentSummary{\powerSeriesOld}
    \powerPsiComment{\powerSeriesOld}
    \timeConversionComment
    Four of the six data-generating models of the \powerSeriesOld series
    are shown;
    please see Figure~S\ref{figPowerPsiOld} for all results.
}{figPowerPsiOld4}

\mFigure{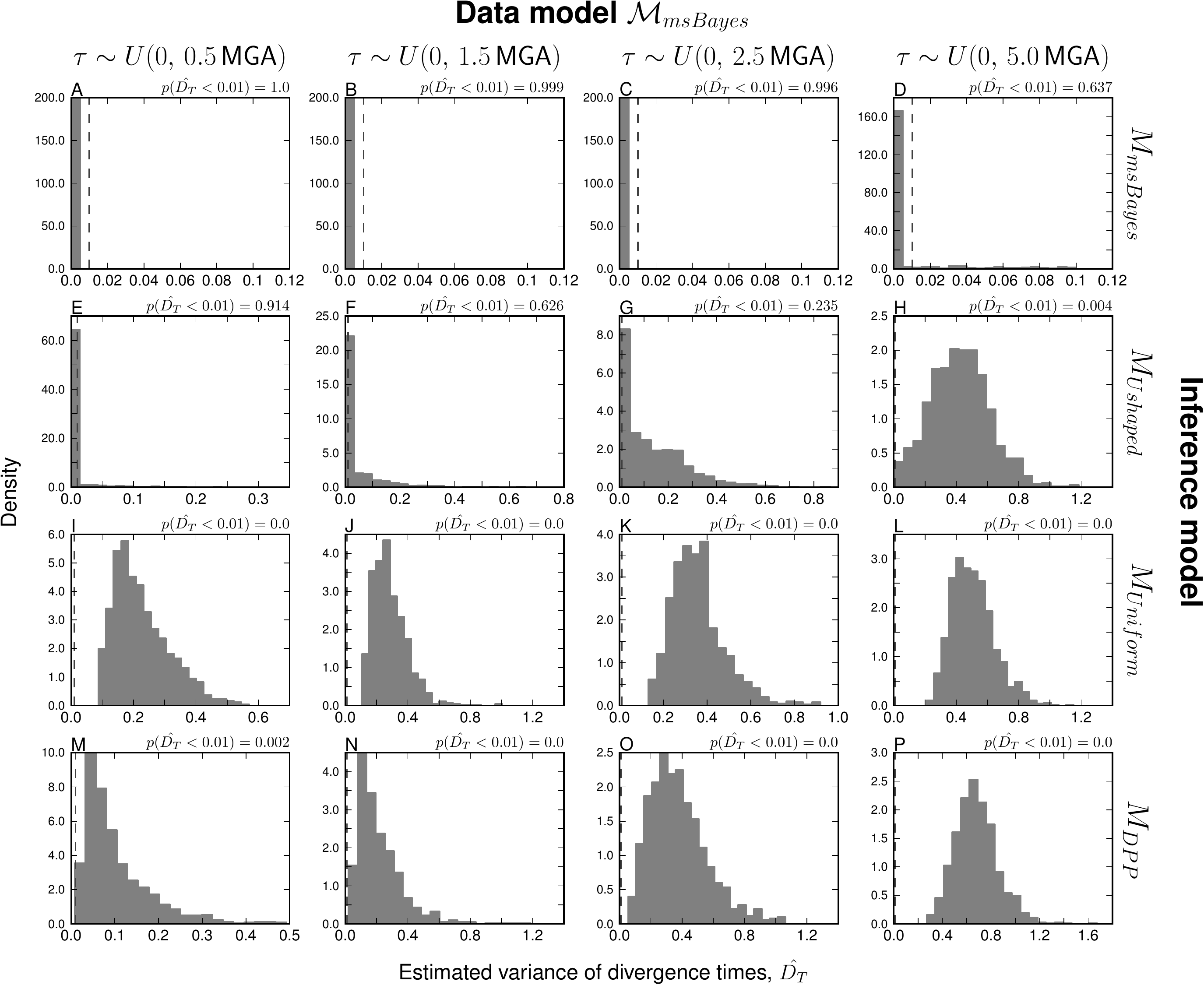}{
    \ifbmc{\csentence{Power to avoid spurious estimation of small temporal
    variance in divergences when divergence times are random.}}{}
    \powerCommentSummary{\powerSeriesOld}
    \powerDispersionComment{\powerSeriesOld}
    \timeConversionComment
    Four of the six data-generating models of the \powerSeriesOld series
    are shown;
    please see Figure~S\ref{figPowerOmegaOld} for all results.
}{figPowerOmegaOld4}

\mFigure{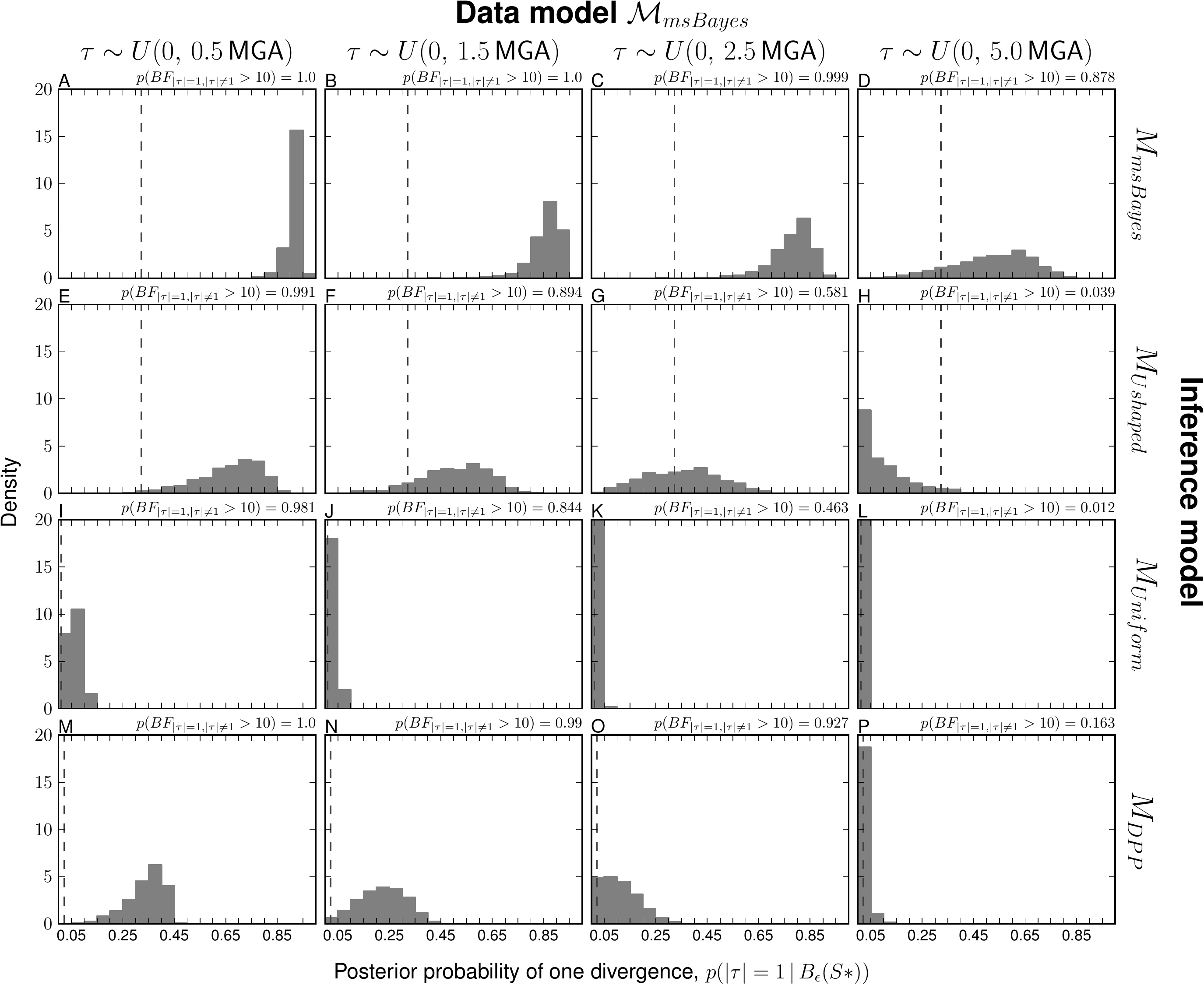}{
    \ifbmc{\csentence{Power to avoid spurious support for one divergence event
    when divergence times are random.}}{}
    \powerSupportCommentSummary{\powerSeriesOld}
    \powerProbComment{$p(\divTimeNum = 1 | \ssSpace)$}{\powerSeriesOld}
    \timeConversionComment
    Four of the six data-generating models of the \powerSeriesOld series
    are shown;
    please see Figure~S\ref{figPowerPsiProbOld} for all results.
}{figPowerPsiProbOld4}

\mFigure{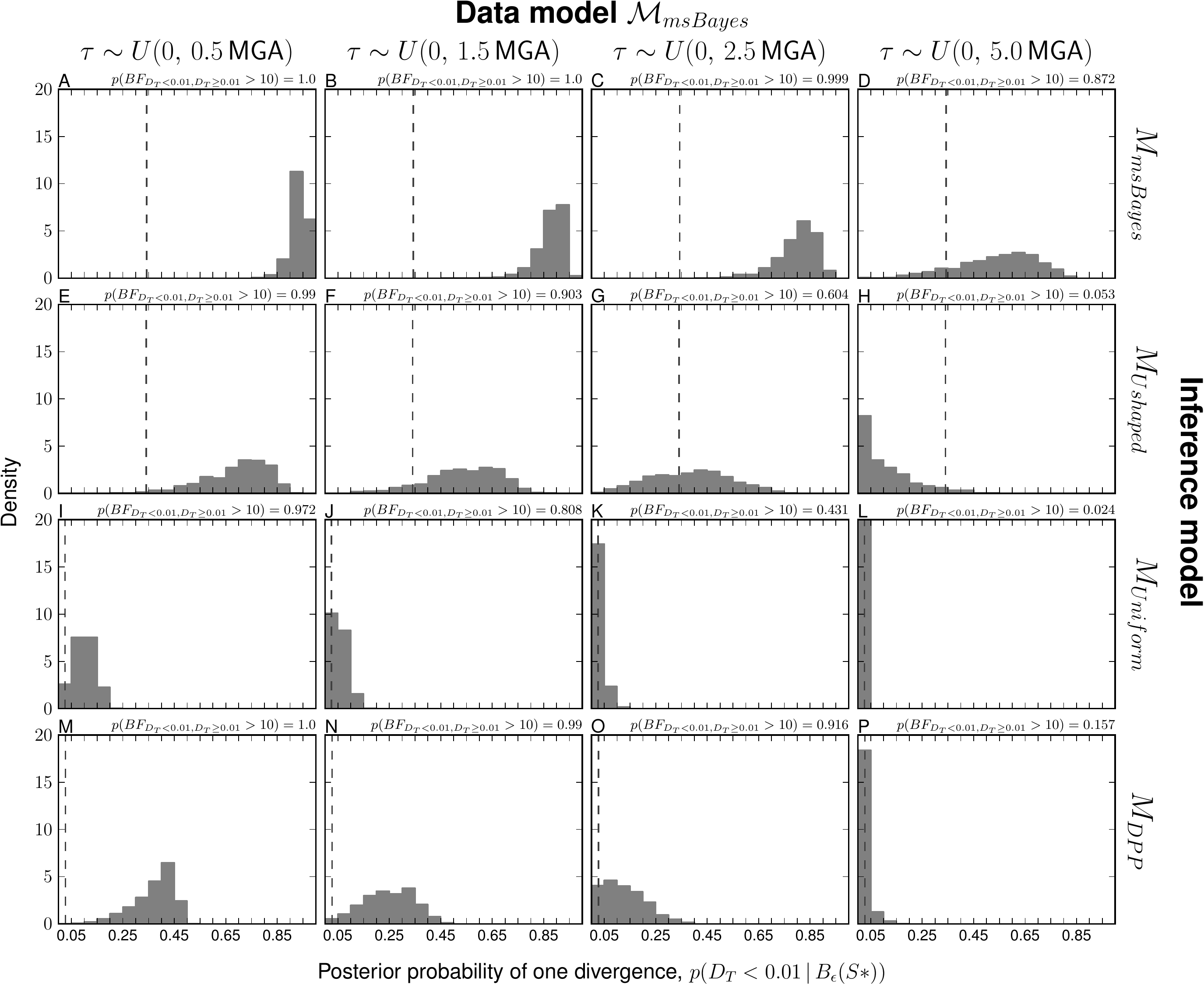}{
    \ifbmc{\csentence{Power to avoid spurious support for no temporal variance
    in divergences (i.e., $\divTimeDispersion < 0.01$) when divergence times
    are random.}}{}
    \powerSupportCommentSummary{\powerSeriesOld}
    \powerProbComment{$p(\divTimeDispersion < 0.01 | \ssSpace)$}{\powerSeriesOld}
    \timeConversionComment
    Four of the six data-generating models of the \powerSeriesOld series
    are shown;
    please see Figure~S\ref{figPowerOmegaProbOld} for all results.
}{figPowerOmegaProbOld4}

\mFigure{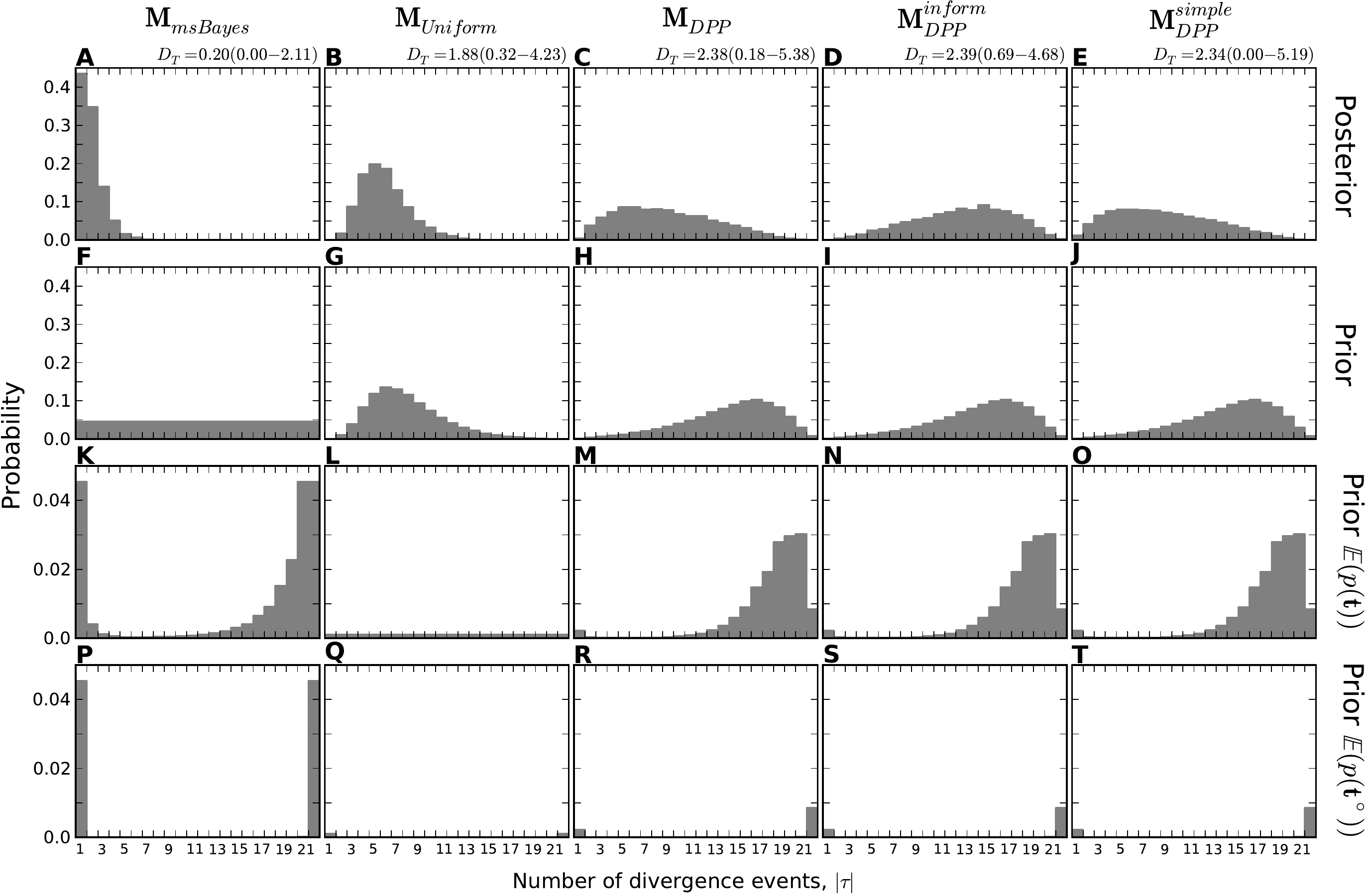}{
    \ifbmc{\csentence{Estimated number of divergence events for 22 taxa from
    the Philippines.}}{}
    The (A--E) posterior and (F--J) prior probabilities of the number of
    divergence events (\divTimeNum) when the data of the 22 pairs of taxa from
    the Philippines are analyzed under the five models indicated at the top of
    each column of plots (Table~\ref{tabEmpiricalModels}).
    The average prior probability of an (K--O) unordered and (P--T)
    ordered model of divergence (\divTimeIndexVector) with \divTimeNum
    divergence-time parameters is also shown.
    The posterior median of the dispersion index of divergence times
    (\divTimeDispersion) is also given for each model, followed by the 95\%
    highest posterior density interval in parentheses.
}{figPhilippines}

\widthFigure{0.4}{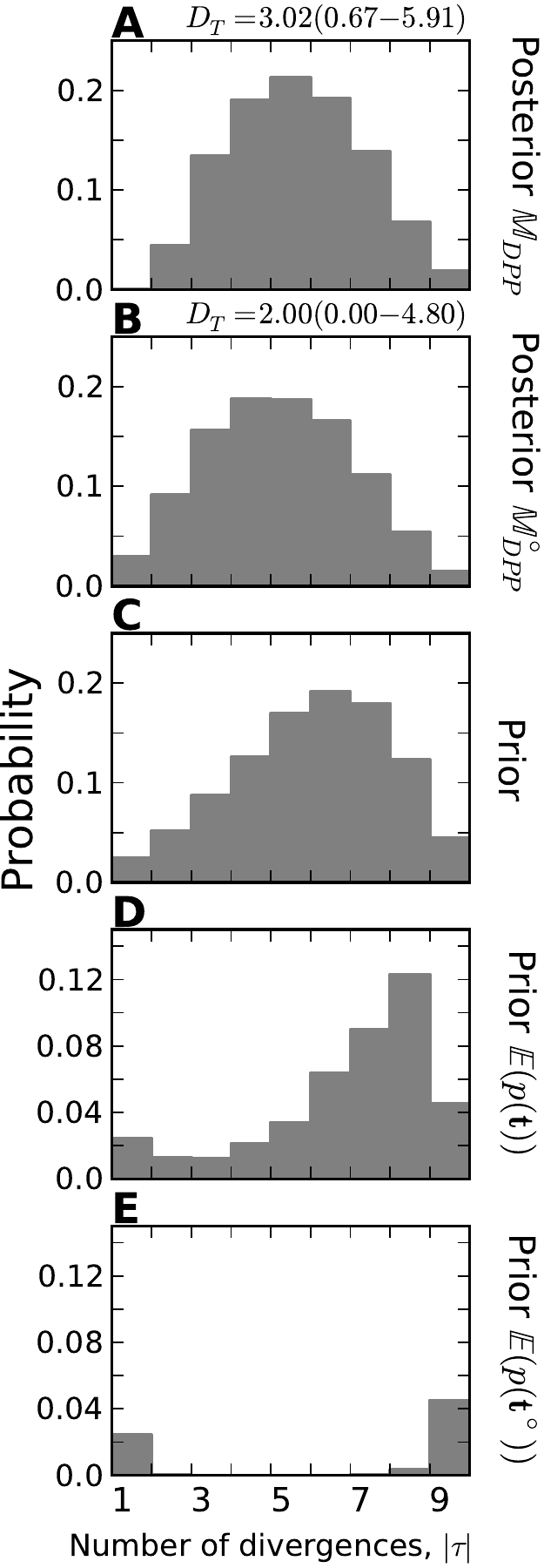}{listformat=figList}{
    \ifbmc{\csentence{Estimated number of divergence events for 9 taxa from the
    Philippines.}}{}
    The posterior probabilities of the number of divergence events,
    \divTimeNum, when the data of the 9 pairs of taxa from Negros and Panay
    Islands are analyzed under the DPP model that samples over (A) unordered
    and (B) ordered models of divergence (Table~\ref{tabEmpiricalModels}).
    Both models share the same (C) prior probability of the number of
    divergence events, and the average prior probability of an (D)
    unordered and (E) ordered model of divergence (\divTimeIndexVector) with
    \divTimeNum divergence-time parameters.
    The posterior median of the dispersion index of divergence times
    (\divTimeDispersion) is also given for each model, followed by the 95\%
    highest posterior density interval in parentheses.
}{figNegrosPanay}
\clearpage

\clearpage
\setcounter{table}{0}
\setcounter{figure}{0}
\setcounter{figure}{0}
\setcounter{table}{0}
\setcounter{page}{1}
\setcounter{section}{0}

\singlespacing

\section*{Supporting Information}
\hangindent=1cm
Oaks, J.\ R. \msTitle.

\newpage
\singlespacing

\begin{table}
    \centering
    \rowcolors{2}{}{myGray} \mcrowcolors
    \captionsetup{name=Table S, labelformat=noSpace}
    \caption{An example showing the number of divergence events (\divTimeNum)
        and the associated sample space of the unordered divergence models
        (integer partitions of \npairs{} pairs) and ordered divergence models
        (partitions of \npairs{} pairs) for $\npairs{} = 4$ pairs of
        populations.}
    \begin{tabular}{ c | l | l }
        \divTimeNum & Unordered divergence models & Ordered divergence models \\
        \hline
        1 & $4$          & 1111                                     \\
        2 & $3+1$; $2+2$ & 1112, 1121, 1211, 2111, 1122, 1212, 1221 \\
        3 & $2+1+1$      & 1123, 1213, 1231, 1223, 1232, 1233       \\
        4 & $1+1+1+1$    & 1234                                     \\
    \end{tabular}
    \label{tabSampleSpace}
\end{table}

\clearpage

\siFigure{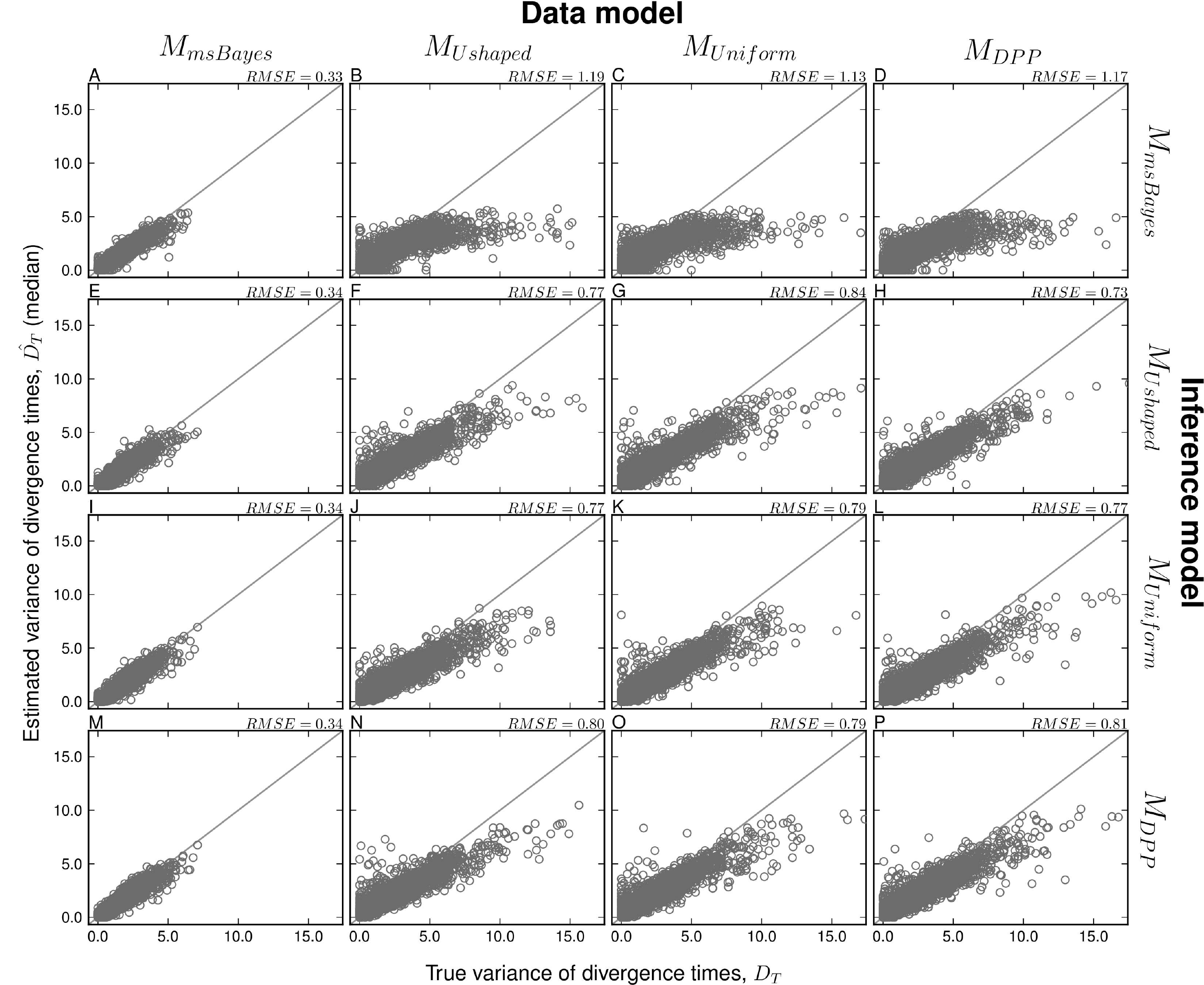}{
    \validationAccuracyComparisonCaption{unadjusted}{\divTimeDispersion}
}{figValidationAccuracyOmega}

\siFigure{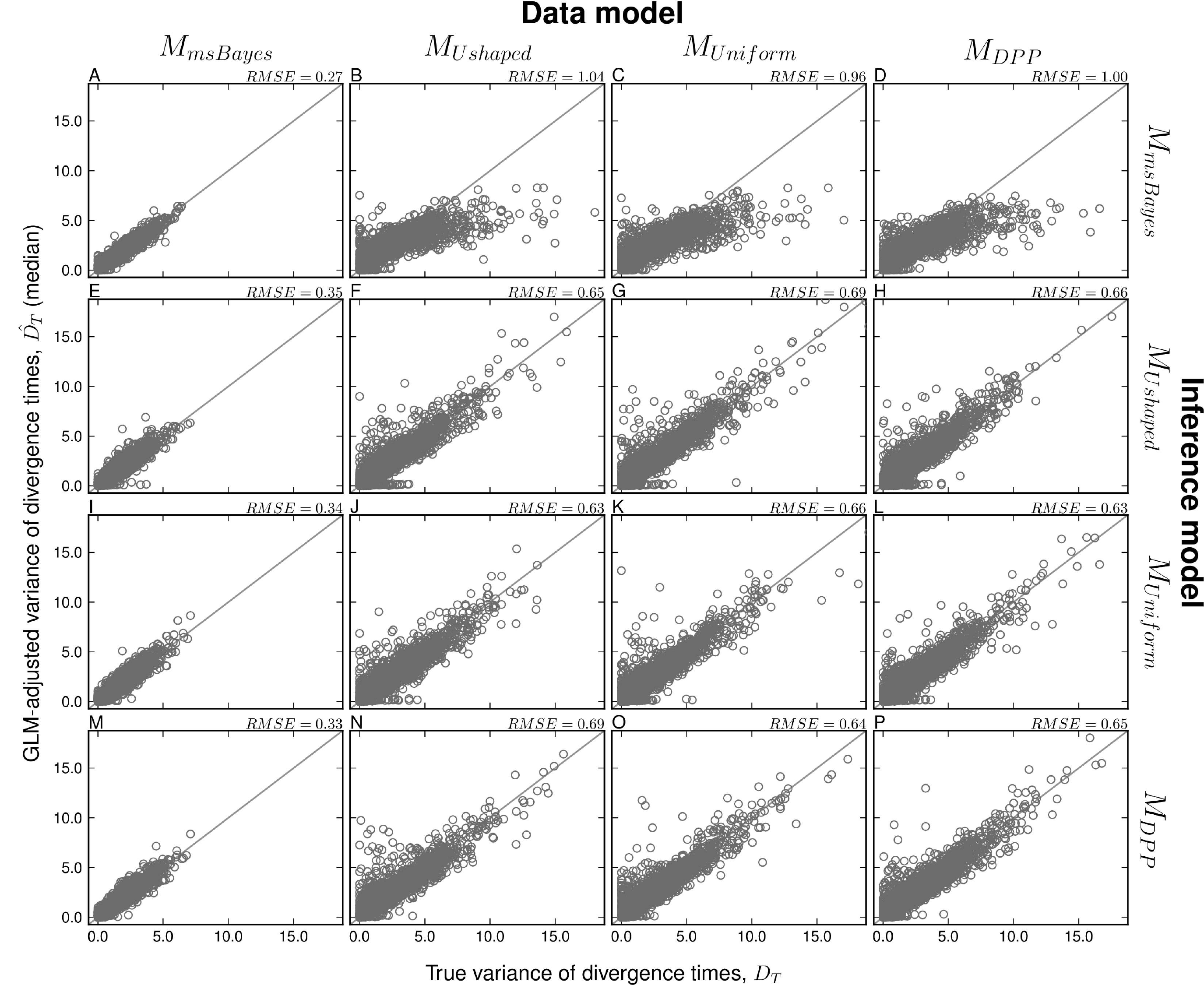}{
    \validationAccuracyComparisonCaption{GLM-adjusted}{\divTimeDispersion}
}{figValidationAccuracyOmegaGlm}

\siFigure{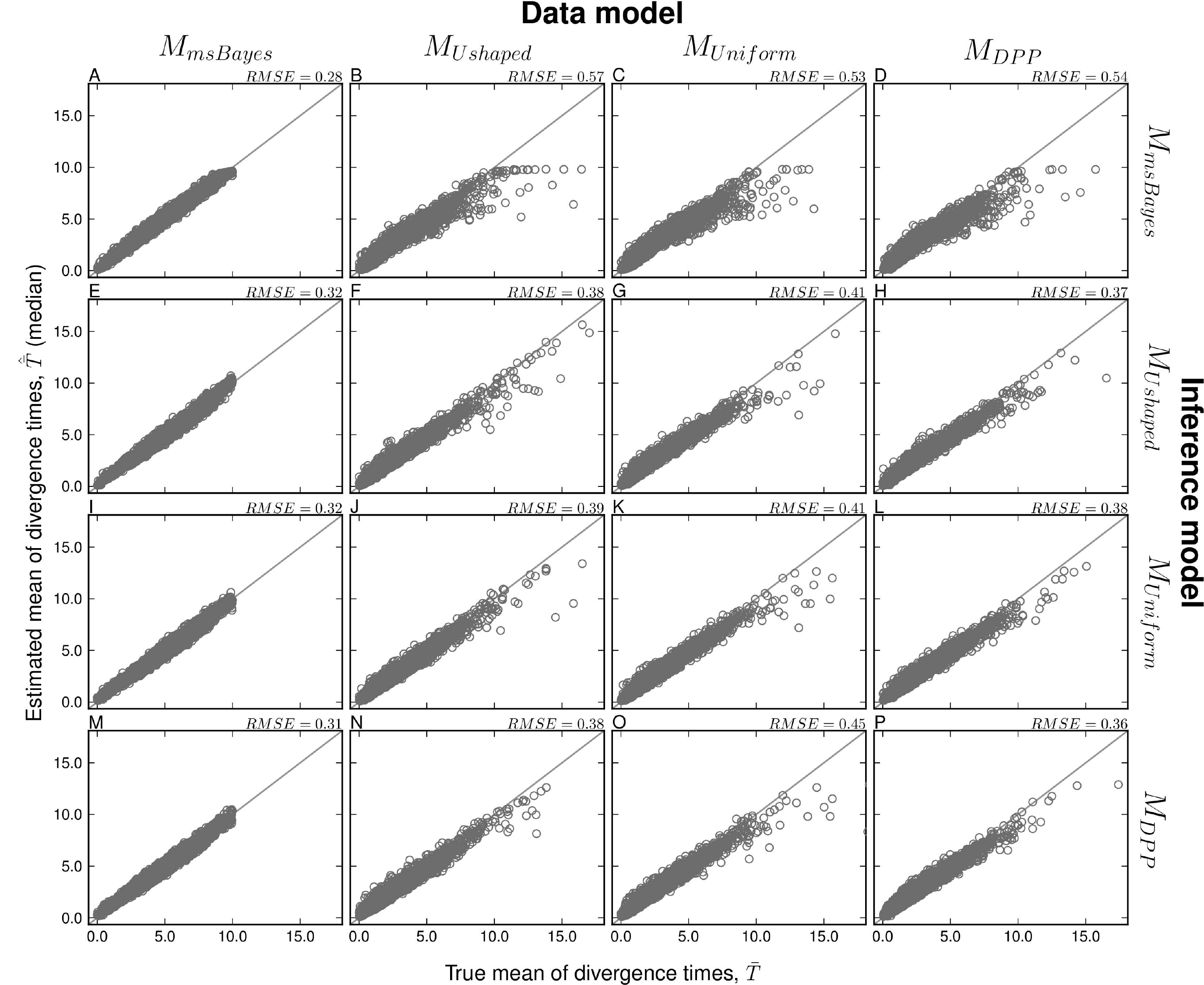}{
    \validationAccuracyComparisonCaption{unadjusted}{\divTimeMean}
}{figValidationAccuracyTime}

\siFigure{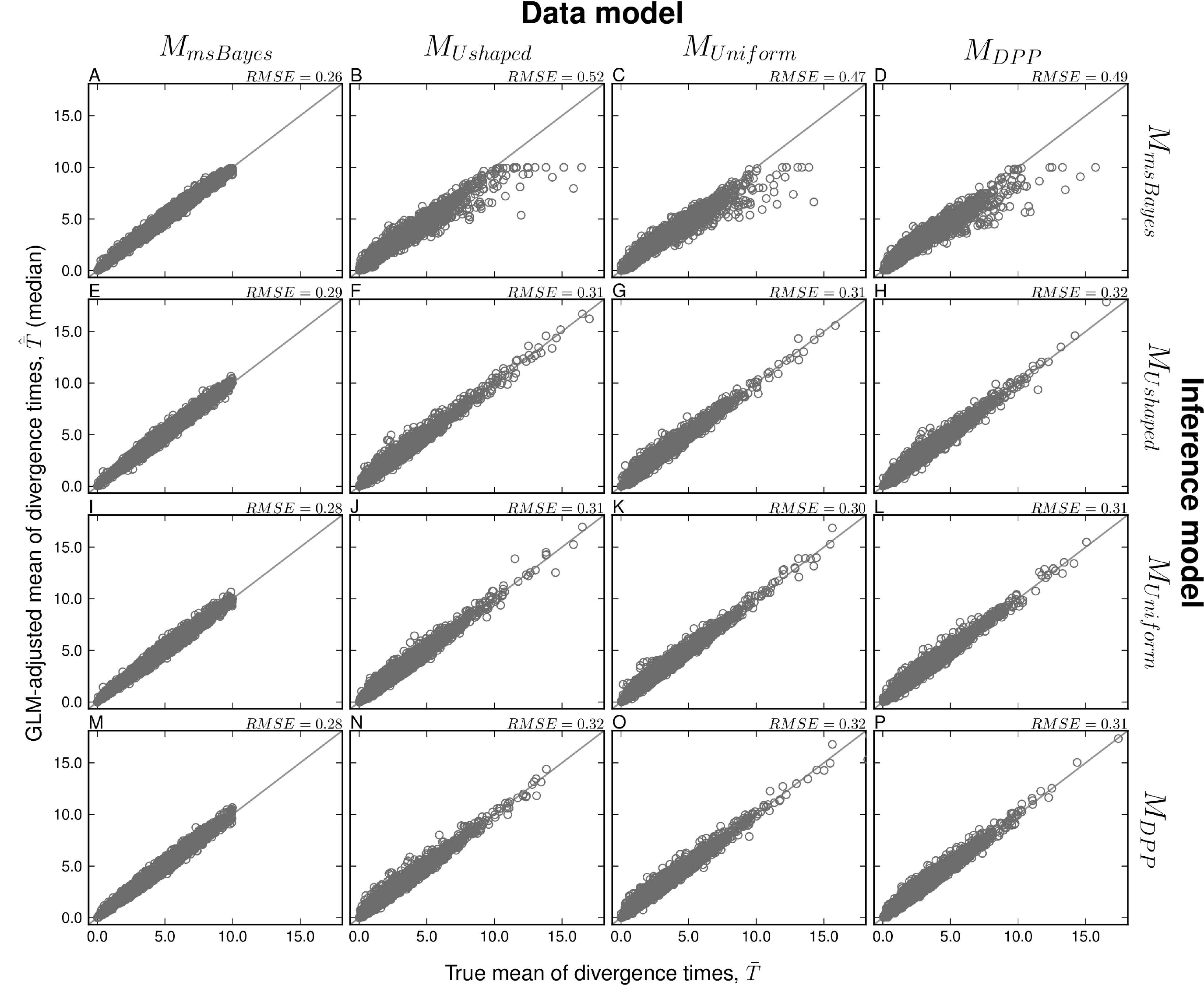}{
    \validationAccuracyComparisonCaption{GLM-adjusted}{\divTimeMean}
}{figValidationAccuracyTimeGlm}

\siFigure{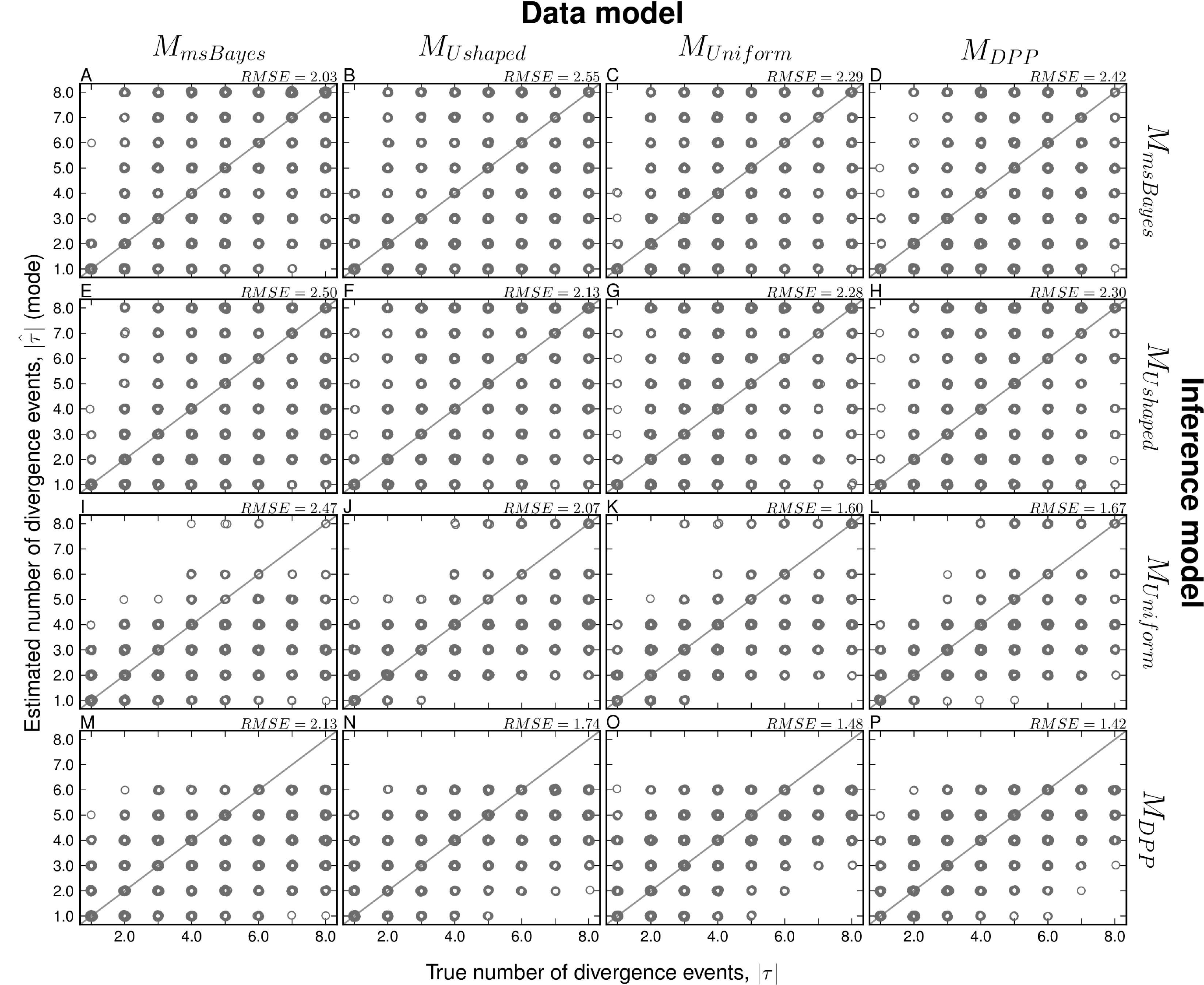}{
    \validationAccuracyComparisonCaption{unadjusted}{\divTimeNum}
    \jitterComment
}{figValidationAccuracyPsi}

\siFigure{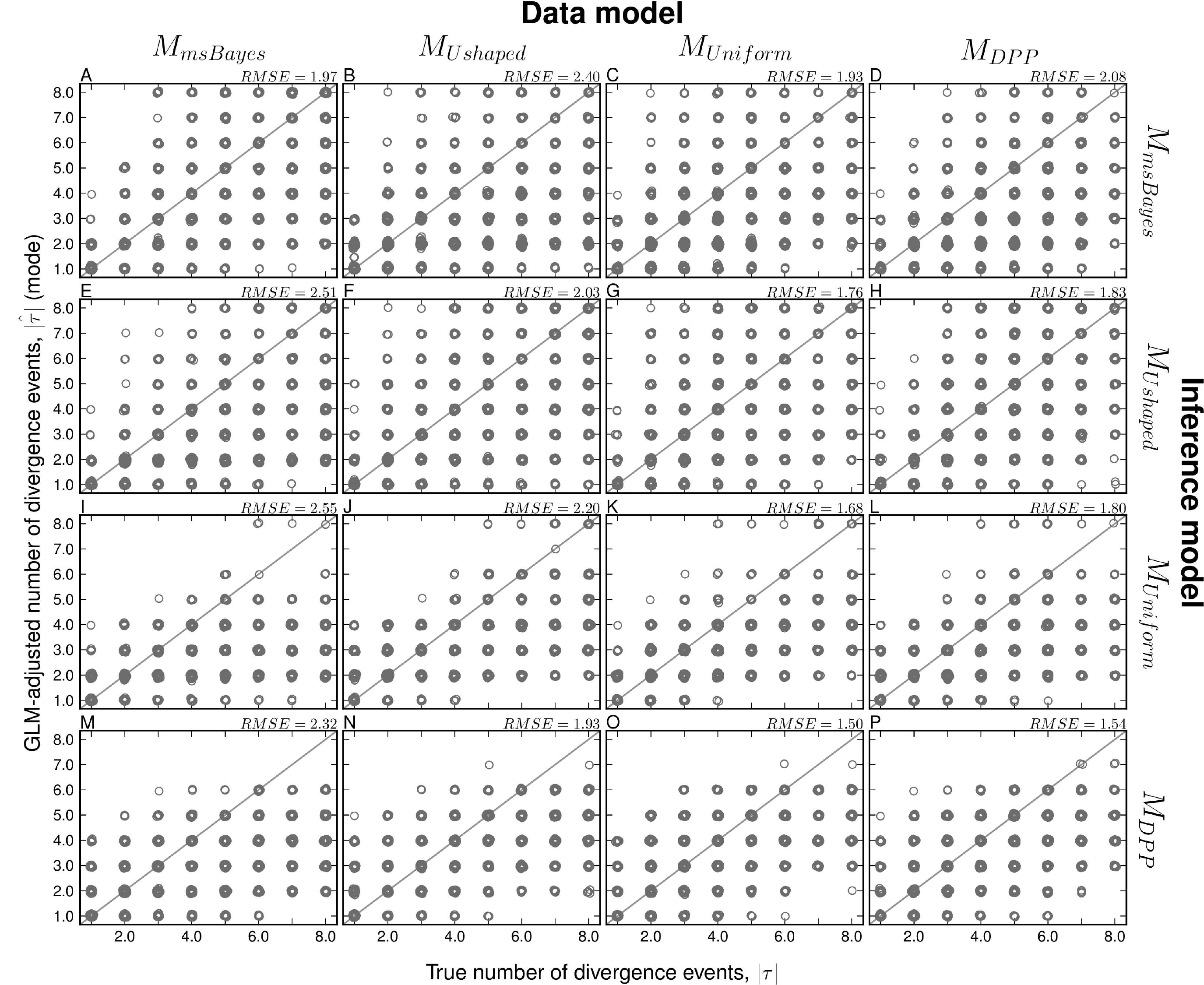}{
    \validationAccuracyComparisonCaption{GLM-adjusted}{\divTimeNum}
    \jitterComment
}{figValidationAccuracyPsiGlm}

\siFigure{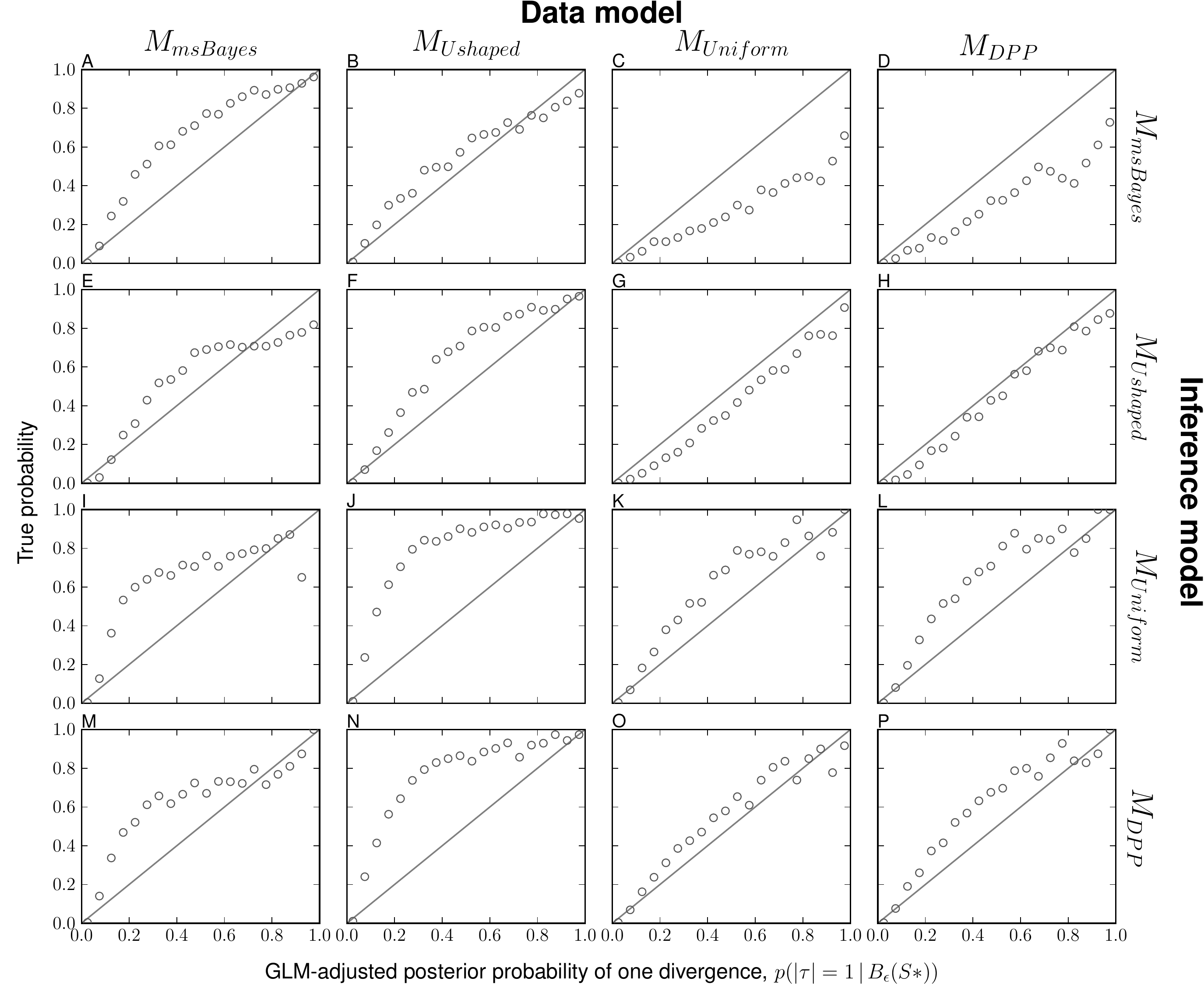}{
    \validationModelChoiceComparisonCaption{GLM-adjusted}{$\divTimeNum = 1$}
}{figValidationModelChoicePsiGlm}

\siFigure{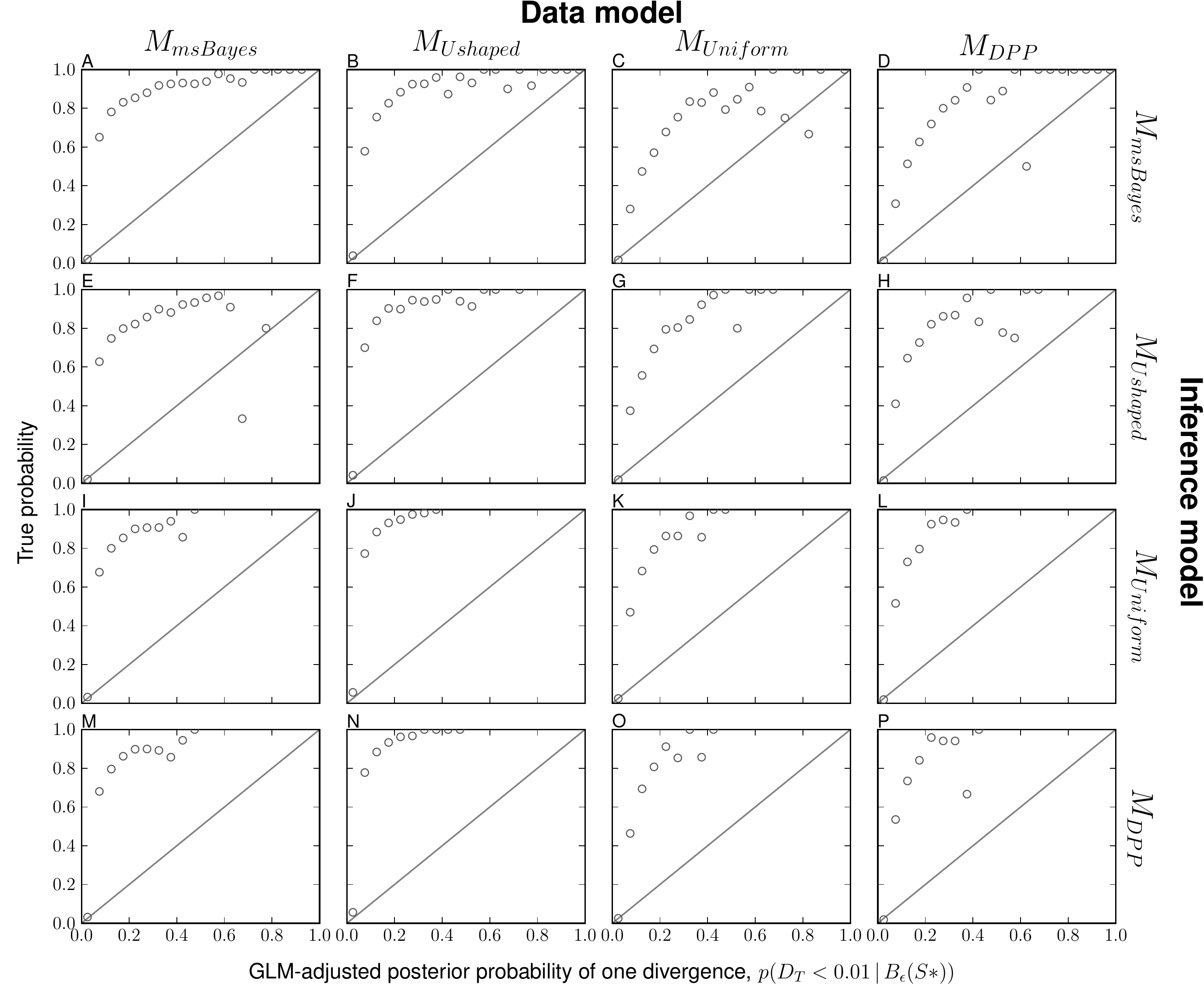}{
    \validationModelChoiceComparisonCaption{GLM-adjusted}{$\divTimeDispersion < 0.01$}
}{figValidationModelChoiceOmegaGlm}

\siFigure{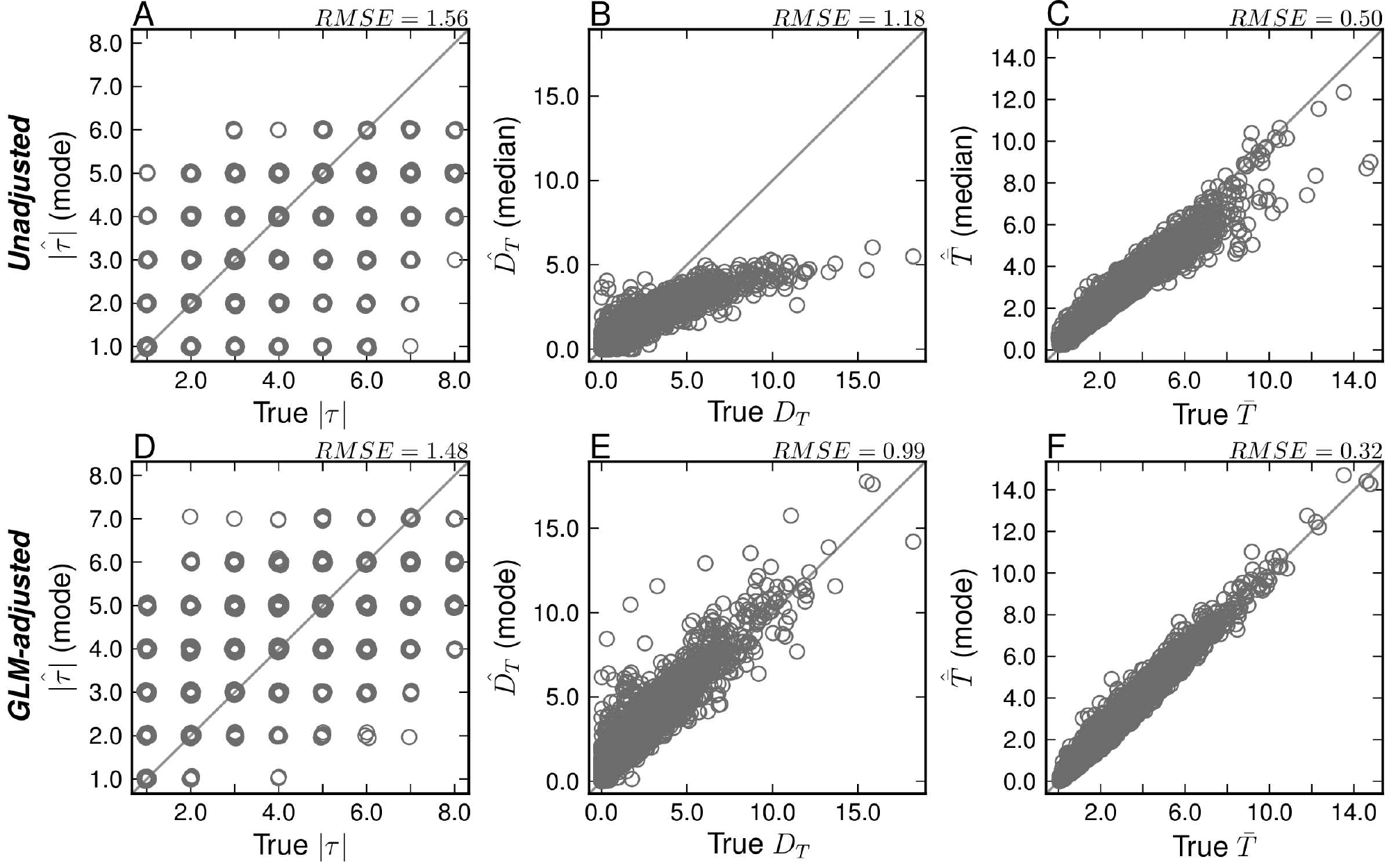}{
    \validationAccuracyCaption{\modelDPPOrdered}{\modelDPPOrdered}
}{figValidationAccuracyDPPOrderedDPPOrdered}

\siFigure{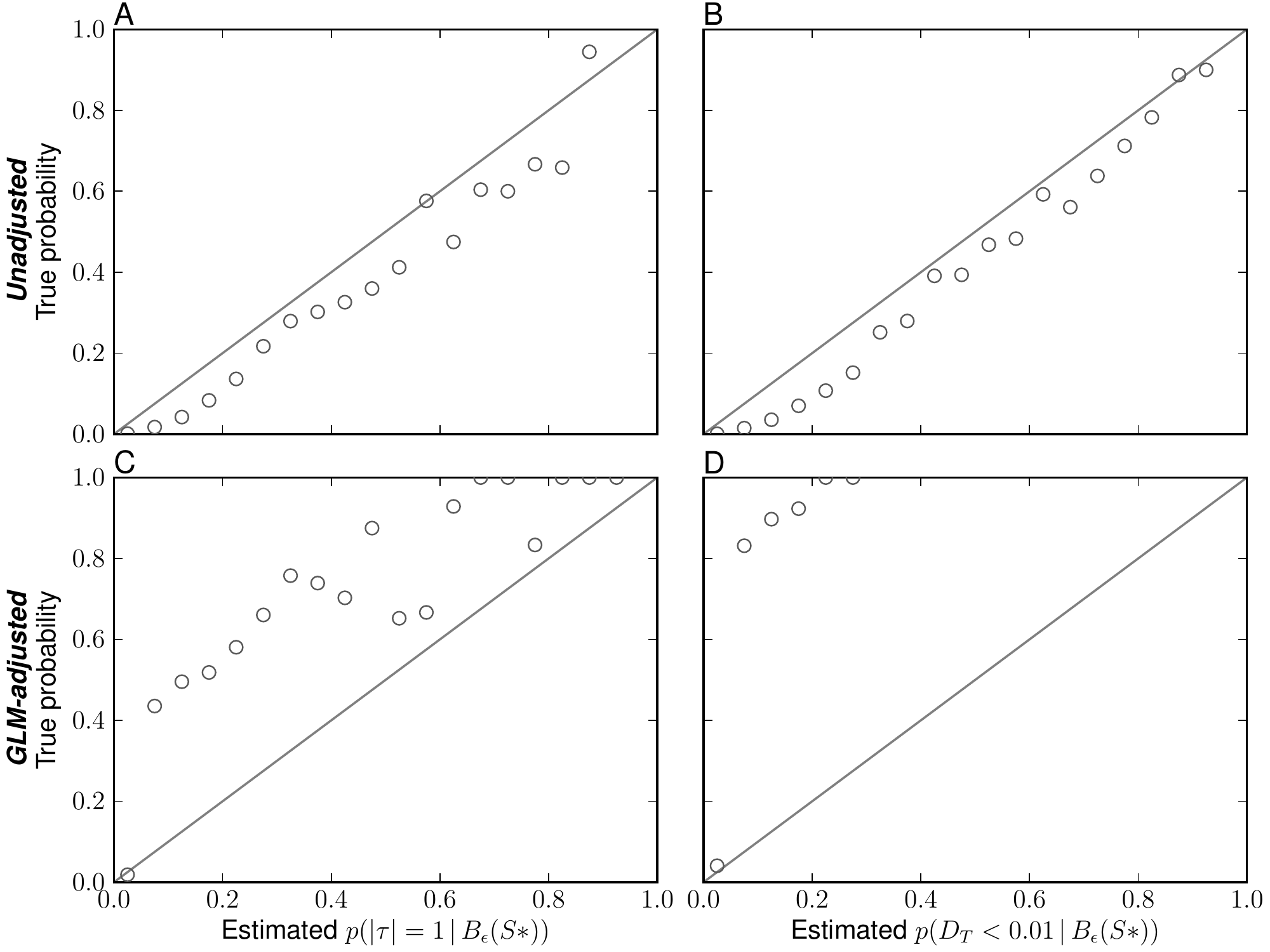}{
    \validationModelChoiceCaption{\modelDPPOrdered}{\modelDPPOrdered}
}{figValidationModelChoiceDPPOrderedDPPOrdered}

\siSidewaysFigure{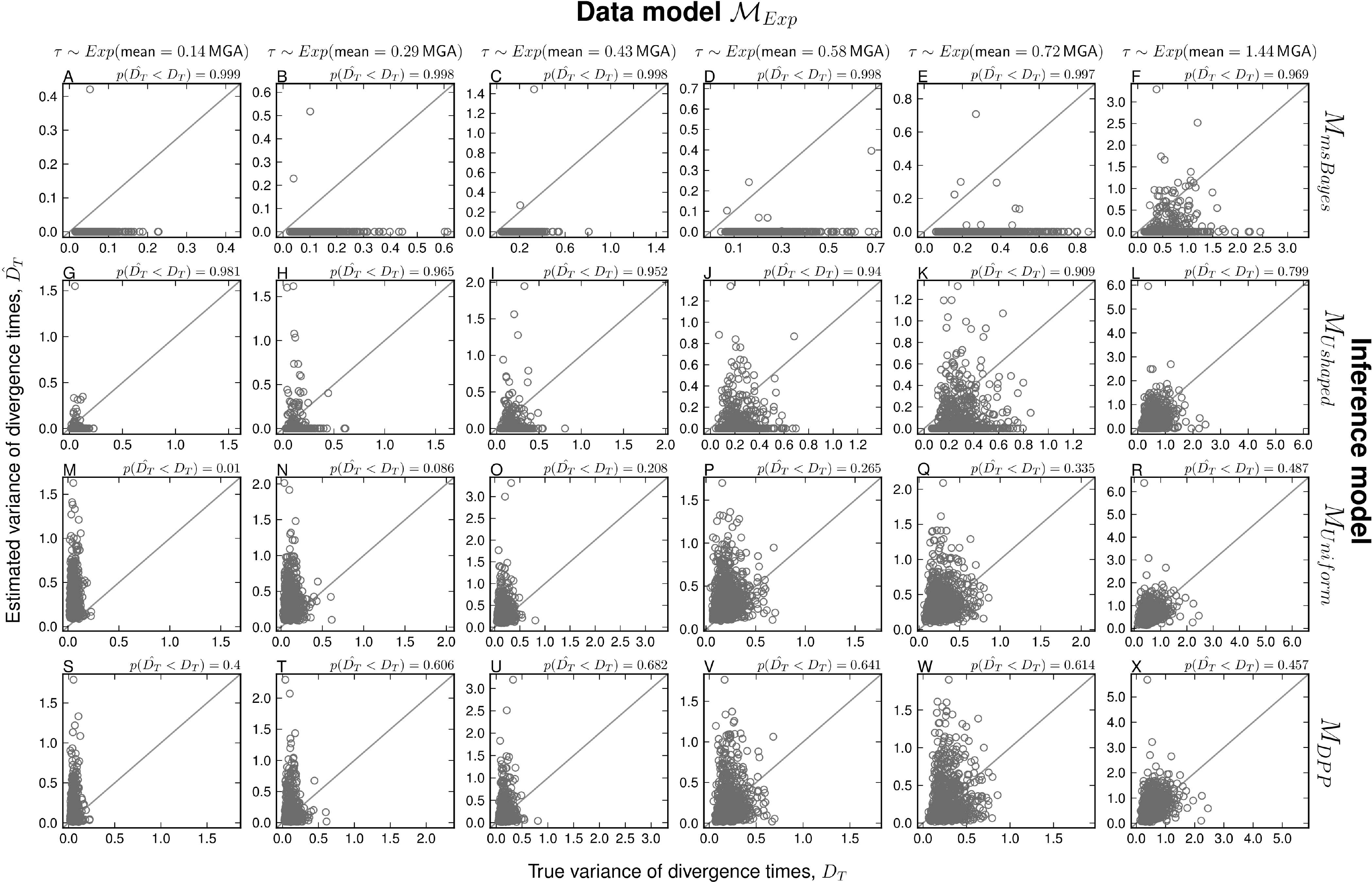}{
    \powerAccuracyComparisonCaption{\powerSeriesExp}
}{figPowerAccuracyExp}

\siSidewaysFigure{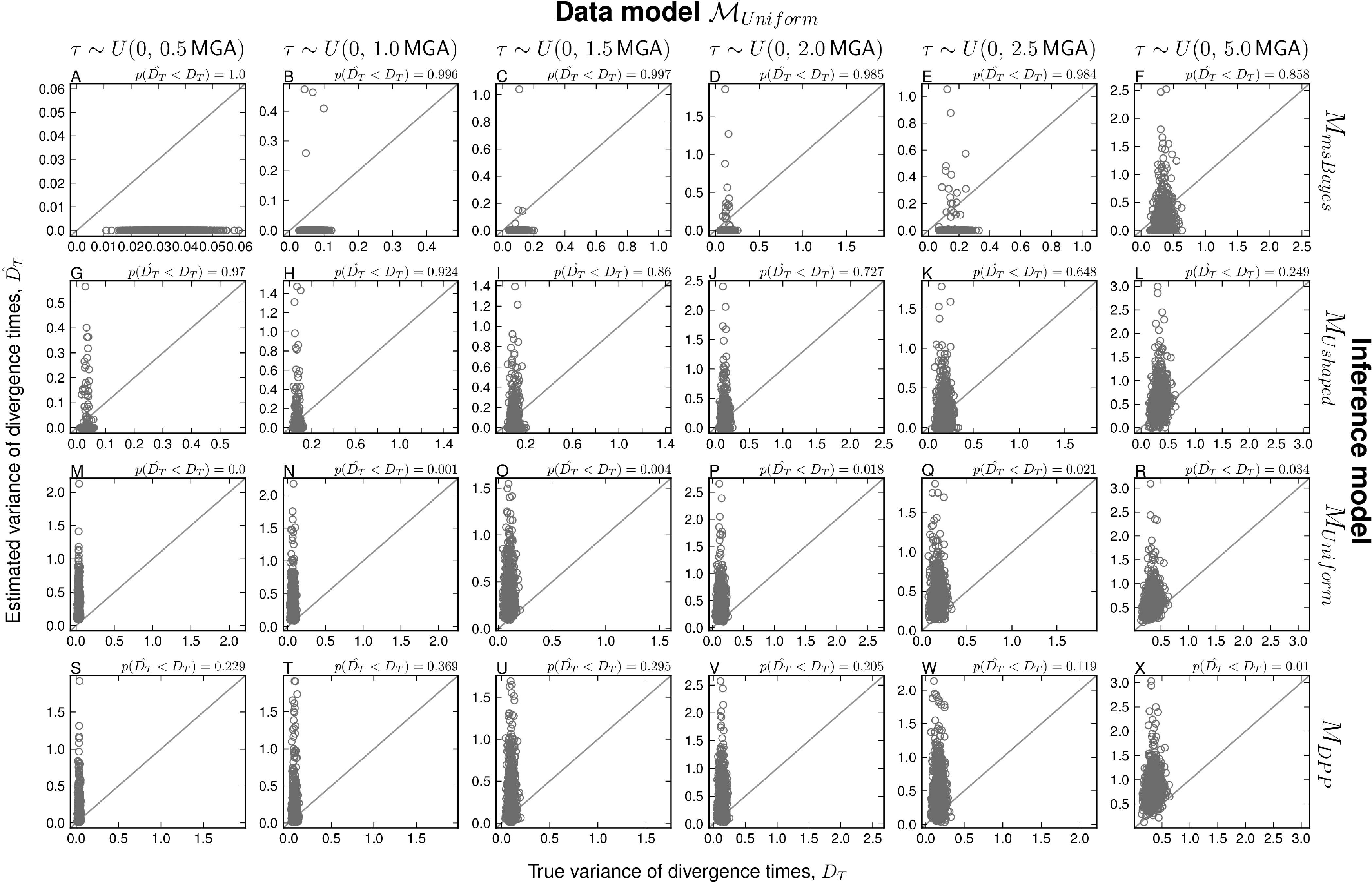}{
    \powerAccuracyComparisonCaption{\powerSeriesUniform}
}{figPowerAccuracyUniform}

\siSidewaysFigure{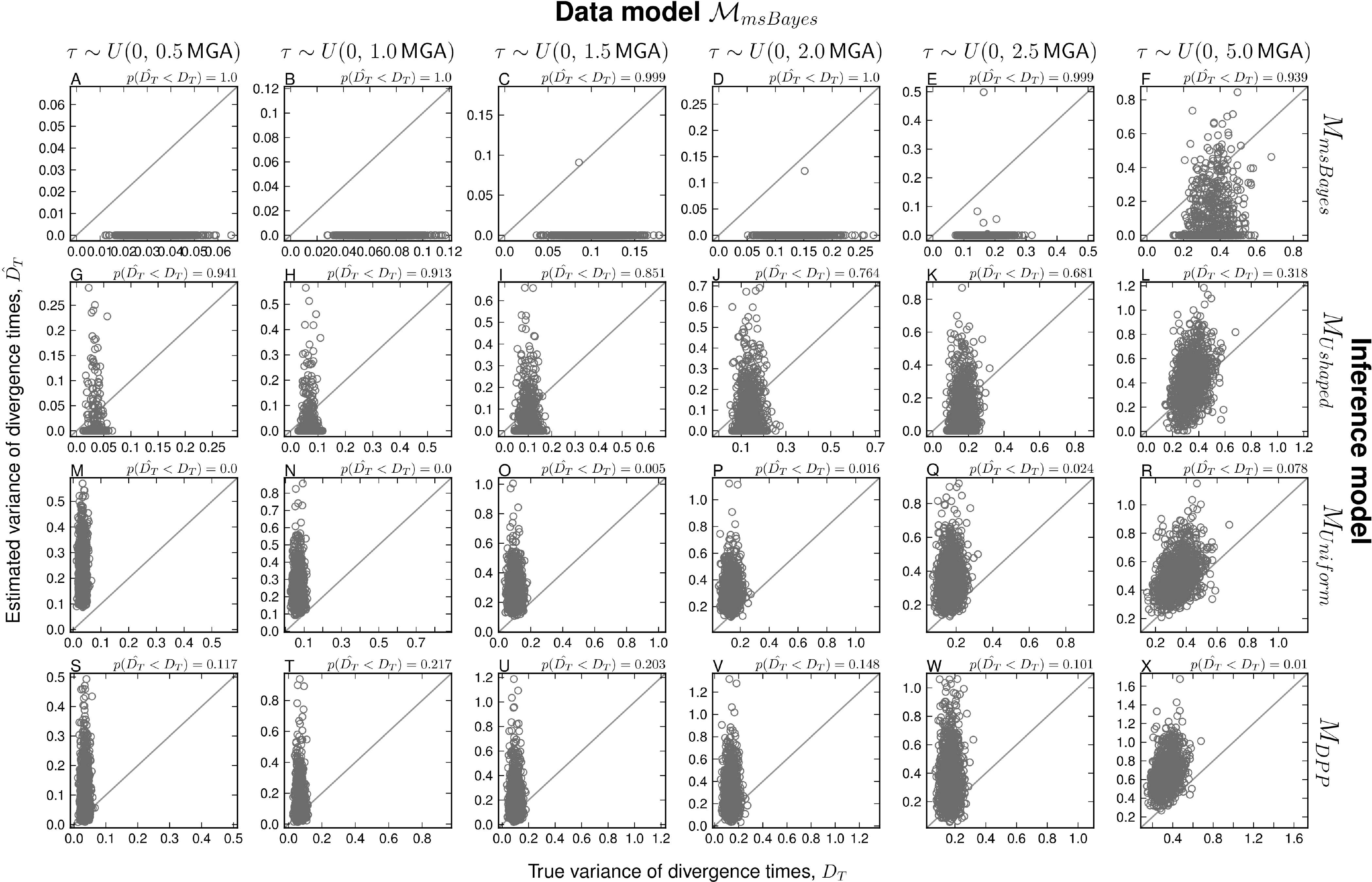}{
    \powerAccuracyComparisonCaption{\powerSeriesOld}
}{figPowerAccuracyOld}

\siSidewaysFigure{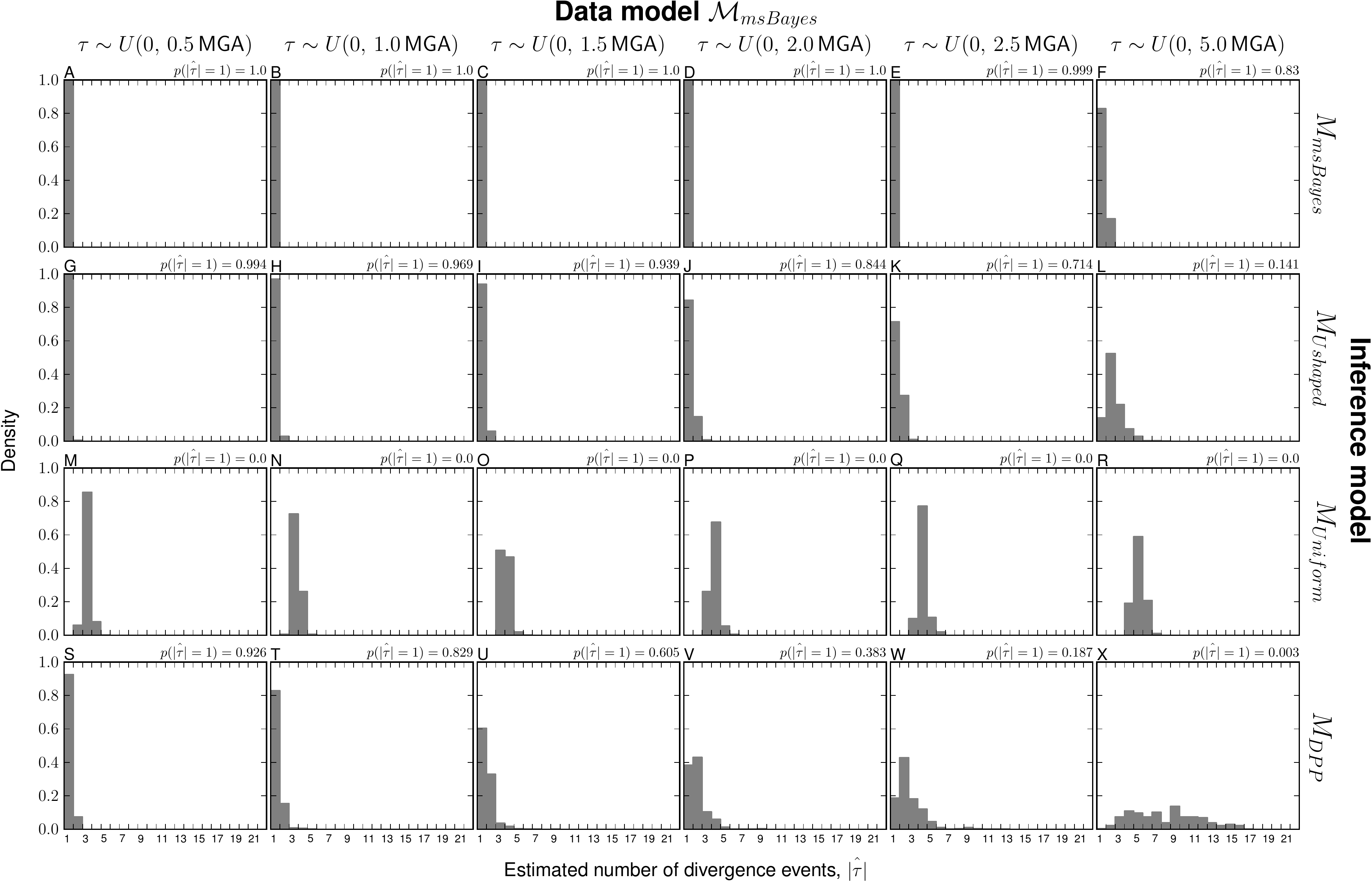}{
    \powerComment{\powerSeriesOld}
    \powerPsiComment{\powerSeriesOld}
    \timeConversionComment
}{figPowerPsiOld}

\siSidewaysFigure{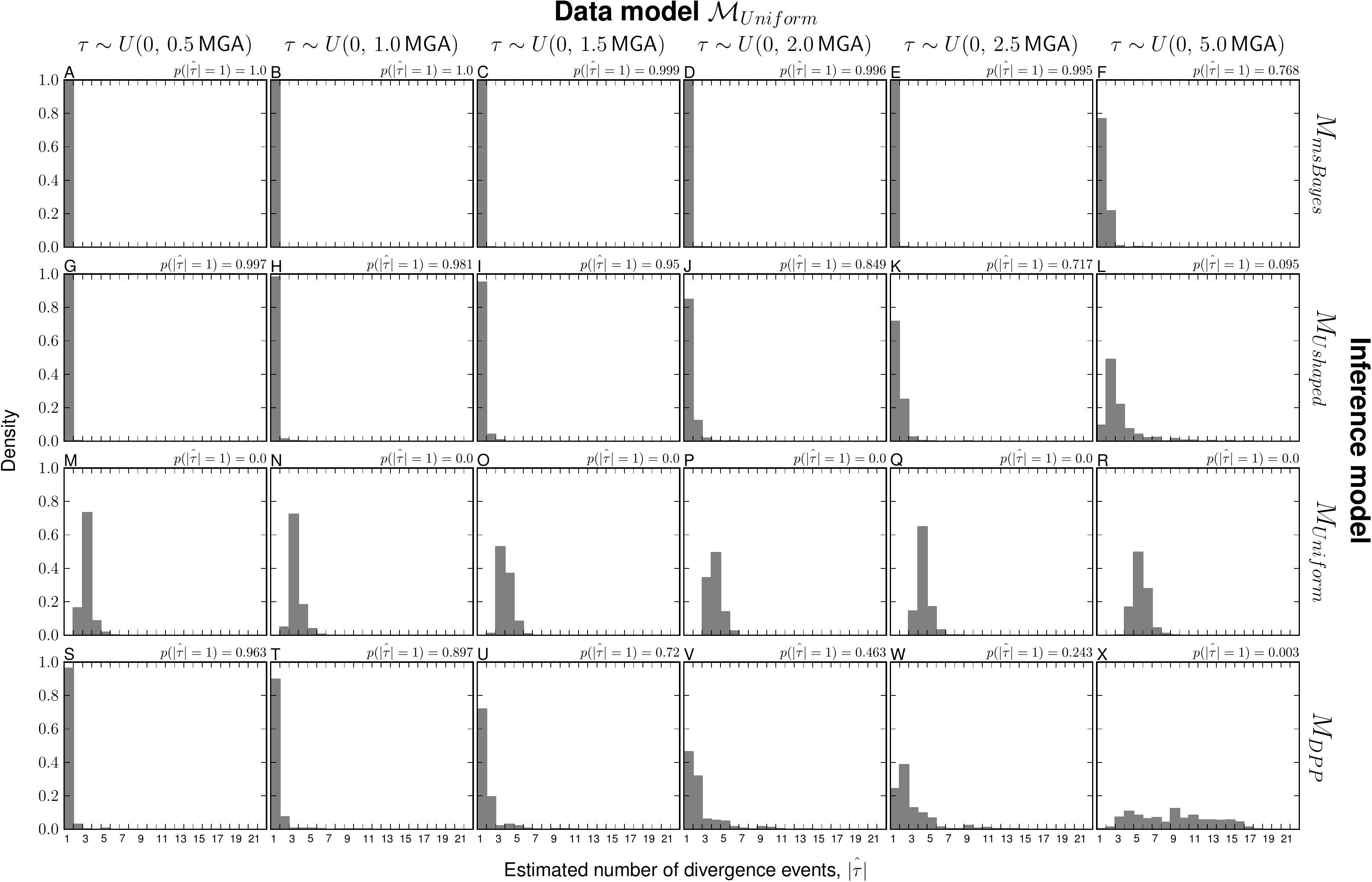}{
    \powerComment{\powerSeriesUniform}
    \powerPsiComment{\powerSeriesUniform}
    \timeConversionComment
}{figPowerPsiUniform}

\siSidewaysFigure{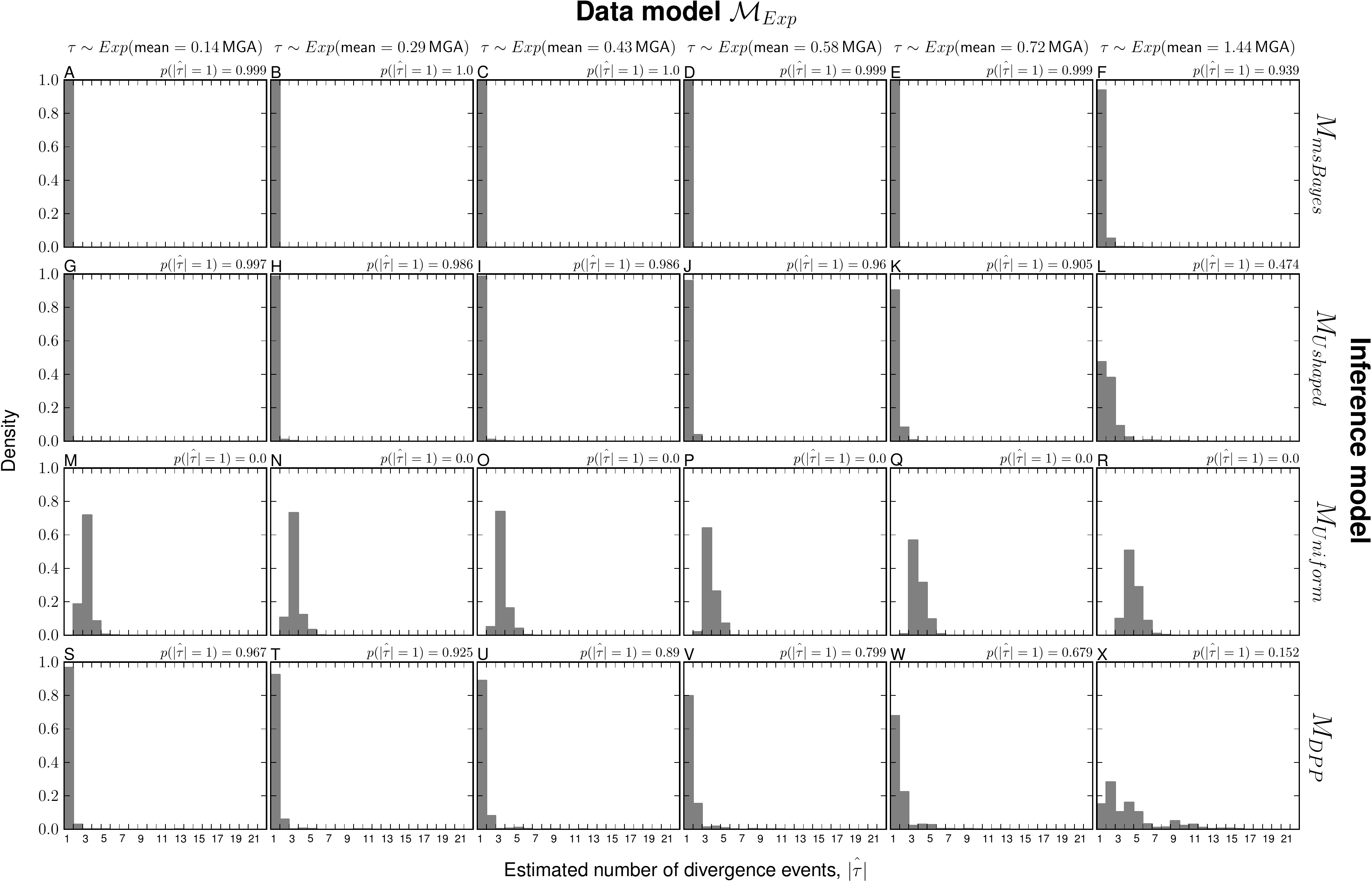}{
    \powerComment{\powerSeriesExp}
    \powerPsiComment{\powerSeriesExp}
    \timeConversionComment
}{figPowerPsiExp}

\siSidewaysFigure{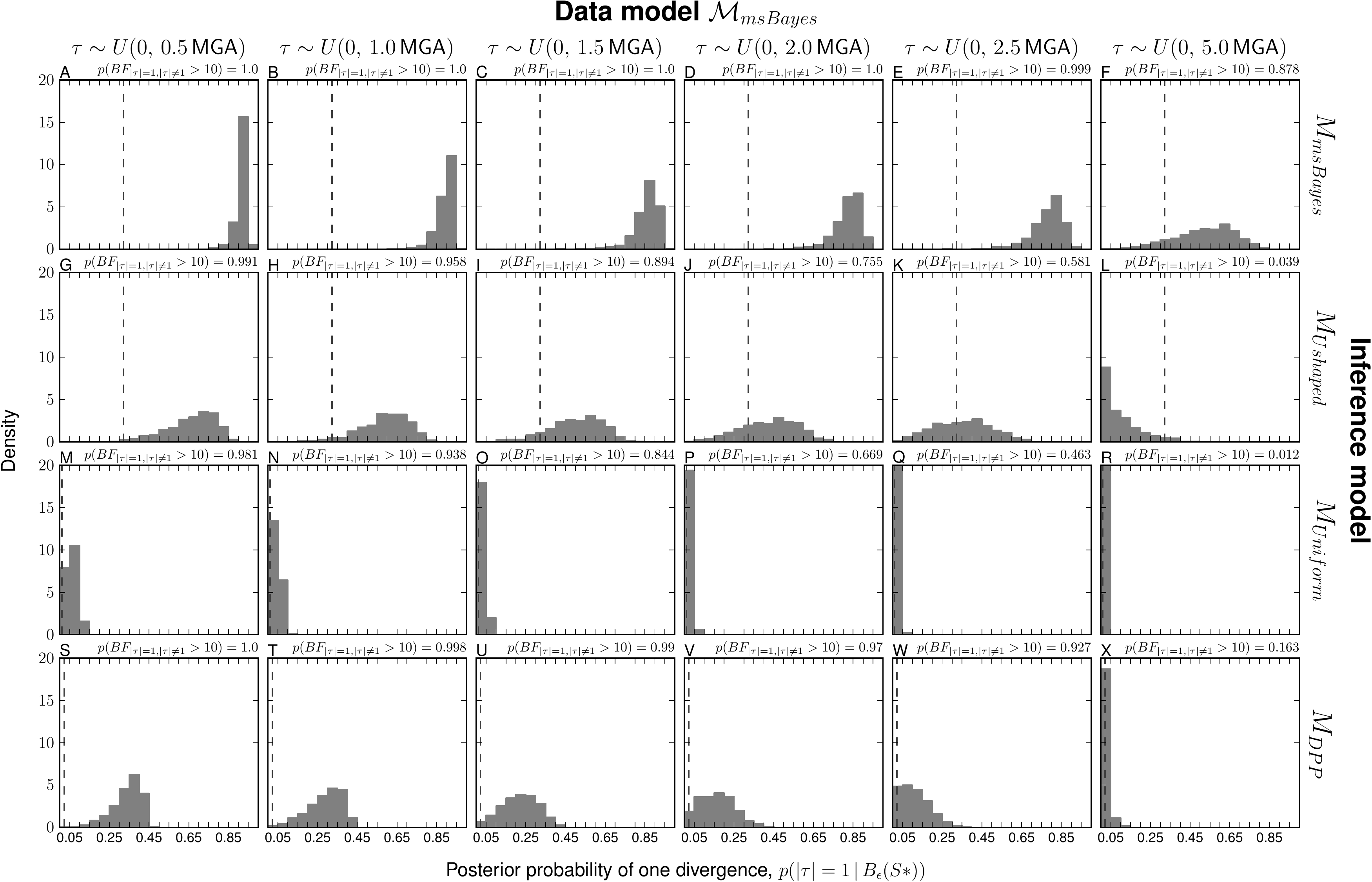}{
    \powerSupportComment{\powerSeriesOld}
    \powerProbComment{$p(\divTimeNum = 1 | \ssSpace)$}{\powerSeriesOld}
    \timeConversionComment
}{figPowerPsiProbOld}

\siSidewaysFigure{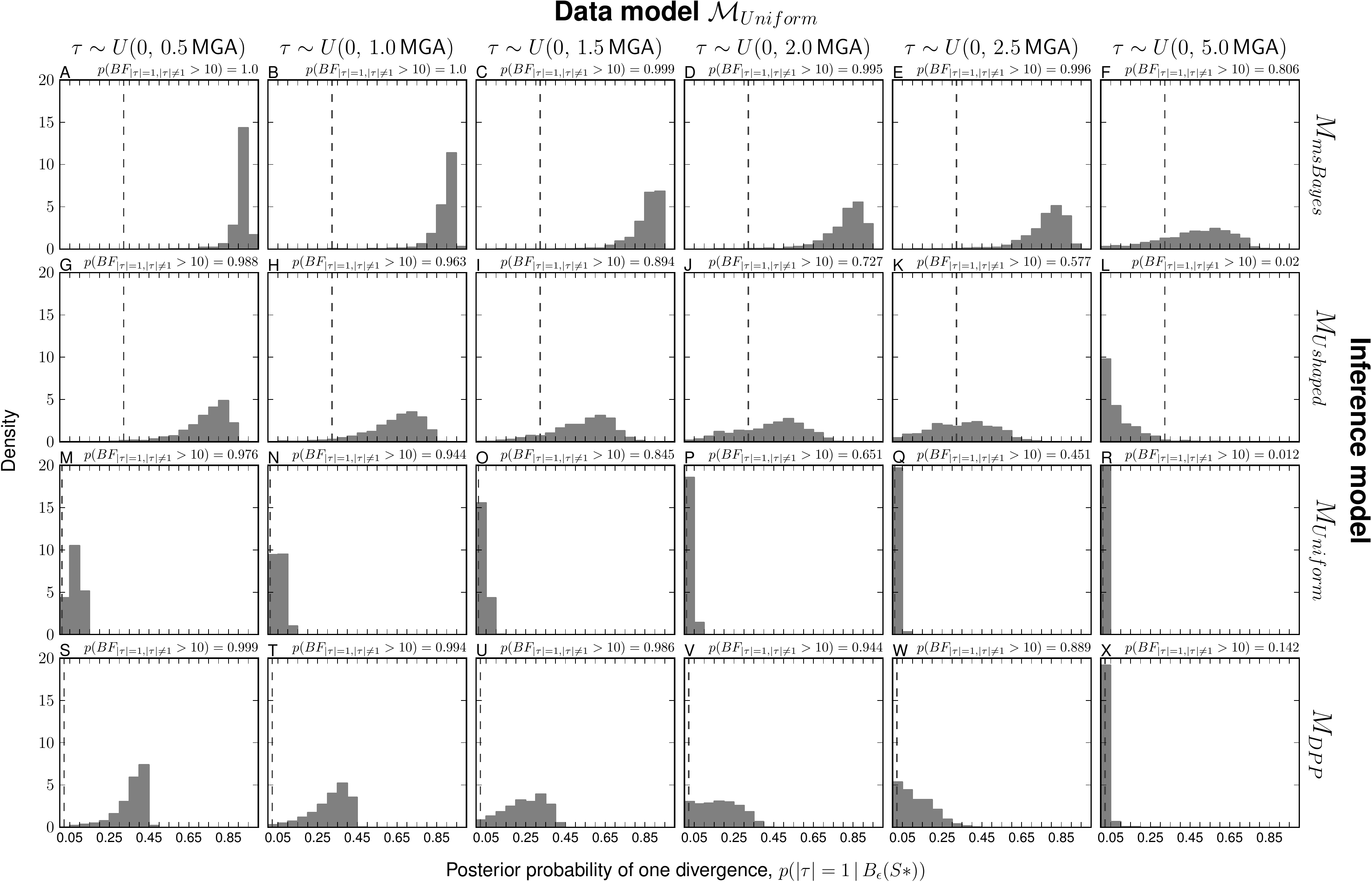}{
    \powerSupportComment{\powerSeriesUniform}
    \powerProbComment{$p(\divTimeNum = 1 | \ssSpace)$}{\powerSeriesUniform}
    \timeConversionComment
}{figPowerPsiProbUniform}

\siSidewaysFigure{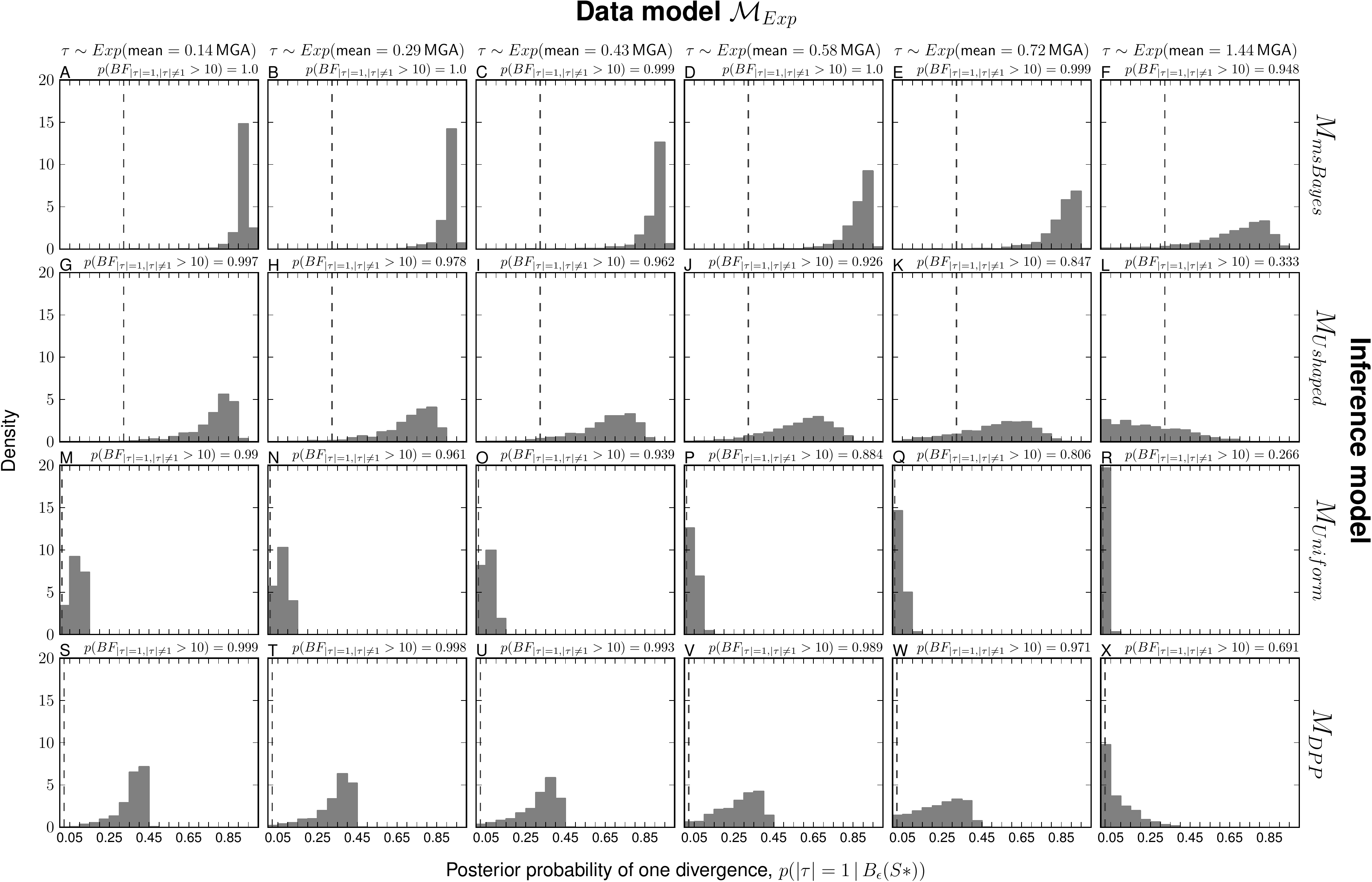}{
    \powerSupportComment{\powerSeriesExp}
    \powerProbComment{$p(\divTimeNum = 1 | \ssSpace)$}{\powerSeriesExp}
    \timeConversionComment
}{figPowerPsiProbExp}

\siSidewaysFigure{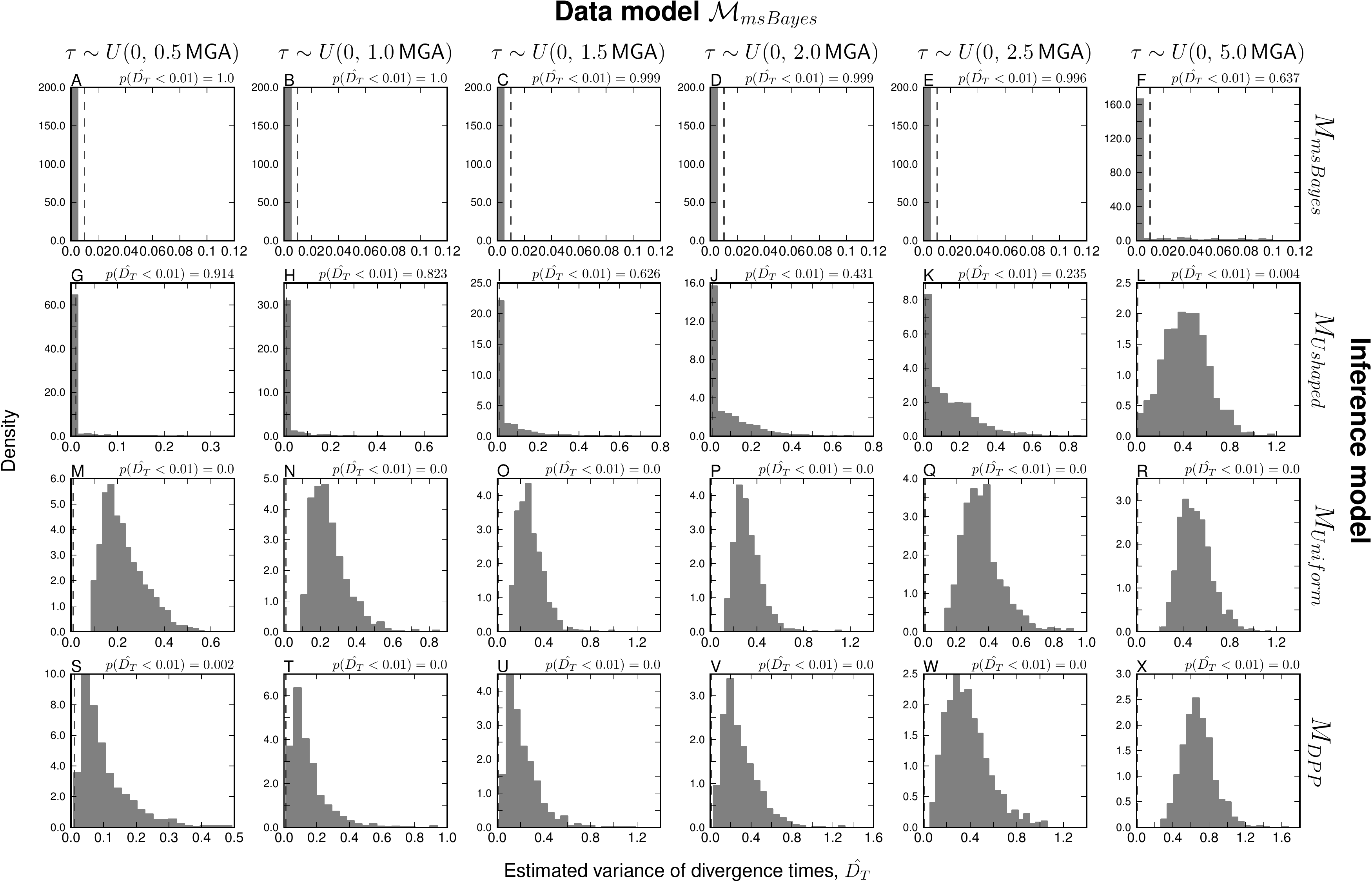}{
    \powerComment{\powerSeriesOld}
    \powerDispersionComment{\powerSeriesOld}
    \timeConversionComment
}{figPowerOmegaOld}

\siSidewaysFigure{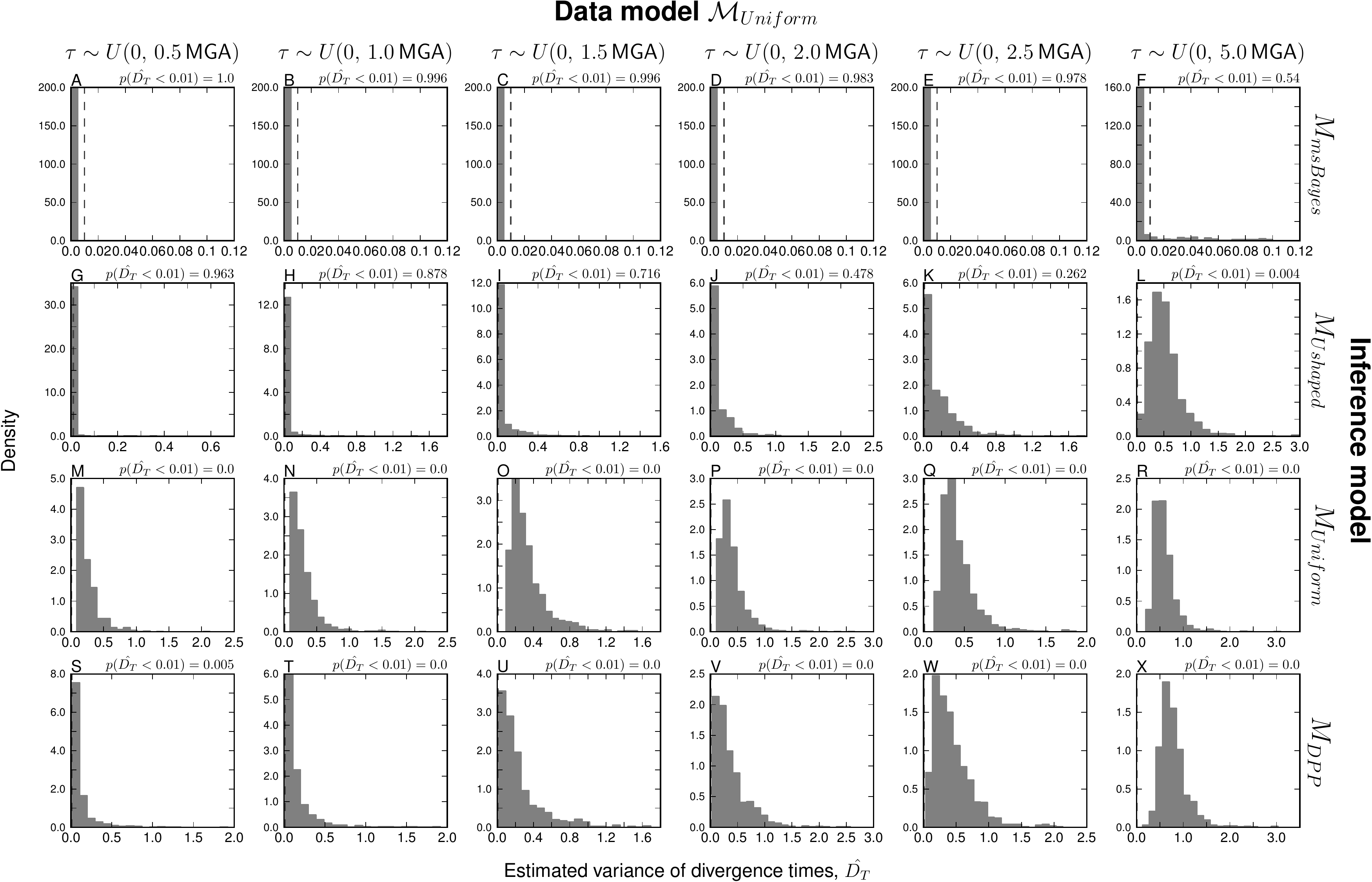}{
    \powerComment{\powerSeriesUniform}
    \powerDispersionComment{\powerSeriesUniform}
    \timeConversionComment
}{figPowerOmegaUniform}

\siSidewaysFigure{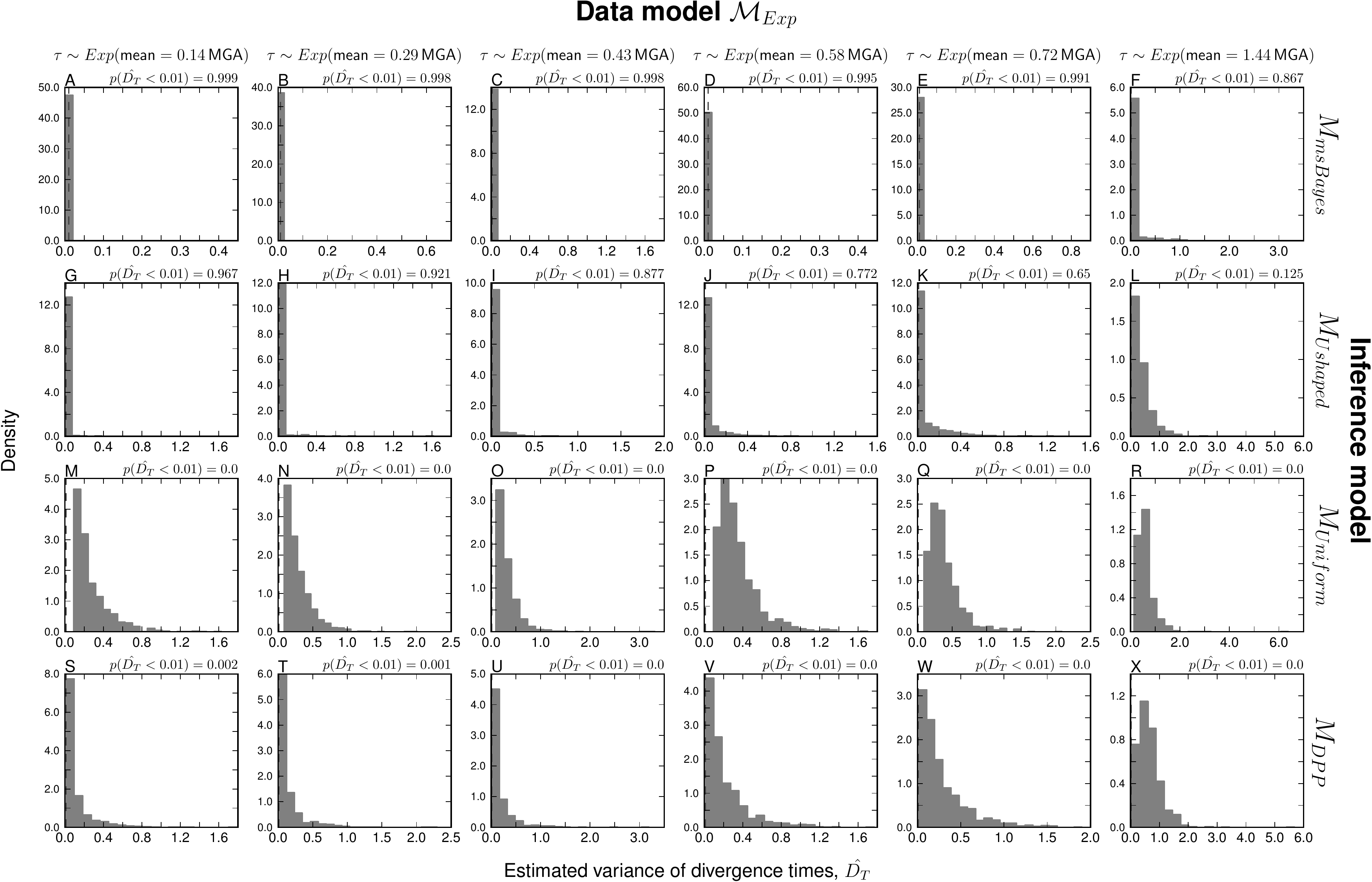}{
    \powerComment{\powerSeriesExp}
    \powerDispersionComment{\powerSeriesExp}
    \timeConversionComment
}{figPowerOmegaExp}

\siSidewaysFigure{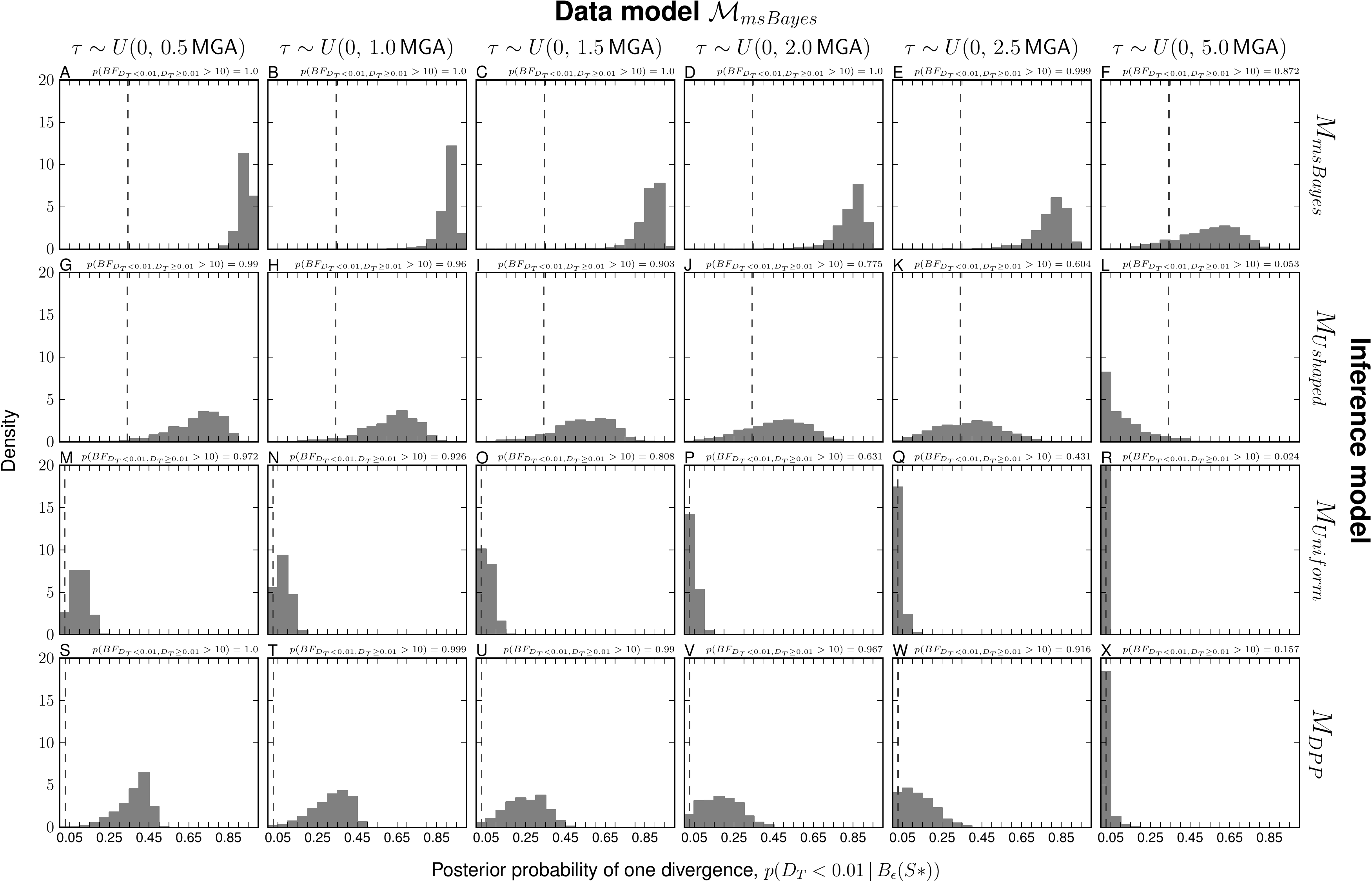}{
    \powerSupportComment{\powerSeriesOld}
    \powerProbComment{$p(\divTimeDispersion < 0.01 | \ssSpace)$}{\powerSeriesOld}
    \timeConversionComment
}{figPowerOmegaProbOld}

\siSidewaysFigure{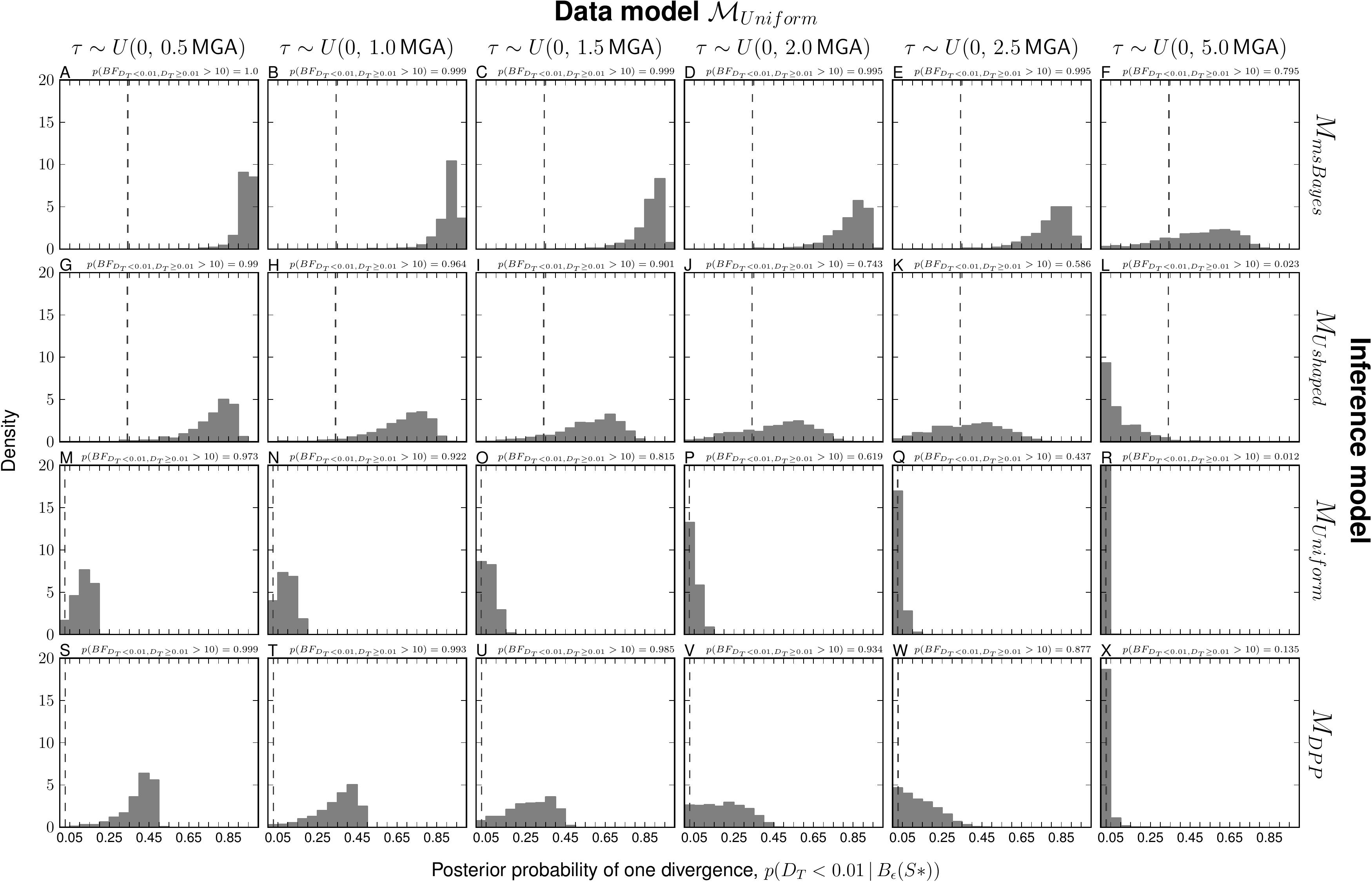}{
    \powerSupportComment{\powerSeriesUniform}
    \powerProbComment{$p(\divTimeDispersion < 0.01 | \ssSpace)$}{\powerSeriesUniform}
    \timeConversionComment
}{figPowerOmegaProbUniform}

\siSidewaysFigure{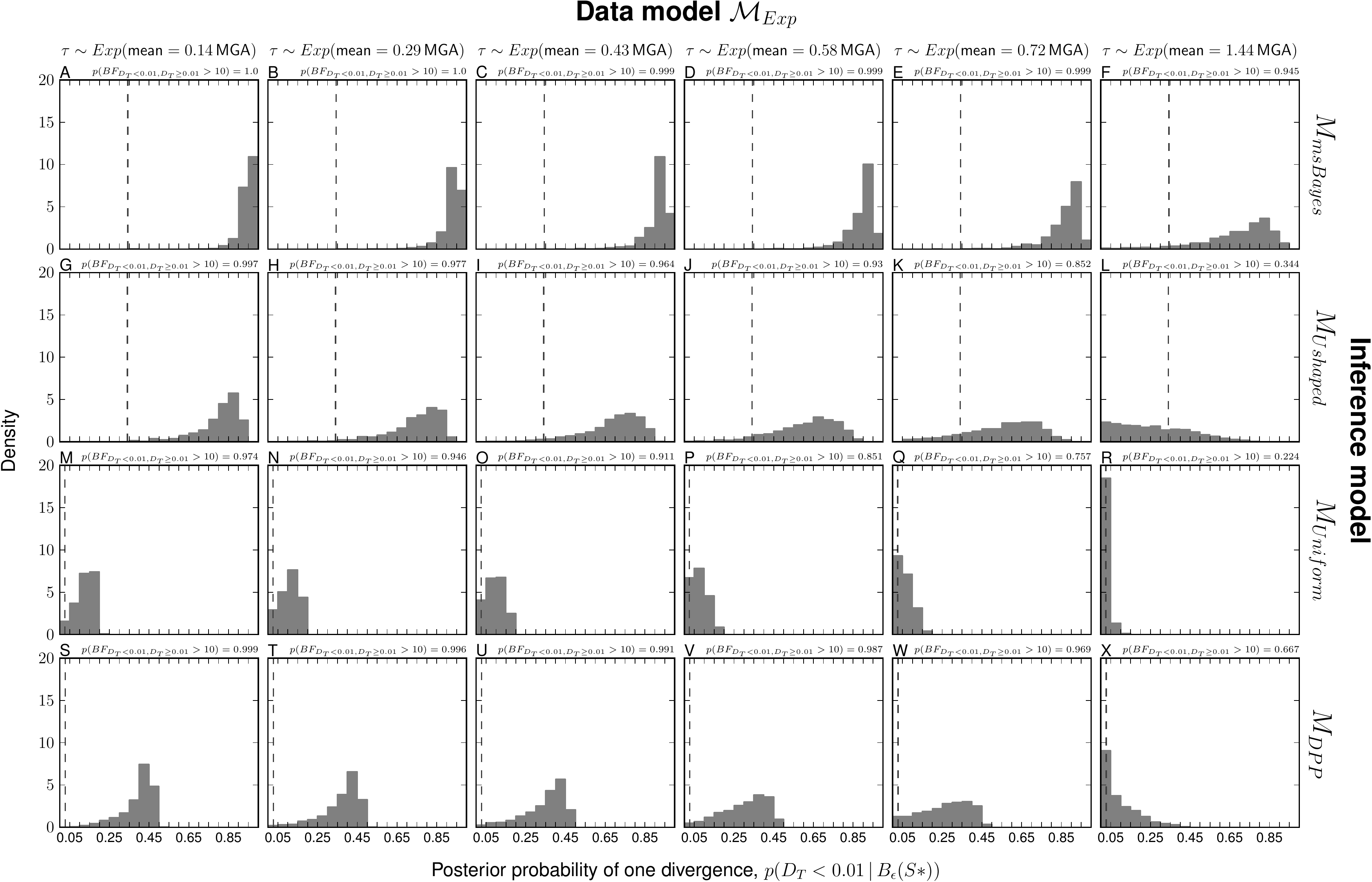}{
    \powerSupportComment{\powerSeriesExp}
    \powerProbComment{$p(\divTimeDispersion < 0.01 | \ssSpace)$}{\powerSeriesExp}
    \timeConversionComment
}{figPowerOmegaProbExp}

\siEightFigure{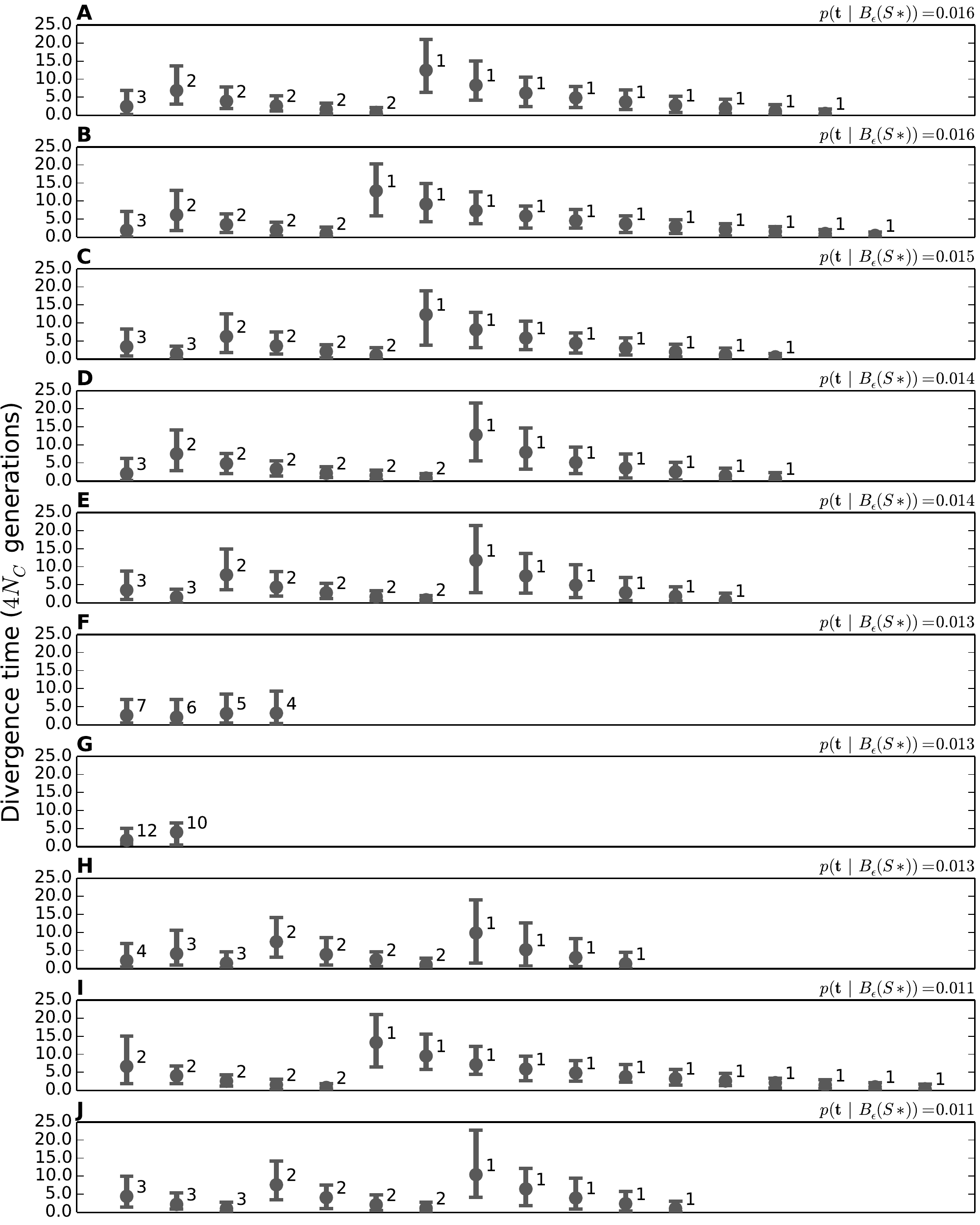}{
    The divergence-model results when the 22 pairs of taxa from the Philippines
    are analyzed under the \empModelDPP model (Table~\ref{tabEmpiricalModels}).
    The 10 unordered divergence models with highest posterior probability
    ($p(\divTimeIndexVector \given \ssSpace)$) are shown, where the numbers
    indicate the inferred number of taxon pairs that diverged at each event.
    The times indicate the posterior median and 95\% highest posterior density
    (HPD) interval conditional on each divergence model.
    For each model, times are summarized across posterior samples by the number
    of taxon pairs associated with each divergence. For models in which there
    are multiple divergence events with the same number of taxon pairs, the
    events are sorted by time to summarize the divergence times in a consistant
    way.
}{figDivModelsDPP}

\siEightFigure{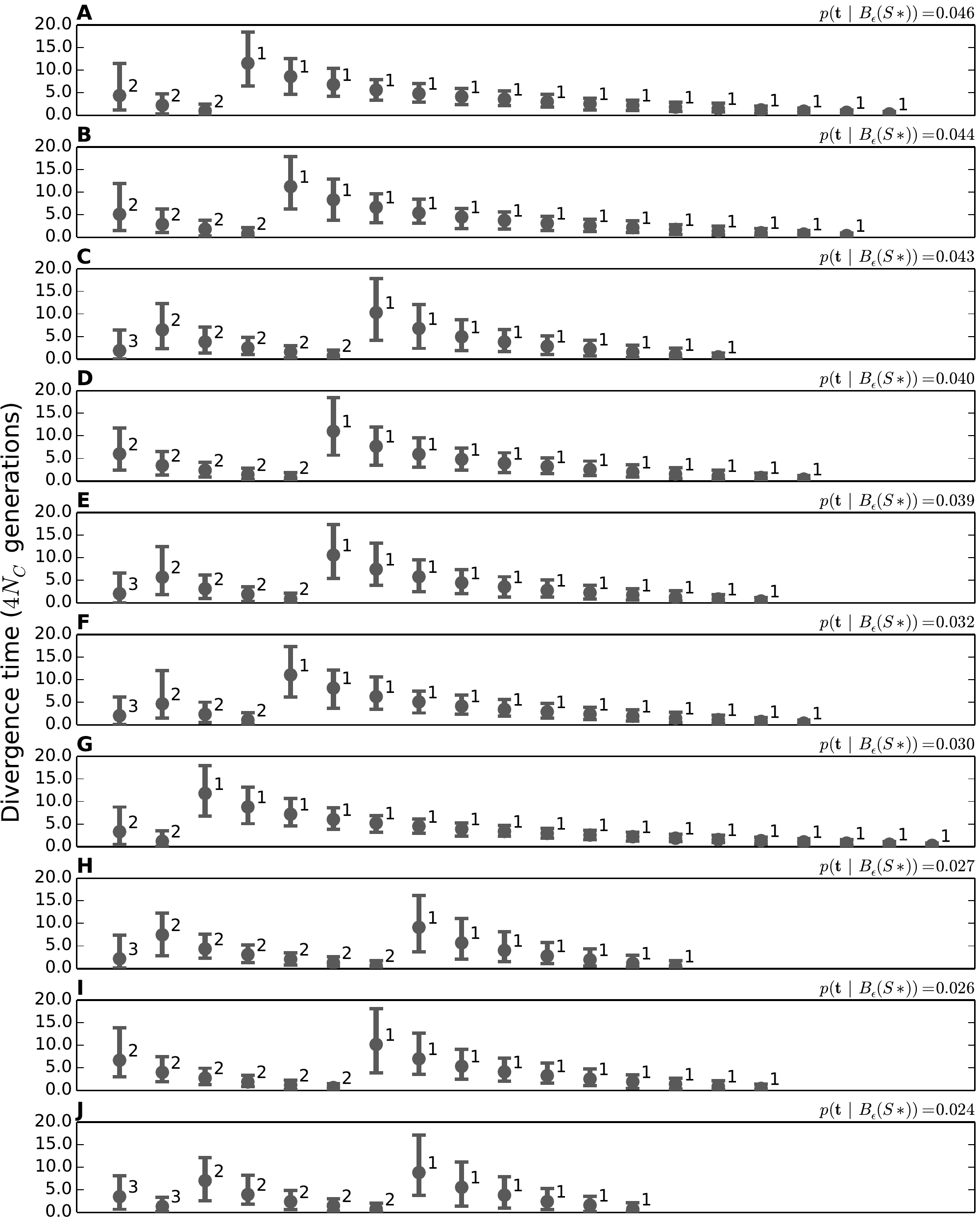}{
    The divergence-model results when the 22 pairs of taxa from the Philippines
    are analyzed under the \empModelDPPInform model
    (Table~\ref{tabEmpiricalModels}).
    The 10 unordered divergence models with highest posterior probability
    ($p(\divTimeIndexVector \given \ssSpace)$) are shown, where the numbers
    indicate the inferred number of taxon pairs that diverged at each event.
    The times indicate the posterior median and 95\% highest posterior density
    (HPD) interval conditional on each divergence model.
    For each model, times are summarized across posterior samples by the number
    of taxon pairs associated with each divergence. For models in which there
    are multiple divergence events with the same number of taxon pairs, the
    events are sorted by time to summarize the divergence times in a consistant
    way.
}{figDivModelsDPPInform}

\siEightFigure{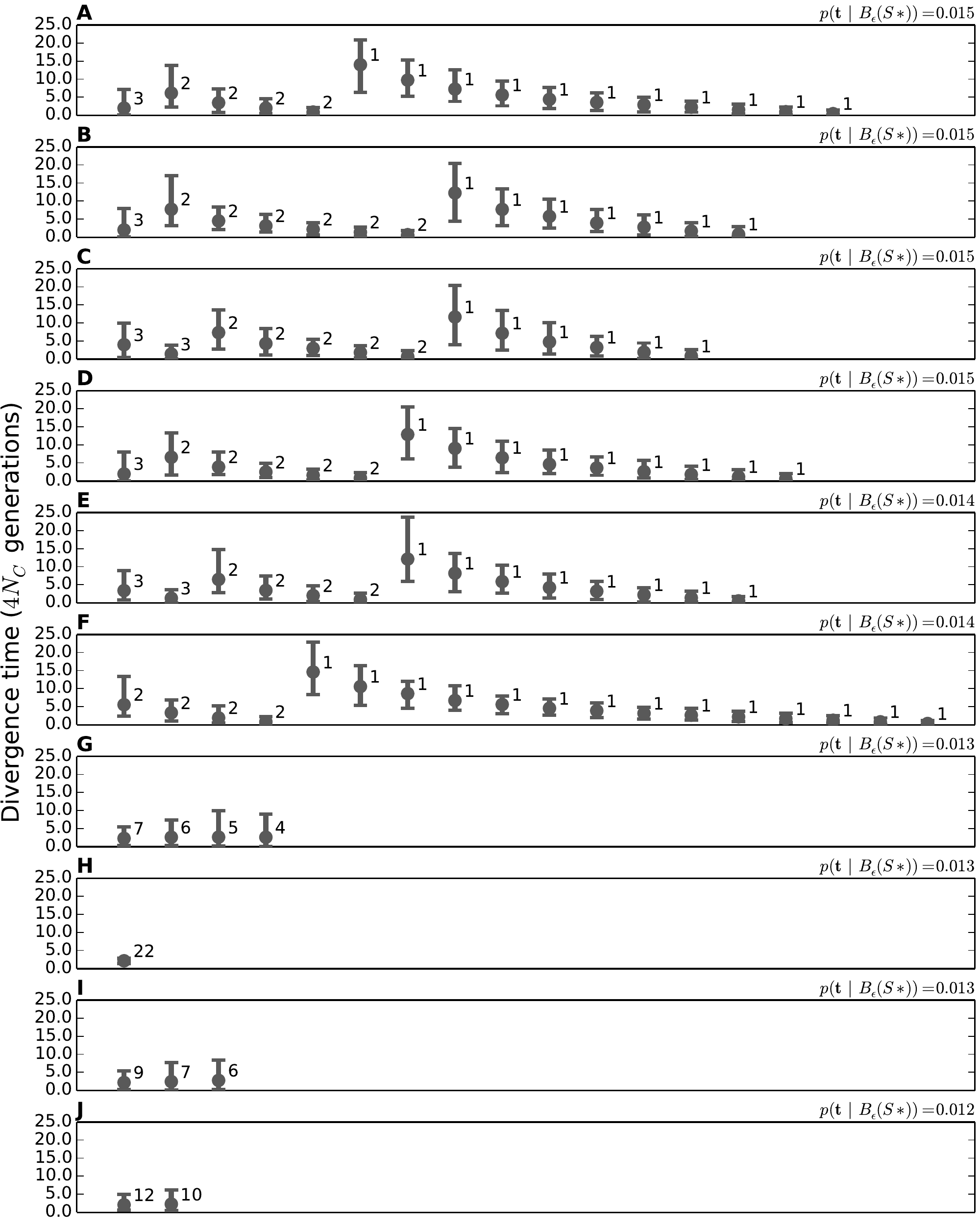}{
    The divergence-model results when the 22 pairs of taxa from the Philippines
    are analyzed under the \empModelDPPSimple model
    (Table~\ref{tabEmpiricalModels}).
    The 10 unordered divergence models with highest posterior probability
    ($p(\divTimeIndexVector \given \ssSpace)$) are shown, where the numbers
    indicate the inferred number of taxon pairs that diverged at each event.
    The times indicate the posterior median and 95\% highest posterior density
    (HPD) interval conditional on each divergence model.
    For each model, times are summarized across posterior samples by the number
    of taxon pairs associated with each divergence. For models in which there
    are multiple divergence events with the same number of taxon pairs, the
    events are sorted by time to summarize the divergence times in a consistant
    way.
}{figDivModelsDPPSimple}

\siEightFigure{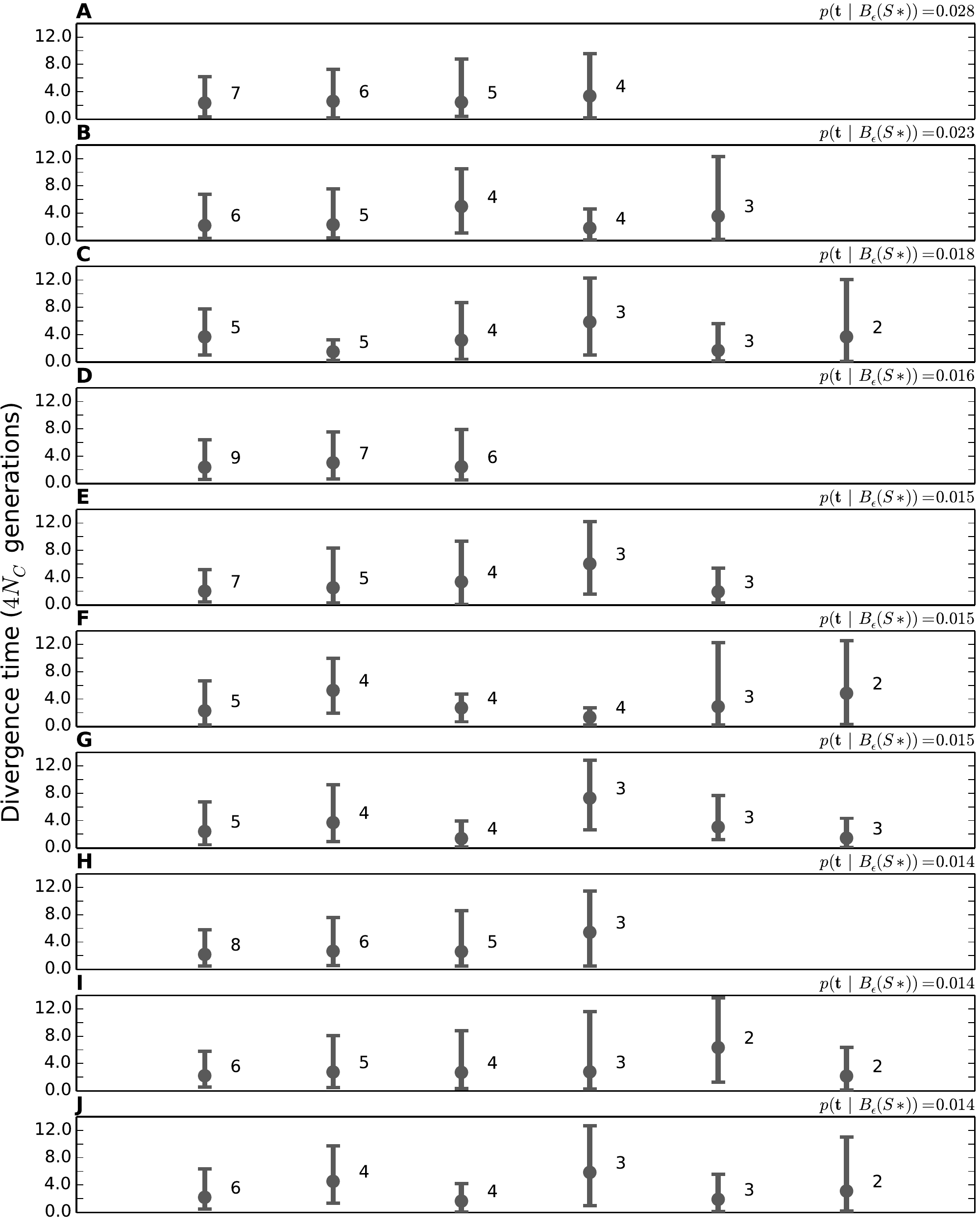}{
    The divergence-model results when the 22 pairs of taxa from the Philippines
    are analyzed under the \empModelUniform model
    (Table~\ref{tabEmpiricalModels}).
    The 10 unordered divergence models with highest posterior probability
    ($p(\divTimeIndexVector \given \ssSpace)$) are shown, where the numbers
    indicate the inferred number of taxon pairs that diverged at each event.
    The times indicate the posterior median and 95\% highest posterior density
    (HPD) interval conditional on each divergence model.
    For each model, times are summarized across posterior samples by the number
    of taxon pairs associated with each divergence. For models in which there
    are multiple divergence events with the same number of taxon pairs, the
    events are sorted by time to summarize the divergence times in a consistant
    way.
}{figDivModelsUniform}

\siEightFigure{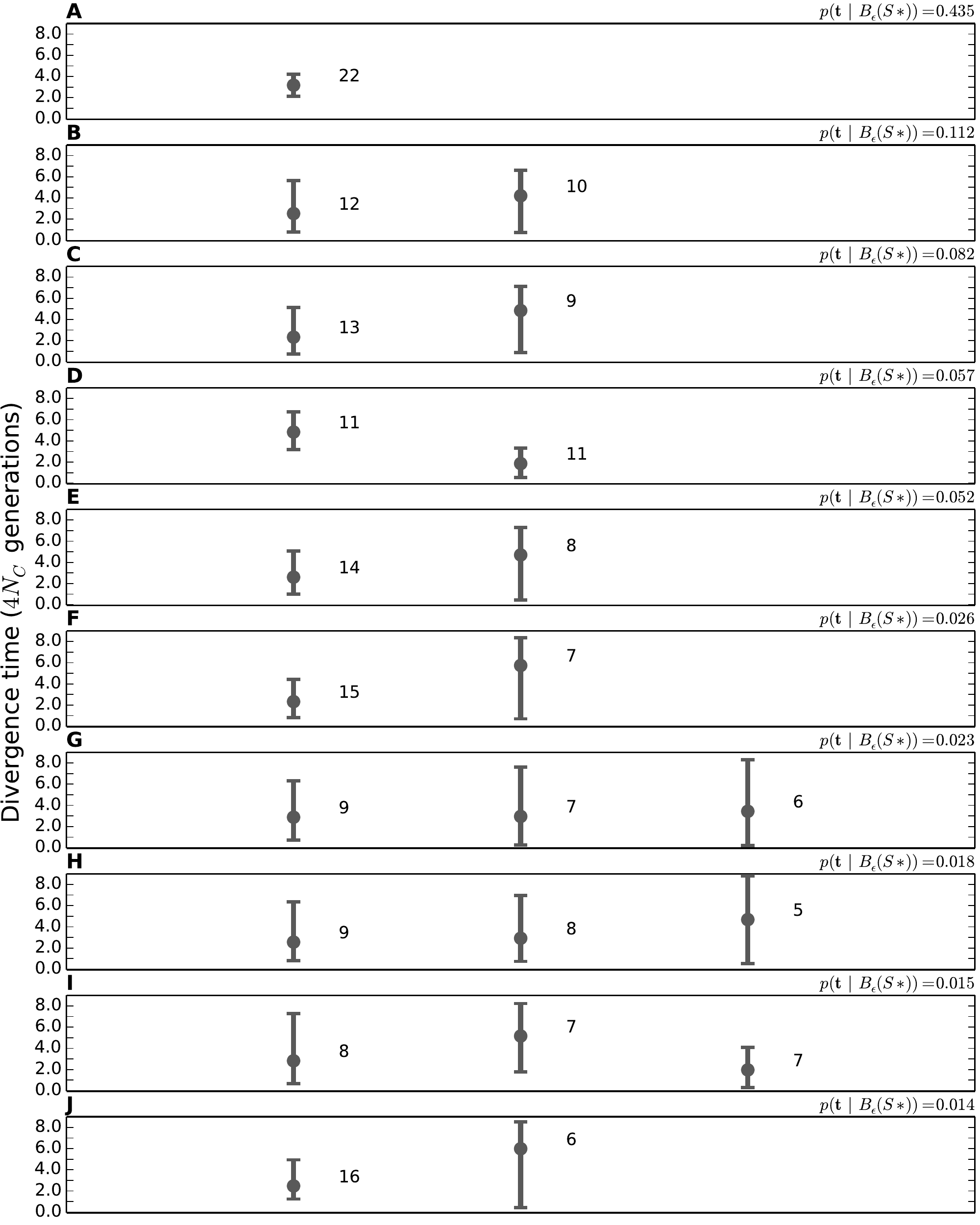}{
    The divergence-model results when the 22 pairs of taxa from the Philippines
    are analyzed under the \empModelOld model
    (Table~\ref{tabEmpiricalModels}).
    The 10 unordered divergence models with highest posterior probability
    ($p(\divTimeIndexVector \given \ssSpace)$) are shown, where the numbers
    indicate the inferred number of taxon pairs that diverged at each event.
    The times indicate the posterior median and 95\% highest posterior density
    (HPD) interval conditional on each divergence model.
    For each model, times are summarized across posterior samples by the number
    of taxon pairs associated with each divergence. For models in which there
    are multiple divergence events with the same number of taxon pairs, the
    events are sorted by time to summarize the divergence times in a consistant
    way.
}{figDivModelsOld}

\siEightFigure{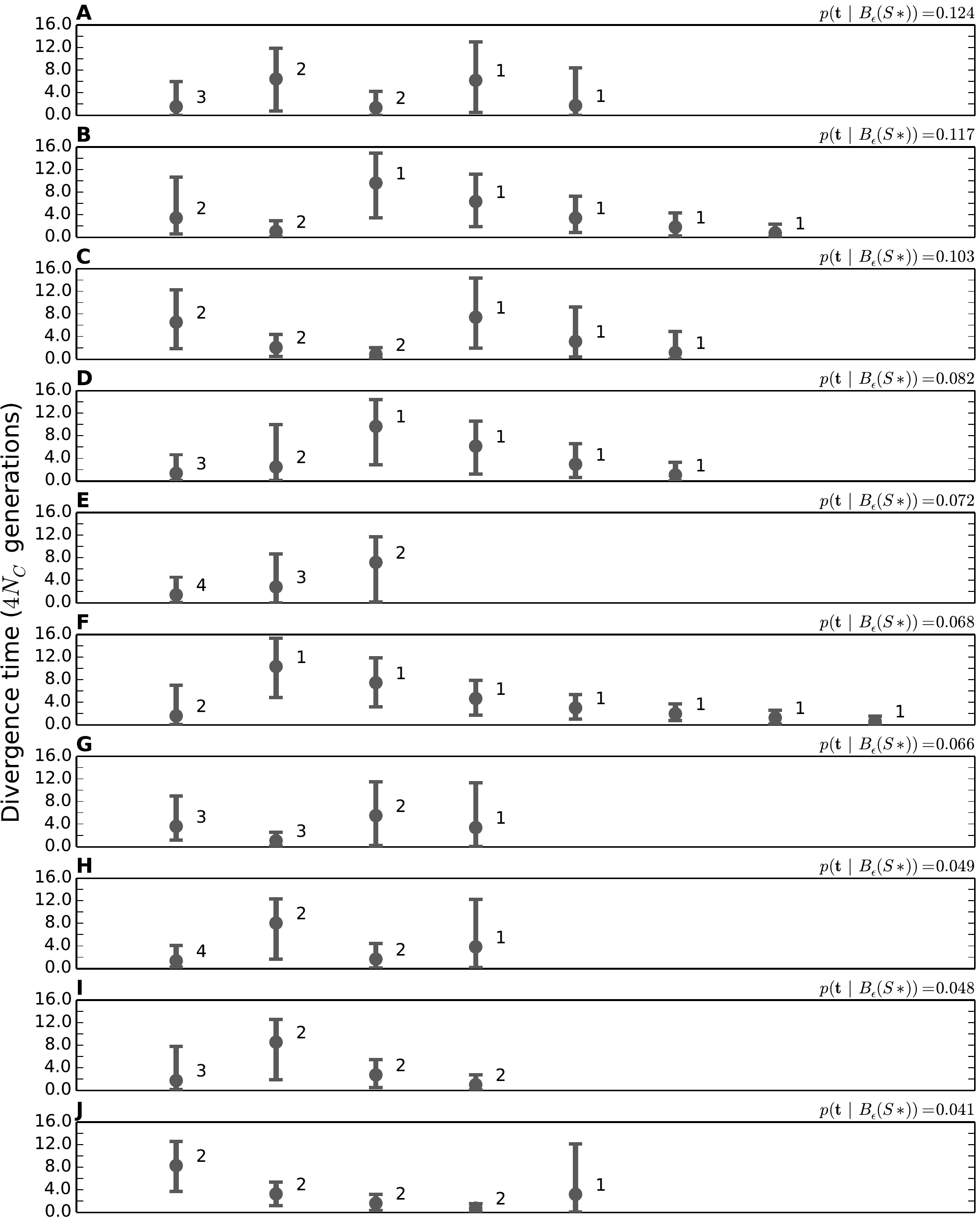}{
    The divergence-model results when the 9 pairs of taxa from the Islands of
    Negros and Panay are analyzed under the \npModelDPP model sampling over
    unordered models of divergence.
    (Table~\ref{tabEmpiricalModels}).
    The 10 unordered divergence models with highest posterior probability
    ($p(\divTimeIndexVector \given \ssSpace)$) are shown, where the numbers
    indicate the inferred number of taxon pairs that diverged at each event.
    The times indicate the posterior median and 95\% highest posterior density
    (HPD) interval conditional on each divergence model.
    For each model, times are summarized across posterior samples by the number
    of taxon pairs associated with each divergence. For models in which there
    are multiple divergence events with the same number of taxon pairs, the
    events are sorted by time to summarize the divergence times in a consistant
    way.
}{figDivModelsNP}

\siEightFigure{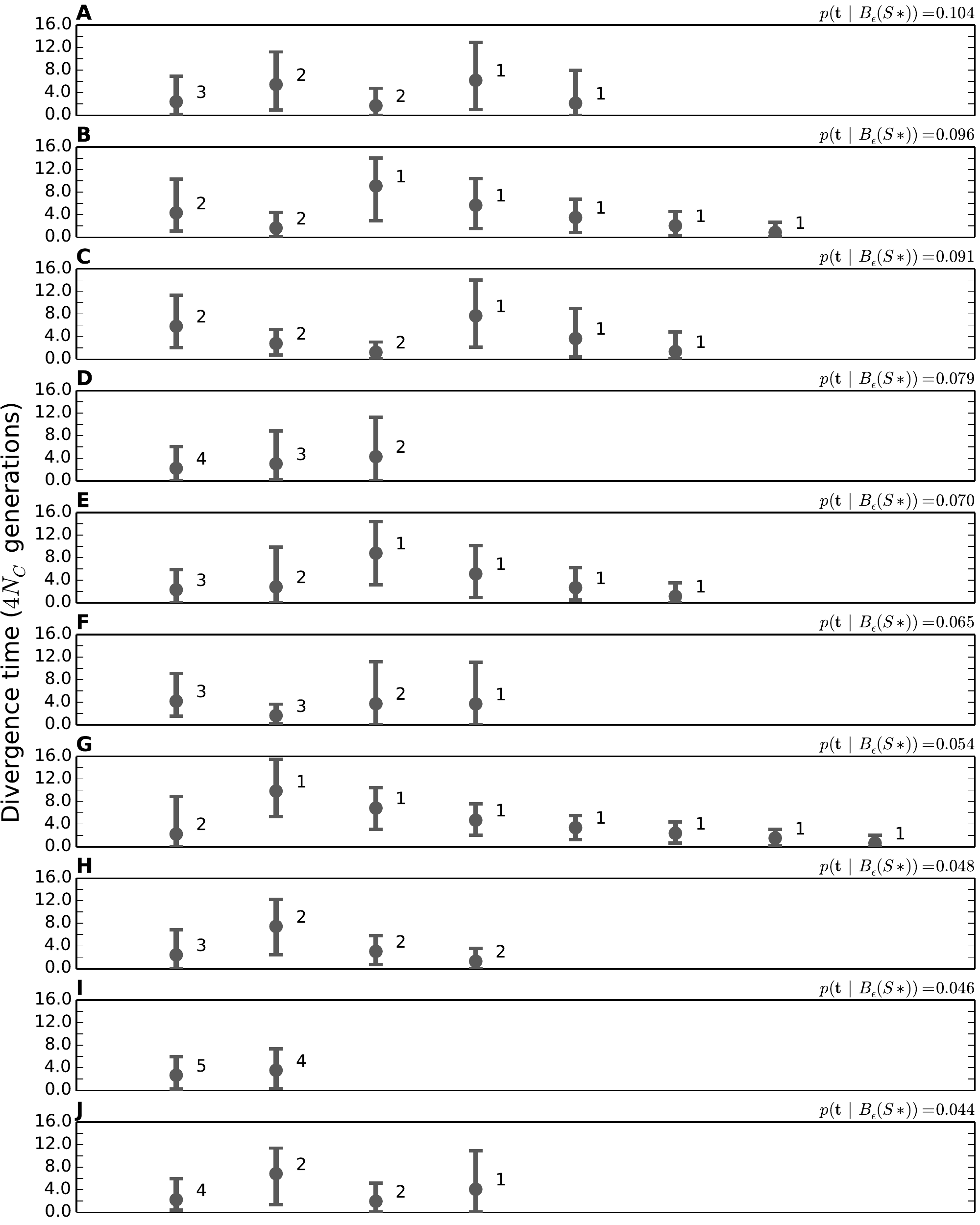}{
    The divergence-model results when the 9 pairs of taxa from the Islands of
    Negros and Panay are analyzed under the \npModelDPPOrdered model sampling
    over ordered models of divergence.
    (Table~\ref{tabEmpiricalModels}).
    The posterior sample of divergence models were summarized while ignoring
    the identity of the taxon pairs in order to compare the results of the
    \npModelDPP model.
    The 10 unordered divergence models with highest posterior probability
    ($p(\divTimeIndexVector \given \ssSpace)$) are shown, where the numbers
    indicate the inferred number of taxon pairs that diverged at each event.
    The times indicate the posterior median and 95\% highest posterior density
    (HPD) interval conditional on each divergence model.
    For each model, times are summarized across posterior samples by the number
    of taxon pairs associated with each divergence. For models in which there
    are multiple divergence events with the same number of taxon pairs, the
    events are sorted by time to summarize the divergence times in a consistant
    way.
}{figDivModelsNPOrdered}

\siFigure{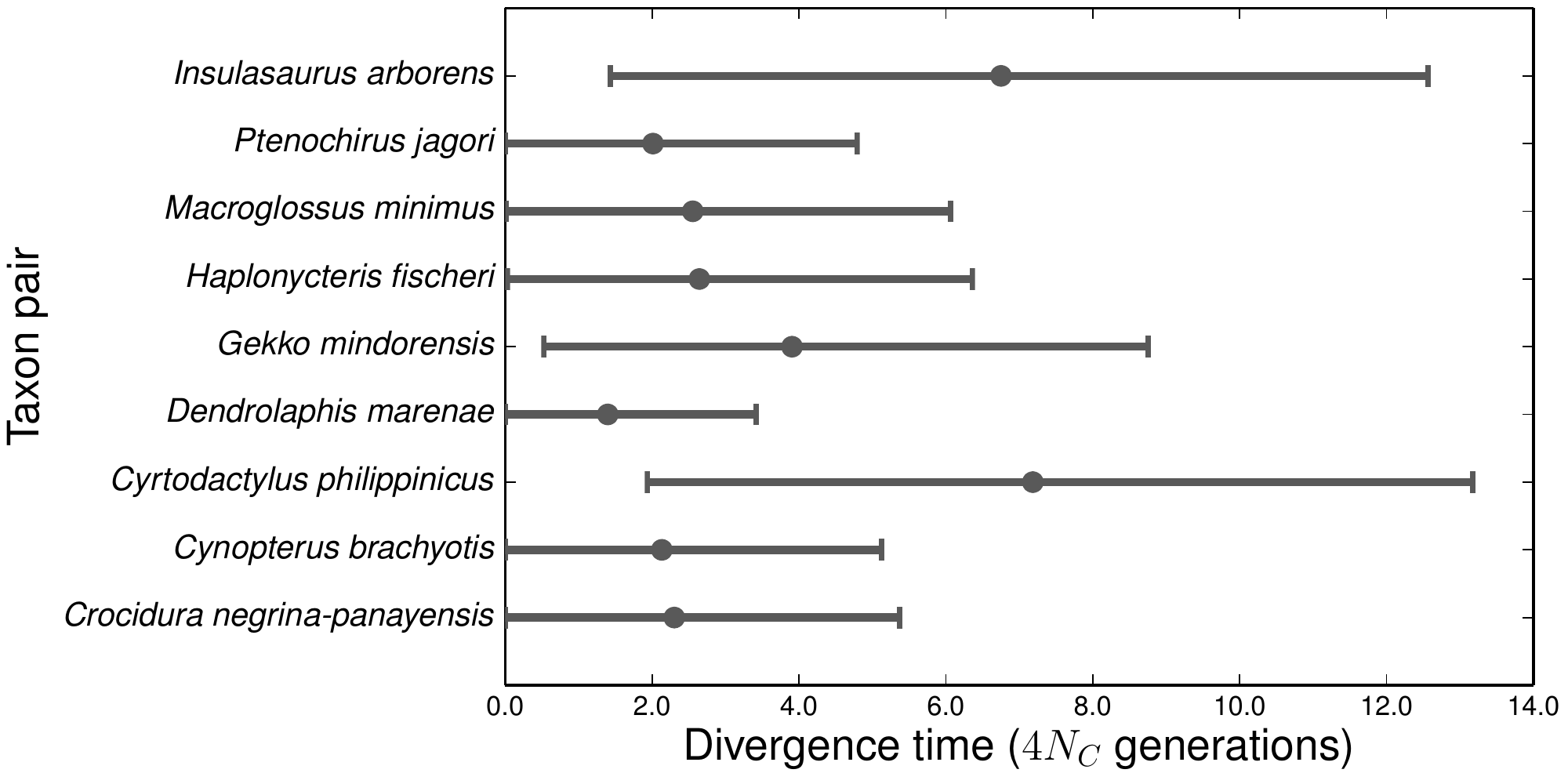}{
    The marginal divergence-time results when the 9 pairs of taxa from the
    Islands of Negros and Panay are analyzed under the \npModelDPPOrdered model
    that samples over ordered models of divergence
    (Table~\ref{tabEmpiricalModels}).
    The times indicate the posterior median and 95\% highest posterior density
    (HPD) interval.
}{figMarginalTimes}

\end{document}